\documentclass[fleqn,usenatbib]{mnras}

\usepackage{newtxtext,newtxmath}

\usepackage[T1]{fontenc}

\DeclareRobustCommand{\VAN}[3]{#2}
\let\VANthebibliography\thebibliography
\def\thebibliography{\DeclareRobustCommand{\VAN}[3]{##3}\VANthebibliography}

\usepackage{graphicx}	
\usepackage{amsmath}	
\usepackage{subfig}  
\usepackage{lineno}
\usepackage{xcolor}
\usepackage{orcidlink} 

\title[Stellar mass growth in massive galaxy clusters]{The Atacama Cosmology Telescope: stellar mass growth in massive galaxy clusters from DR5 over the past 7 billion years}

\author[D. C. Ragavan et al.]{Damien C. Ragavan$^{\orcidlink{0000-0003-0670-8387}}$,$^{1}$\thanks{E-mail: damiencole02@gmail.com}
Unnikrishnan Sureshkumar$^{\orcidlink{0000-0002-2210-0681}}$,$^{1,2}$
Matt Hilton$^{\orcidlink{0000-0002-8490-8117}}$,$^{1,3}$
John P. Hughes$^{\orcidlink{0000-0002-8816-6800}}$,$^{5}$
\newauthor
Kavilan Moodley$^{\orcidlink{0000-0001-6606-7142}}$,$^{3,4}$
Tony Mroczkowski$^{\orcidlink{0000-0003-3816-5372}}$,$^{6}$
Bruce Partridge$^{\orcidlink{0000-0001-6541-9265}}$,$^{7}$ 
Maria Salatino$^{\orcidlink{0000-0003-4006-1134}}$,$^{8}$
Cristóbal Sifón$^{\orcidlink{0000-0002-8149-1352}}$,$^{9}$
\newauthor
Eve M. Vavagiakis$^{\orcidlink{0000-0002-2105-7589}}$$^{10,11}$ and
Edward J. Wollack$^{\orcidlink{0000-0002-7567-4451}}$$^{12}$
\\
\\
$^{1}$Wits Centre for Astrophysics, School of Physics, University of the Witwatersrand, Private Bag 3, 2050, Johannesburg, South Africa\\
$^{2}$ National Centre for Nuclear Research, ul. Pasteura 7, 02-093 Warsaw, Poland\\
$^{3}$ Astrophysics Research Centre, University of KwaZulu-Natal, Westville Campus, Durban 4041, South Africa\\
$^{4}$ School of Mathematics, Statistics \& Computer Science, University of KwaZulu-Natal, Westville Campus, Durban 4041, South Africa\\
$^{5}$Department of Physics and Astronomy, Rutgers, the State University of New Jersey, 136 Frelinghuysen Road, Piscataway, NJ 08854-8019, USA\\
$^{6}$Institute of Space Sciences (ICE-CSIC), Carrer de Can Magrans, s/n, 08193 Cerdanyola del Vall\`es, Barcelona, Spain\\
$^{7}$Department of Physics and Astronomy, Haverford College, Haverford, PA, 19041, USA\\
$^{8}$Kavli Institute for Particle Astrophysics and Cosmology, Stanford University, Stanford, CA, 94035, USA\\
$^{9}$Instituto de F\'isica, Pontificia Universidad Cat\'olica de Valpara\'iso, Casilla 4059, Valpara\'iso, Chile\\
$^{10}$Department of Physics, Duke University, Durham, NC, 27704, USA\\
$^{11}$Department of Physics, Cornell University, Ithaca, NY, 14580, USA\\
$^{12}$NASA/Goddard Space Flight Center, Greenbelt, MD, 20771, USA
}

\date{Accepted XXX. Received YYY; in original form ZZZ}

\pubyear{2026}

\begin{document}
\label{firstpage}
\pagerange{\pageref{firstpage}--\pageref{lastpage}}
\maketitle

\begin{abstract}
We probe the stellar mass growth in a sample of 568 Sunyaev-Zel'dovich (SZ) selected galaxy clusters with masses greater than $2.9 \times 10^{14} \mathrm{M_{\odot}}$ and redshifts in the range $0.2 < z < 0.8$, drawn from the fifth data release of the Atacama Cosmology Telescope (ACT DR5). By utilising deep photometry from the tenth data release of the Dark Energy Camera Legacy Survey (DECaLS DR10), we construct redshift- and cluster mass-binned composite cluster stellar mass functions (SMFs), down to  $M_* = 10^{9.5} \mathrm{M_{\odot}}$. This work presents the first analysis of the cluster SMF for this cluster sample at this epoch. We find that the characteristic stellar mass ($M^*$) of the cluster SMF evolves marginally from $0.55 \leq z < 0.8$, with most of the measurable growth occurring at $ 0.2 < z < 0.55$. This suggests that most of the massive galaxy population in clusters ($M_* \gtrsim 10^{10.75} \mathrm{M_{\odot}}$) is largely established by $z \sim 0.8$, with subsequent evolution driven by late-time assembly processes. The low-mass slope ($\alpha$) of the composite cluster SMF is flat at high-$z$ ($z \sim 0.8$) but steepens at $z < 0.55$, suggesting an abundance of massive galaxies in high-$z$ clusters compared to low-$z$ clusters. We measure the evolution of cluster stellar mass fractions contained within galaxies with $M_* > 10^{9.5} \mathrm{M_{\odot}}$ between $ 0.2 < z < 0.8$, and find evidence of significant growth, by a factor of $2.5$, after accounting for the growth in cluster halo mass over this epoch.

\end{abstract}

\begin{keywords}
galaxies: clusters: general -- galaxies: evolution  -- galaxies: luminosity function, mass function
\end{keywords}



\section{Introduction}
\label{sec:introduction}

An important question in galaxy formation theory is how galaxies build up their stellar mass over cosmic time. In the current Lambda cold dark matter ($\Lambda$CDM) cosmological paradigm, hierarchical structure formation predicts that smaller dark matter haloes form first via gravitational collapse and then grow through accretion and merger events, to form progressively larger haloes (\citealp{Press&Schechter1974}; \citealp{White&Rees1978}; \citealp{Springel2005}). Galaxies form within these haloes from the condensation of baryonic gas. The stellar mass within galaxies is initially accumulated through in-situ star formation, and then later through accretion and mergers. However, observational evidence for an "anti-hierarchical" or downsizing scenario \citep{Cowie1996} for the growth of the stellar baryonic component of the haloes, in which more massive galaxies formed the bulk of their stars earlier and over a shorter timescale compared to low-mass galaxies, has become increasingly prevalent (e.g., \citealp{Brammer2011}; \citealp{Carnall2019}; \citealp{Borghi_2022}; \citealp{Cedres2025}; \citealp{Zhou2025}). This observed scenario does not contradict the hierarchical growth of dark matter haloes, but rather highlights the intricate relationship between galaxy evolution and halo assembly.

A valuable observational tool for addressing the question of how galaxies assemble mass is the galaxy stellar mass function (SMF), which describes the stellar mass distribution of galaxies. Precise measurements of the galaxy SMF help us understand the crucial physical processes that shape stellar mass growth and identify the types of galaxies affected by these mechanisms. Early studies on the SMF were aimed at probing the local Universe (e.g., \citealp{Cole20012df}; \citealp{Bell2003optical}; \citealp{Blanton_2003}). Using stellar masses estimated from galaxy photometry, some of these studies were able to compute the galaxy SMF down to $M_* = 10^{8} \mathrm{M_{\odot}}$ (\citealp{Li&White2009}; \citealp{Baldry2012}), which has provided good constraints on the evolution of the local galaxy SMF with cosmic time. Many of these early studies focused on the importance of the SMF for galaxies in the field rather than in galaxy clusters (e.g., \citealp{Drory_2005}; \citealp{Bundy2006}; \citealp{Baldry2008}). Some studies observed a moderate redshift evolution of the field galaxy SMF for galaxies with $M_* \geq 10^{11} \rm{M_{\odot}}$ since $z = 1$ (e.g., \citealp{Fontana2004}; \citealp{Bundy2006}). However, studies such as \cite{Fontana2006} found a differential, mass-dependent evolution of the galaxy SMF since $z \simeq 1.0 -1.5$, with low-mass galaxies evolving faster than massive galaxies.  

Recent advances over the past couple of decades have allowed for the construction of deeper optical and near-infrared (NIR) extra-galactic surveys, such as the Cosmic Evolution Survey (COSMOS; \citealp{Scoville2007}), the Dark Energy Spectroscopic Instrument (DESI) Legacy Surveys \citep{Dey2019}, JWST Advanced Deep Extragalactic Survey (JADES; \citealp{Eisenstein2026}),  and many more. These surveys have made it possible to probe and accurately measure the SMF at high redshifts. This has resulted in a large volume of recent studies which have provided measurements of the galaxy SMF up to $z \approx 8.5$, mostly using photometric redshifts (e.g., \citealp{vanderBurg2018, vanderBurg2020}; \citealp{McLeod2021}; \citealp{COSMOS2020}; \citealp{Navarro-Carrera_2024}). The advent of the Euclid telescope \citep{Mellier2025} and the Vera C. Rubin Observatory \citep{Ivezic2019} will revolutionise SMF measurements by increasing precision thanks to accurate photometry, and enhancing completeness due to deep coverage.

In the context of galaxy clusters, which are the most massive virialised systems in the Universe, the SMF plays an important role in understanding the distribution of stellar mass within clusters and the role of the cluster environment in galaxy evolution. Physical processes, such as feedback from active galactic nuclei (AGN) (\citealp{Binney&Tabor1995}; \citealp{SpringelDiMatteoHernquist2005}; \citealp{Bower2006}), ram pressure stripping (\citealp{Gunn&Gott1972}; \citealp{Bekki2009}) and strangulation or starvation (\citealp{Larson1980}; \citealp{Kawata2008}), within high-density regions such as galaxy clusters and groups, can often lead to the environmental quenching of star formation and an excess in the number of quiescent galaxies (e.g., \citealp{Lewis2002}; \citealp{Blanton2005}; \citealp{Wetzel2012}; \citealp{Cleland2021}). These physical mechanisms ultimately shape the evolution of galaxies.

For this reason, numerous studies have been devoted to studying the environmental dependence of the SMF (e.g., \citealp{Vulcani2013}; \citealp{Mortlock2014}; \citealp{vanderBurg2018,vanderBurg2020}). For instance, \cite{Annunziatella2014} explored the environmental dependence of the SMF and found that the cluster galaxy population is dominated by quiescent galaxies (for stellar masses $M_{*} \geq 10 ^{10} \mathrm{M_{\odot}}$) in their study of the massive MACS J1206.2-0847 galaxy cluster at $z = 0.44$. Similarly, \cite{vanderBurg2013} measured the SMF of 10 clusters from the Gemini Cluster Astrophysics Spectroscopic Survey (GCLASS) at $z \sim 1$ and found a higher fraction of quiescent galaxies (for $M_* > 10^{10} \mathrm{M_{\odot}}$) in these systems compared to the field. More recently, \cite{vanderBurg2020} investigated the environmental dependence of the quenching process in 11 galaxy clusters from $1.0 < z < 1.4$, drawn from the Gemini Observations of Galaxies in Rich Early Environments (GOGREEN; \citealp{Balogh2017}) survey, by measuring their SMFs. They found that the quenched fraction increases strongly with stellar mass, and environmental effects on galaxy quenching in high-mass galaxies are already pronounced at early cosmic times ($z >1$). Recent interest has also been focused on studying the cluster SMF at intermediate redshifts ($z <1$). For instance, \cite{vanderBurg2018} performed a study of the cluster SMF for 21 massive Planck-detected clusters at $0.5 < z < 0.7$. They found that the SMF shape of quiescent galaxies differs significantly between cluster and field environments, and that the fraction of passive galaxies is higher in clusters. Recently, \cite{Shaikh2025} examined the SMF dependence on environmental factors using multi-band photometry on a large sample of 105 galaxy clusters from the eROSITA Final Equatorial Depth Surveys (eFEDS; \citealp{Liu2022}), at $0.385 < z < 0.8$. They found that the cluster environment has a strong effect on the galaxy stellar mass distribution, particularly in the inner regions of massive clusters.

In this paper, we examine the galaxy stellar mass assembly within 568 galaxy clusters from $0.2 < z < 0.8$, detected by the Atacama Cosmology Telescope (ACT; \citealp{Marriage2009}). ACT was a 6-m millimetre-wave telescope (\citealp{Henderson2016}; \citealp{Thornton_2016}) located on Cerro Toco in the Atacama Desert in Chile. It was optimised for measuring the sky-high multipole ($l > 1000$) angular power spectra and galaxy clusters using the thermal Sunyaev-Zel'dovich effect (SZ; \citealp{Sunyaev1970}). 

The SZ effect arises from the inverse Compton scattering of photons from the cosmic microwave background (CMB) when they interact with electrons in the hot gas (i.e. the intracluster medium, or ICM) in galaxy clusters (\citealp{Birkinshaw1999}; \citealp{Carlstrom2002}; \citealp{Mroczkowski2019}). As a result, the CMB blackbody spectrum is shifted in the direction of a cluster. The SZ signal of clusters depends largely on mass and not on redshift, which allows for the construction of mass-limited samples, such as the ACT DR5 cluster catalogue \citep{Hilton_2021}. 

Using optical data from the Dark Energy Camera Legacy Survey (DECaLS; \citealp{Dey2019}), we construct redshift and cluster mass binned stacked/composite cluster SMFs. Additionally, we investigate the stellar content in cluster galaxies by considering the scaling relationship between cluster stellar mass and halo mass, and the evolution of the stellar mass fractions within these clusters. A study to investigate the stellar content of brightest cluster galaxies and intracluster light in a sample of ACT-DES clusters at $0.2 < z < 0.8$ was performed by \cite{Golden-Marx2023}. Our cluster selection differs from theirs, resulting in different sample sizes. Unlike our study, they do not perform SMF measurements for their cluster sample, nor do they examine the total stellar content in the DR5 clusters. Hence, this type of study has never been performed on this specific ACT cluster sample in the redshift range of $0.2 < z < 0.8$.

The structure of this paper is as follows. In Section~\ref{sec:data}, we describe our galaxy cluster sample and the photometric data. In Section~\ref{sec:SampleSelection}, we describe the selection methods used to define our cluster galaxy sample, and the methods to determine galaxy stellar mass and completeness. Section~\ref{sec:method} presents the stacking method used to construct the redshift- and cluster mass-binned cluster SMFs. In Section~\ref{sec:Results}, we present the results and details of how we model the composite cluster SMFs. We also examine the relationship between cluster stellar mass and cluster mass in this section. In Section~\ref{sec:Discussion}, we discuss the implications of our findings in the context of galaxy evolution and with respect to previous SMF studies, and present the evolution of the cluster galaxy stellar mass. We finally present our summary and conclusions in Section~\ref{sec:conclusion}.

Throughout this paper, we represent stellar mass with the symbol $M_{*}$ and the characteristic stellar mass in log scale as $M^* = \log_{10}[M^*_*/\mathrm{M_{\odot}]}$, unless otherwise stated. We also assume a $\Lambda$CDM cosmology with $H_{\rm 0} =~70~\mathrm{km \, s}^{-1} \mathrm{Mpc}^{-1}$, $\Omega_{\mathrm{\Lambda}}$ = 0.7 and $\Omega_{\mathrm{M}}$ = 0.3.

\section{DATA}
\label{sec:data}

\subsection{ACT DR5 cluster sample}
\label{subsec:actdr5} 

In this work, we use galaxy clusters from the fifth data release of ACT \citep[ACT DR5;][]{Hilton_2021}. The DR5 catalogue consists of 4,195 galaxy clusters and is one of the largest SZ selected cluster samples assembled to date. DR5 covers the redshift range $0.04 < z < 1.91$ and includes 868 newly discovered systems, all within a sky coverage of $13,211$ deg$^{2}$. \cite{Hilton_2021} obtained cluster mass estimates derived from the SZ signal for each cluster, using the mass scaling relation from \cite{Arnaud2010}. The total DR5 cluster sample has a $90$ per cent mass completeness limit of $M_{\mathrm{500c}} > 3.8 \times 10^{14} \mathrm{M_{\odot}}$ (at $z = 0.5$) for clusters detected with a signal-to-noise ratio (S/N) $> 5$.

\begin{figure}
	\includegraphics[width=\columnwidth]{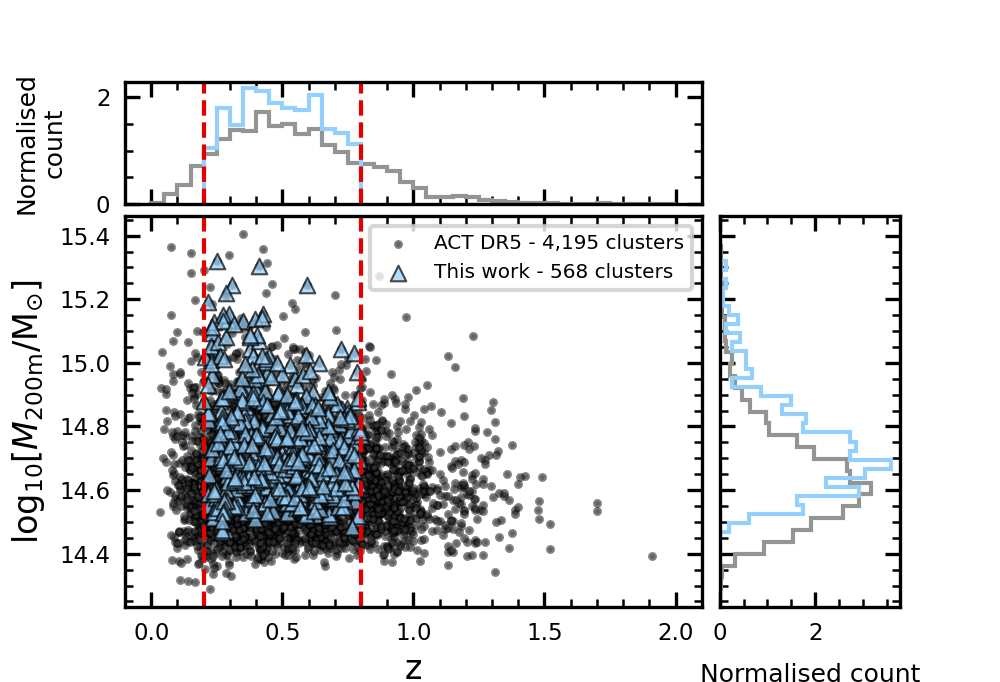}
    \caption{The distribution of log cluster mass ($\log _{10}[M_{\mathrm{200m}}/\mathrm{M_{\odot}}]$) with redshift for the ACT DR5 cluster sample (shown as grey circles), and the cluster sample selected in this work (shown as light blue triangles). Our cluster sample has been selected based on the optical depth in DECaLS DR10 (see Section \ref{subsec:Completeness}). The redshift (\textit{Top panel}) and cluster mass (\textit{Right panel}) normalised distributions for each sample are also displayed.}
    \label{fig:CMvsz}
\end{figure}

We select galaxy clusters from DR5 with S/N $> 5$ measured at a fixed $2.4'$ filter scale (S/N$_{2.4}$ in \citealt{Hilton_2021}), in the redshift range $0.2 < z < 0.8$. The selection of this redshift range is based on the available stellar mass depth we can probe using DECaLS photometry (see Section \ref{subsec:Completeness}). The S/N $>5$ cut ensures a purer sample of high-mass clusters. Below S/N $\approx 5$ detections are more likely to be spurious, i.e. caused by noise fluctuations. Additionally, we exclude 9 clusters that were found to have significantly different redshifts ($\Delta z \geq 0.05$) in the recently released ACT DR6 cluster catalogue \citep{Hilton2025}. Inclusion of these clusters could impact and bias the cluster SMF in numerous ways, such as incorrect cluster member assignment. We provide a list with the details of the removed clusters in Table \ref{tab:Removed_clus_table}. After applying the S/N cut and excluding the 9 clusters, we have a remaining sample of 1,654 clusters. This sample is further reduced to 568 clusters based on the depth available in optical/IR data from DECaLS (see Section \ref{subsec:Completeness}). Of the total sample of galaxy clusters, 264 have spectroscopic redshifts ($46.5$ per cent), and 304 have only photometric redshifts ($53.5$ per cent). The spectroscopic redshifts for clusters are obtained from the ACT DR5 catalogue \citep{Hilton_2021}. The redshift and mass distributions of our final cluster sample, overlayed on the total distribution of the DR5 cluster sample, are presented in Fig. \ref{fig:CMvsz}.

\subsection{DECaLS DR10 photometry}
\label{subsec:decalsdr10}

To study the stellar mass distribution of the ACT DR5 cluster sample, we make use of optical/IR data from the tenth data release of DECaLS (DECaLS DR10). DECaLS is one of three surveys that form part of the Dark Energy Spectroscopic Instrument (DESI) Legacy Imaging Surveys \citep{Dey2019}. DECaLS uses the Dark Energy Camera (DECam; \citealp{Flaugher_2015}), which is mounted on the 4-m Blanco telescope at the Cerro Tololo Inter-American Observatory (CTIO). DECam supplies targeted imaging for DESI in the $g$, $r$, and $z$ bands. DECaLS provides photometry for most of the ACT DR5 cluster search area footprint, with a $10,822$ deg$^{2}$ overlap \citep{Hilton_2021}. Source catalogues constructed using an algorithm called \textsc{The Tractor}\footnote[1]{\url{https://github.com/dstndstn/tractor}} \citep{Lang2016} were used to provide the photometric data. \textsc{The Tractor} is a source extraction tool that implements forced photometry on multi-wavelength data to produce accurate flux/magnitude measurements for galaxies \citep{Blum2016}. For more details on \textsc{The Tractor} code, see \cite{Lang2016b} and \cite{Dey2019}.

In particular, DR10 includes objects from DECaLS in newly observed regions and also new $i$-band data covering a sky area of $\approx 15,300$ deg$^{2}$ in 6 optical/IR bands ($grizW1W2$). $W1$ and $W2$ are the two shortest mid-infrared wavelength bands (3.4 $\mu$m and 4.6 $\mu$m) of the 4 bands used by the Wide-field Infrared Survey Explorer (WISE; \citealp{Wright2010}).

\section{SAMPLE SELECTION}
\label{sec:SampleSelection}

\subsection{Photometric redshifts}
\label{subsec:photoz}

The estimation of photometric redshifts (photo-$z$s) using SED-fitting is a widely adopted technique for inferring the redshifts of galaxies (e.g., \citealp{Benitez2000}; \citealp{Brammer2008}; \citealp{Rudnick_2009}). This method relies on modelling observed broadband photometry with a library of template SEDs, which are constructed from empirical galaxy spectra. By comparing the observed fluxes with redshifted template spectra, a photometric redshift probability distribution ($p(z)$) is computed. The best-fit redshift corresponds to the minimum $\chi^2$ value of the template fit \citep{Rudnick_2009}. The SED-fitting approach is particularly useful for large photometric surveys (e.g., DES) where obtaining spectroscopic redshifts for all galaxies is challenging due to time and resource constraints.

To estimate the photo-$z$s of galaxies, we make use of the publicly available photometric redshift code \textsc{zCluster} \footnote[2]{\url{https://github.com/ACTCollaboration/zCluster}} (\citealp{Hilton_2018}). This code has a built-in SED-fitting method to measure cluster and galaxy photo-$z$s. First, we use \textsc{zCluster} to query DECaLS DR10 and retrieve a catalogue of galaxies found within $36'$ of each cluster position. \textsc{zCluster} is then able to measure the $p(z)$ of each galaxy in the direction of each cluster. By fitting each galaxy broadband SED against an appropriate set of provided spectral templates, \textsc{zCluster} is able to measure individual galaxy $p(z)$ distributions \citep{Hilton_2018}. We employ a subset of the spectral templates from the COSMOS survey (\citealp{Ilbert_2009}; \citealp{Salvato2011}), which encompasses a variety of typical galaxies and AGNs. The maximum likelihood photo-$z$ of each galaxy is given by the peak of the $p(z)$ distribution.

\subsection{Cluster membership}
\label{subsec:ClusterMembership}

Accurate determination of which galaxies are physically associated with a given cluster (i.e. cluster members) is essential for deriving the cluster SMF. For cluster members, we consider galaxies found within the approximate virial radius $R_{\mathrm{200m}}$ of the cluster. $R_{\mathrm{200m}}$ is defined as the radius within which the average density is equal to 200 times the mean density of the Universe at that particular redshift, and is given by the equation,

\begin{equation}
    R_{\mathrm{200m}} = \sqrt[3] {\frac{3M_{\mathrm{200m}}}{4 \pi 200 \rho_\mathrm{m}(z)}},
	\label{eqn:R200m}
\end{equation}
where $M_{\mathrm{200m}}$ is the cluster mass within the approximate virial radius $R_{\mathrm{200m}}$, and $\rho_\mathrm{m}(z)$ is the mean matter density of the Universe at the cluster redshift. As confirmed by \cite{Andreon2010}, the use of $R_{\mathrm{200}}$ instead of $R_{500}$ (which is a smaller radius that encloses the cluster core) is more advantageous for stellar mass estimations and reduces systematic bias. Since $R_{200}$ encompasses most of the cluster, it allows for a more complete and reliable measure of the stellar mass content of the cluster. 

The masses of the clusters ($M_{\mathrm{200m}}$) are obtained from the ACT DR5 catalogue. \cite{Hilton_2021} converted $M_{\mathrm{200m}}$ from $M_{\mathrm{500c}}$ using the concentration-mass (c-M) relation from \cite{Bhattacharya2013}. The $M_{\mathrm{500c}}$ masses provided in DR5 were estimated assuming the Universal Pressure Profile (UPP) and the scaling relation by \cite{Arnaud2010} to convert the SZ signal into mass \citep{Hilton_2021}.

The most accurate method for identifying galaxy cluster members is by using accurate and reliable spectroscopic redshifts (spec-$z$s). However, the task of obtaining spec-$z$s for large galaxy samples at relatively high-$z$ ($z >0.5$) is time-intensive. Meanwhile, using photometric data allows us to probe deeper magnitude limits for large surveys and enable large galaxy samples. For this reason, photo-$z$s are commonly used to determine cluster membership (e.g., \citealp{Pello2009}; \citealp{Rozo2015}; \citealp{CastignaniBenoist2016}).

Early techniques on photometric redshift membership involve simply selecting any galaxy found within a chosen redshift slice around the cluster redshift (e.g., \citealp{Kodama1999}). However, the usage of a large redshift slice (e.g., $\pm 0.3$) can lead to field galaxy contamination (e.g., \citealp{Toft2004}). Furthermore, this simplistic approach makes use of the maximum likelihood redshift and neglects the information stored in the galaxy $p(z)$ distribution \citep{Pello2009}. Using the full galaxy $p(z)$ distribution also mitigates biases caused by photo-$z$ scatter since it encodes the full uncertainty of the redshift estimate \citep{Sheldon2012}. Thus, we employ a method that makes use of the full $p(z)$ distributions, instead of the maximum-likelihood photo-$z$s. 

Once galaxies found within $R_{\mathrm{200m}}$ of each respective cluster are selected, we assign photo-$z$ membership using a modified version of a technique first developed by \cite{Brunner_2000}, and introduced by \cite{Pello2009}. This method involves finding the probability that a galaxy is a member of a cluster by integrating over each galaxy's $p(z)$ distribution within a certain range/slice of the cluster redshift. This approach has been adopted in many cluster luminosity function (LF) and SMF studies (e.g., \citealp{Rudnick_2009}; \citealp{Vulcani2013}; \citealp{Bhatawdekar2019}). The probability that a galaxy is found to be a cluster member, $P_{\textrm{mem}}$, is given by:

\begin{equation}
    P_{\mathrm{mem}} = \int^{z_{\mathrm{cl}} +\Delta z_{\mathrm{cl}}}_{z_{\mathrm{cl}} -\Delta z_{\mathrm{cl}}}{p(z) \ \mathrm{d}z},
	\label{eqn:Pz}
\end{equation}
where $z_{\mathrm{cl}}$ is the cluster redshift, and $\Delta z_{\mathrm{cl}} = 0.05$ is the redshift slice around the cluster redshift, which should be on the order of the uncertainty in redshift for the galaxies in question \citep{Rudnick_2009}. By using this method, we probabilistically assign cluster membership to galaxies. We use galaxy $P_{\rm{mem}}$ values as a weighting for the construction of our cluster SMFs (see Section \ref{sec:method}).

To test the robustness of our selected redshift slice, we repeat our analysis for varying values: $\Delta z_{\mathrm{cl}} = 0.02, 0.03, 0.1$.  However, we find that changing $\Delta z_{\rm{cl}}$ does not make any significant difference ($< 1 \sigma$) in the fit parameters of the SMFs we obtain with a slice of $\Delta z_{\rm{cl}} = 0.05$.

A probability prior is commonly used to select reliable cluster members (e.g., \citealp{Rudnick_2009}; \citealp{Vulcani2013}). However, this implies that membership depends strongly on the chosen threshold, which can ultimately bias the SMF shape (e.g., excluding low-mass galaxies with a less certain photo-$z$). Hence, we do not apply a probability threshold and include all galaxies that could be possible members (i.e. $P_{\textrm{mem}} > 0$). This means that if a galaxy has $P_{\textrm{mem}} = 0.05$ then it has a $5$ per cent chance of being a cluster member and will still contribute partially to the SMF, instead of being discarded.

\subsection{Inclusion of the brightest cluster galaxies}
\label{subsec:InclusionofBCGs}

The dominance of the quiescent galaxy population in clusters strongly impacts the shape of the cluster SMF, particularly at the high-mass end. This sometimes leads to a ``bump'' or ``excess'' in stellar mass, compared to the rest of the galaxy population, observed at the high-mass end (e.g., \citealp{Vulcani2013}; \citealp{vanderBurg2018}). In galaxy clusters, this excess is mainly driven by the inclusion of brightest cluster galaxies (BCGs), which are the most massive and luminous galaxies found in clusters. BCGs represent a distinct population of galaxies that is different to the general galaxy population (\citealp{vonderLinden2007}; \citealp{Yen-Ting2010}; \citealp{Lidman2012}). As such, a typical procedure in cluster SMF studies is to exclude BCGs by implementing a radial cut around the cluster centre, as they may alter trends in the observed SMF (e.g., \citealp{Vulcani2013}; \citealp{vanderBurg2013, vanderBurg2018}). However, excluding BCGs could lead to an underestimation of the number of high-mass galaxies and biased measurements of the cluster SMF. Furthermore, the exclusion of BCGs has not been found to have a significant difference in the shape of the cluster SMF \citep{Calvi2013}. Hence, we include BCGs in the cluster SMFs in order to better assess the overall stellar mass growth of clusters.

\subsection{Stellar masses}
\label{subsec:StellarMasses}

Stellar masses are measured for each cluster galaxy using a similar SED-fitting method adopted by \cite{Burke2015}. Stellar population libraries from \cite{Bruzual&Charlot2003} are used to produce synthetic SEDs for various stellar population parameters. For these parameters, we assume a \cite{Chabrier2003} Initial Mass Function (IMF), the \cite{Calzetti2000} dust attenuation model, and metallicity values $Z = [0.005, 0.02, 0.2, 0.4, 1.0, 2.5] \ \mathrm{Z_{\odot}}$. Each synthetic SED represents a galaxy with a specific age, metallicity, star formation history (SFH), and dust attenuation. The observed galaxy SED is then compared to a grid of the synthetic SEDs. A $\chi^2$ minimisation method is used to determine the best-fitting model. The normalisation factor of the best-fit model is taken as the galaxy's stellar mass.

Stellar masses derived from SED-fitting depend on various assumed parameters, such as the choice of stellar population synthesis (SPS) model (e.g., \citealp{Bruzual&Charlot2003}; \citealp{Maraston2005}; \citealp{Conroy2009}) and IMF used (e.g., \citealp{Salpeter1955}; \citealp{Kroupa2001}). These assumptions lead to systematic changes to the SMF, with individual galaxy stellar masses differing by a factor of $\sim 2-3$ depending on the method used (e.g., \citealp{Marchesini2009}; \citealp{Longhetti2009}). Nevertheless, the IMF chosen in this work has also been used in many previous studies (e.g., \citealp{Annunziatella2014, Annunziatella2016}, \citealp{vanderBurg2018, vanderBurg2020}), allowing for a consistent comparison with these studies. A further study of the effects of varying assumptions in stellar mass estimation is beyond the scope of this paper.

\subsection{Completeness}
\label{subsec:Completeness}

A common issue that arises in SMF/LF studies is incompleteness at the low-mass/faint end. At high-$z$, faint/low-mass galaxies are more challenging to detect in optical surveys and often fall below the photometric detection threshold. 
Completeness corrections typically involve injecting mock/synthetic galaxies into cluster images and recovering them using a detection algorithm (e.g., \citealp{Hilton2013}). To correct for incompleteness, galaxies are then weighted according to the average incompleteness measured as a function of magnitude or stellar mass \citep{Mancone2010}. In this work, we define a volume-limited sample in terms of galaxy stellar mass, which avoids the need to perform completeness corrections.

\begin{figure}
	\includegraphics[width=\columnwidth]{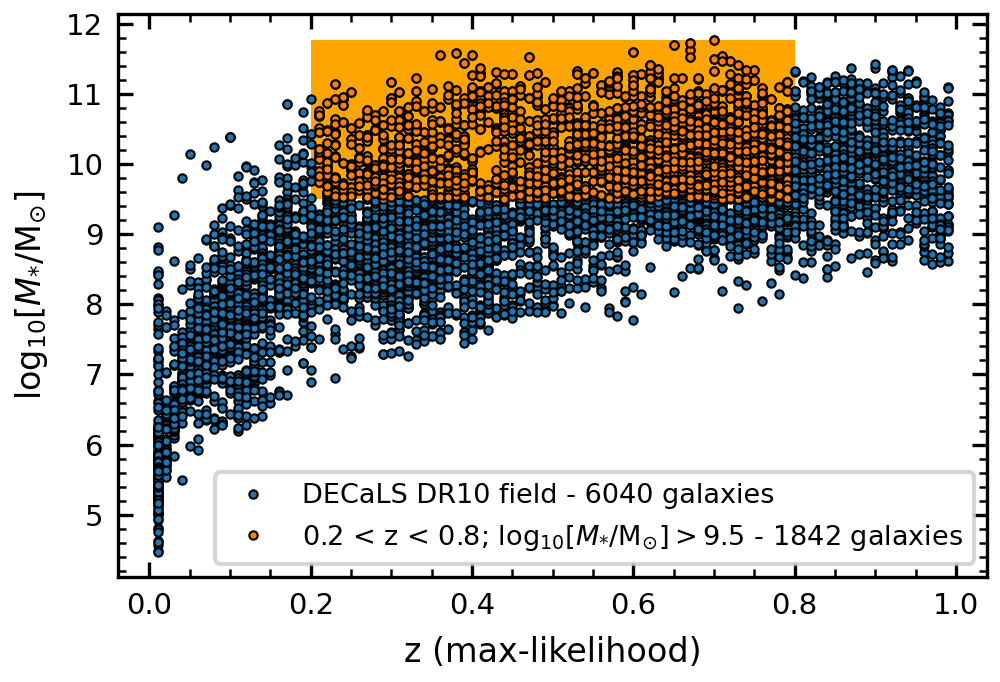}
    \caption{A mass-limited sample obtained for a galaxy field (RA $= 45^{\circ}.0$; Dec $= -51^{\circ}.0$) in DECaLS DR10 that also has DES coverage. This galaxy sample is used to gauge a consistent stellar mass depth in DECaLS DR10. Blue points represent all galaxies within the field region. The orange points in the shaded area show the galaxies above the stellar mass limit (log$_{10}[M_{*}/\mathrm{M_{\odot}}] = 9.5$) and within our selected redshift range for this work ($0.2 < z < 0.8$). We select cluster galaxies above this mass limit for completeness.}
    \label{fig:stellarmassdist}
\end{figure}

We determine the stellar mass limit as follows. First, we construct volume-limited samples from galaxy field regions within DECaLS DR10. This was done using the \textsc{zField} code \citep{Kesebonye2023}, which is incorporated into \textsc{zCluster}. We query five randomly selected galaxy field regions in DR10, each within a search radius of $12'$. Stellar masses and photo-$z$s are estimated using the \textsc{zCluster} SED-fitting procedure (see Section \ref{subsec:photoz} and \ref{subsec:StellarMasses}) for all galaxies found within each field region. We construct a stellar mass distribution for each field region, from which we can infer a galaxy sample within our redshift range ($0.2 < z < 0.8$) and above a minimum stellar mass threshold, i.e. a mass-limited sample. To ensure we gauge a consistent depth in DECaLS DR10, we additionally query field regions that also have DES coverage. DES imaging was focused on a well-defined contiguous region ($5,000$ deg$^2$) of the southern sky in the $grizY$ bands, allowing for more uniform depth coverage \citep{Abbott_2021}. The mass-limited samples for regions with DES coverage revealed that stellar masses are complete for $M_* > 10^{9.5} \mathrm{M_{\odot}}$ up to $z = 0.8$ (see Fig. \ref{fig:stellarmassdist}). 

To ensure that the sample of galaxies included in the SMF is complete down to a consistent stellar mass limit across all galaxy clusters, we also consider the Point Spread Function (PSF) depth in DECaLS DR10. The PSF depth affects how well galaxy fluxes are measured, particularly for faint galaxies. The $z$-band is often used as a proxy for stellar mass due to its weaker sensitivity to dust extinction and relatively stable mass-to-light ratio for old stellar populations (\citealp{Tremonti_2004}; \citealp{Johnson_2007}). Thus, selecting clusters in regions with uniform PSF depth in $z$-band helps reduce bias in the cluster SMF measurement.

To determine the appropriate $z$-band PSF depth cut to implement, we first use \textsc{zCluster} to fetch DECaLS DR10 galaxy catalogues for each of our field regions in DECaLS DR10. These catalogues contain measured PSF depth values for each galaxy (in AB magnitude) for a $5\sigma$ point source detection limit in each band ($grizW1W2$). We then take the median value of the PSF depth in the $z$-band magnitude column. We find that a median $z$-band PSF depth of $25.5$ mag. in DECaLS allows us to detect galaxies down to $M_* = 10^{9.5} \rm{M_{\odot}}$ at the highest redshift considered in this work ($z = 0.8$). At low-$z$, the same stellar mass limit corresponds to brighter magnitudes, meaning galaxy stellar masses are complete above this threshold. For this reason, we only select galaxy clusters that are found in regions where the DECaLS DR10 PSF depth in magnitude is $m_z > 25.5$ mag. This reduces our initial cluster sample from 1,654 to 568 clusters.

\section{THE COMPOSITE STELLAR MASS FUNCTION}
\label{sec:method}

\begin{figure}
	\includegraphics[width=\columnwidth]{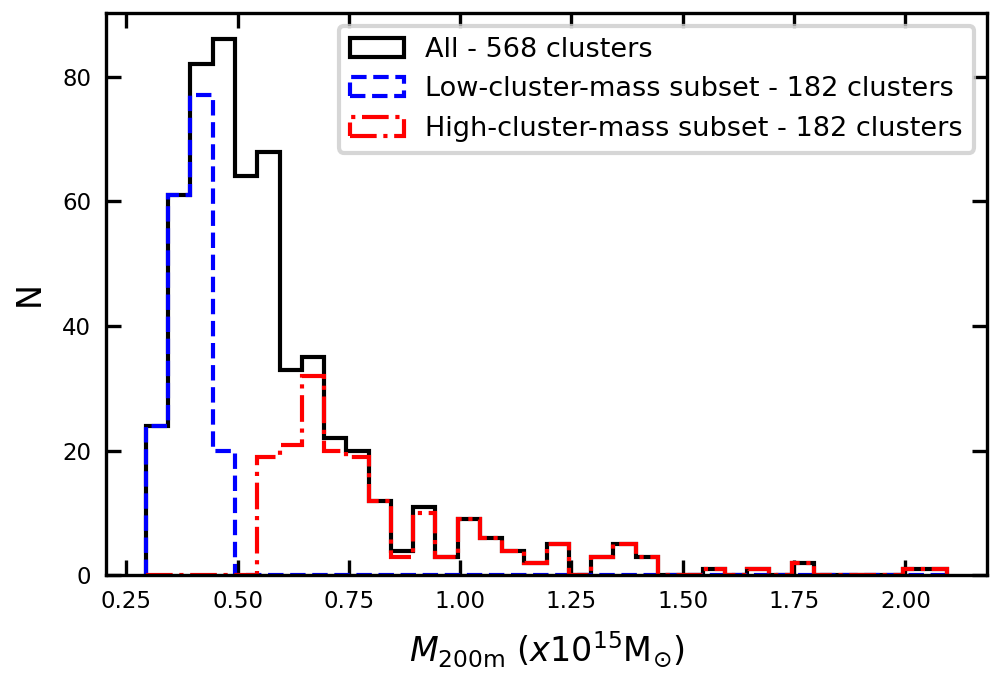}
    \caption{The distribution of cluster mass ($M_{\mathrm{200m}}$) for the total cluster sample (shown in black). Also represented are the total top (shown in red) and bottom (shown in blue) terciles of the cluster mass distribution for which we measure the composite SMF. Cluster mass subsets are defined based on the halo mass distribution within each redshift bin (see Section \ref{sec:method}).} 
    \label{fig:CMdist}
\end{figure}

A common approach in large sample SMF studies is to bin by redshift (e.g., \citealp{Guo_2018}; \citealp{COSMOS2020}).
Due to our large cluster sample, we can bin clusters into 12 redshift bins, with a bin width of $\Delta z = 0.05$, over the range $0.2 < z < 0.8$. Additionally, to investigate mass-dependent trends, we subdivide clusters in each redshift bin into high- and low-cluster-mass subsets by selecting the top and bottom terciles of the cluster mass distribution in that particular redshift bin. 
A total of 182 galaxy clusters with $M_{\mathrm{200m}}$ in the range $ (2.94 -4.84) \times 10^{14} \mathrm{M_{\odot}}$ constitute the low-cluster-mass subset of the entire cluster sample. In contrast, the high-cluster-mass subset of the total cluster sample consists of 182 massive clusters with $M_{\mathrm{200m}}$ in the mass range $5.48 \times 10^{14} -2.09 \times 10^{15} \mathrm{M_{\odot}}$. The cluster mass distribution for our total cluster sample and both cluster-mass terciles are shown in Fig. \ref{fig:CMdist}.

A disadvantage of measuring individual cluster LFs/SMFs is that their quality often varies based on redshift completeness and the number of cluster members \citep{dePropris2003}. Instead of presenting individual cluster SMFs, we measure composite/stacked cluster SMFs and examine differences in subsets binned by cluster mass and redshift. The composite/stacked method makes it possible to identify variations that are difficult to detect due to small sample statistics in individual cluster LFs/SMFs (\citealp{dePropris2003}; \citealp{Vulcani2011}; \citealp{Hilton2013}; \citealp{WenHan2015}; \citealp{vanderBurg2018}). 

Our choice for stellar mass binning is dependent on our derived stellar mass limit (see Section \ref{subsec:Completeness}). We adopt a bin width of $\Delta \log_{10} [M_{*}/\mathrm{M_{\odot}}] = 0.12$ dex for 26 stellar mass bins for the stellar mass range: $9.5 \leq$ log$_{10} [M_{*}/\mathrm{M_{\odot}}] \leq 12.5$. We follow the widely used method for composite LFs/SMFs given by \cite{Colless1989}. The number of galaxies in the $j$th stellar mass bin of the composite cluster SMF is,

\begin{equation}
    N_{\mathrm{c}j} = \frac{N_{\mathrm{c0}}}{m_{j}}\sum_{i} \frac{N_{ij}}{N_{i\mathrm{0}}},
	\label{eq:SMF}
\end{equation}
where $m_{j}$ is the number of clusters contributing to the $j$th bin of the composite SMF. $N_{ij}$ is the number of galaxies contributing to the $j$th bin of the $i$th cluster SMF, i.e. the individual cluster SMF. Each cluster galaxy is counted in the cluster SMF with a weight equal to its probability ($P_{\mathrm{mem}}$).

$N_{i\mathrm{0}}$ is a normalisation parameter, or a cluster richness proxy, for the $i$th cluster SMF. This richness weighting accounts for the richer clusters, which consist of a higher density of galaxies than the poorer clusters, which could dominate the composite SMF. For every cluster, we calculate $N_{i\mathrm{0}}$ by summing the $P_{\mathrm{mem}}$ of all cluster galaxies.

$N_{\mathrm{c0}}$ is the sum of all normalisations,

\begin{equation}
    N_{\mathrm{c0}} = \sum_{i} N_{i\mathrm{0}}.
	\label{eqn:norm}
\end{equation}

The errors in the composite SMF, $N_{\mathrm{c}j}$, are calculated using the bootstrap resampling method \citep{Efron1986}. Within each stellar mass bin, we randomly resample cluster galaxies with replacement. We perform 1,000 bootstrap resamples to build a distribution of SMFs at each stellar mass bin. To obtain the bootstrap error bars for each SMF bin, we compute the standard deviation of the distributions across all resamples. Bootstrap resampling provides a more realistic estimate of the total uncertainty on galaxy number counts than assuming Poisson statistics, as it incorporates sample variance and does not make any assumptions about the error distribution of the data \citep{Andrae2010}.

\section{RESULTS}
\label{sec:Results}

The computations given by Equations \ref{eq:SMF} and \ref{eqn:norm} were carried out for all redshift and cluster mass bins. We measured the composite SMFs in the log stellar mass range of $9.5 \leq$ log$_{10}[M_*/\mathrm{M_{\odot}}] \leq 12.5$ for 12 redshift bins and additionally measured the composite SMF for low- and high-cluster-mass samples within each redshift bin. Each redshift and cluster mass bin consisted of a minimum of 9 clusters. The measured composite SMFs for two redshift bins are shown in Fig. \ref{fig:ExampleSinglSchechterSMFs}, while the SMFs for the remaining 10 bins are shown in Fig. \ref{fig:SinglSchechterSMFs}.

In order to extract relevant information from the cluster SMF, we need to fit an appropriate model to our SMF measurements. In this section, we describe the models we use to fit the composite cluster SMF. Additionally, we study the relationship between cluster stellar mass and halo mass.

\subsection{The Single Schechter model}
\label{subsec:SingleSchechter}

The SMF is dominated by two galaxy populations: the high-mass end, which is characterised by massive, early-type galaxies, and the low-mass end, which is characterised by fainter, late-type galaxies (e.g., \citealp{Bundy2006}; \citealp{Vulcani2011}; \citealp{Calvi2013}). This signature in stellar mass is generally well described by the Schechter function \citep{Schechter1976}, 
 
\begin{equation}
    \phi_\mathrm{s} (M) = \ln (10)\Phi^{*}10^{(M - M^{*})(1 + \alpha)} \ \exp(-10^{(M - M^{*})}),
	\label{eq:SingleSchechter}
\end{equation}
where $\Phi^{*}$ is the normalisation parameter, $M =$ log$_{10}[M_*/\mathrm{M_{\odot}}]$ is the log stellar mass, $M^{*} =$ log$_{10} [M_*^*/\mathrm{M_{\odot}}]$ is the characteristic stellar mass and $\alpha$ is the low-mass-end slope. 

We fit a single Schechter function to each composite cluster SMF using the Markov-Chain Monte Carlo (MCMC) technique. We use the Python package \textsc{emcee} \citep{Foreman-Mackey2013}, with a total of 20,000 MCMC iterations, to obtain the best-fitting Schechter function parameters ($M^*$ and $\alpha$) and their $1\sigma$ uncertainties. Convergence of the MCMC to the posterior distributions of each parameter is monitored using the Gelman-Rubin statistic ($\hat{R}$; \citealp{Gelman&Rubin1992}), and is considered satisfied for the condition $\hat{R} -1 < 0.01$. This convergence condition for $\hat{R}$ is based on the improved diagnostic proposed by \cite{Vehtari2021}.

The normalisation parameter ($\Phi^*$) is fixed such that the value of the integral of the fitted Schechter function over the chosen stellar mass range, $10^{9.5} - 10^{12.5} \mathrm{M_{\odot}}$, equals the total number of galaxies in the composite cluster SMF (\citealp{vanderBurg2018}; \citealp{vanderBurg2020}). A fixed normalisation allows us to retain meaningful information of the total integrated stellar mass and makes comparisons of cluster SMFs across redshift intervals consistent. However, since $\Phi^*$ is not a free parameter, it does not provide any important information about the true galaxy density. Hence, we only consider $M^*$ and $\alpha$ because they describe the shape of the cluster SMF.

\begin{figure*}
    \centering
    \subfloat{\includegraphics[width=1.0\columnwidth]{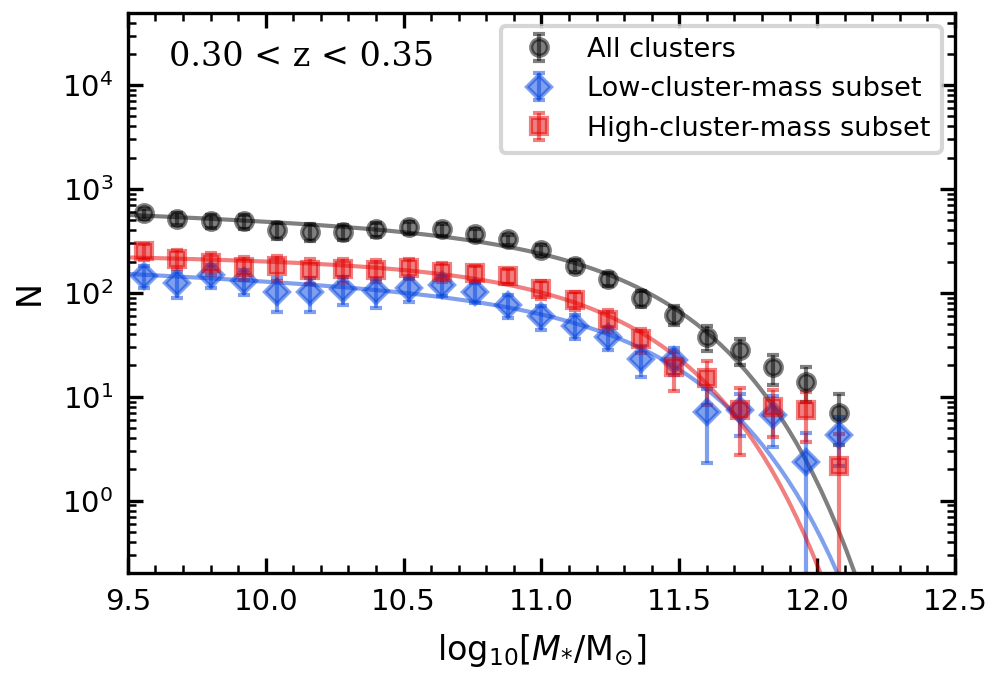}}
    \subfloat{\includegraphics[width=1.0\columnwidth]{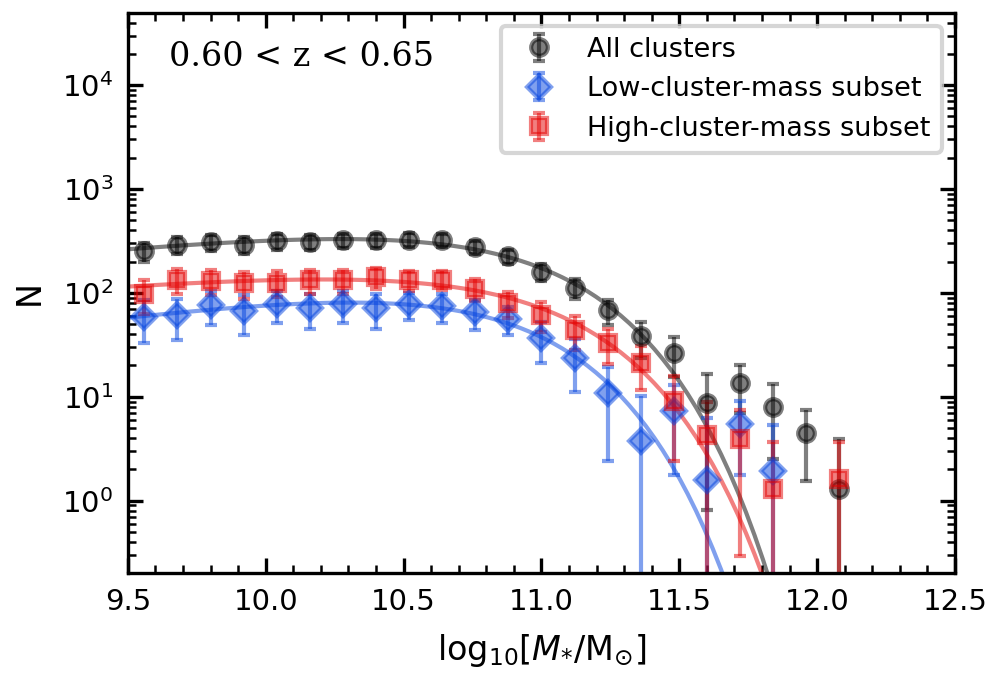}}\\
    
    \caption{The best-fit composite cluster SMFs for two redshift bins, $\langle z \rangle = 0.325$ and $\langle z \rangle= 0.625$. Black circles represent the SMF for all cluster galaxies in the redshift bin. Blue diamonds and red squares represent the SMF of cluster galaxies in the low- and high-mass cluster mass subsets, respectively. Error bars show the $1\sigma$ uncertainties calculated using bootstrap resampling. A solid line shows a best-fit single Schechter model for each SMF. The composite SMFs for all 12 redshift bins are shown in Fig. \ref{fig:SinglSchechterSMFs}. Parameter values of the best-fit are presented in Table \ref{tab:smf_table}. Deviations in stellar mass above the Schechter function are observed at the high-mass end. The shape of both cluster mass subset SMFs is highly comparable to the composite SMF of all clusters in the redshift bin.}
    \label{fig:ExampleSinglSchechterSMFs}
\end{figure*}

To determine the "goodness of fit", we use the chi-squared value given by,

\begin{equation}
    \chi^2 = \sum_{i} \frac{(N_i - N_i^\mathrm{e})^2}{\sigma_i^2},
	\label{eq:Chi-squared}
\end{equation}
where $N_i$ is the number of galaxies in the $i$th bin of the observed composite SMF, $N^\mathrm{e}_i$ is the expected number of galaxies in the $i$th bin from the Schechter model, and $\sigma_i$ is the bootstrap error for $N_i$. The reduced chi-squared value is given by $\chi^2_\nu =\chi^2/ \nu$, where $\nu$ is the degrees of freedom. For model comparisons, we use the Akaike Information Criterion (AIC; \citealp{Akaike1974}) to evaluate how well a single Schechter function fits the composite SMF compared to other models (see Section \ref{subsec:Schechter+Gaussianmodel}). The best-fitting single Schechter function fit parameters for each respective redshift and cluster mass bin are presented in Table \ref{tab:smf_table}. We show composite cluster SMFs, fitted by a single Schechter model, for two redshift bins in Fig. \ref{fig:ExampleSinglSchechterSMFs}. The remainder of the fitted composite SMFs can be found in Fig. \ref{fig:SinglSchechterSMFs}.

\begin{figure}
	
	\includegraphics[width=\columnwidth]{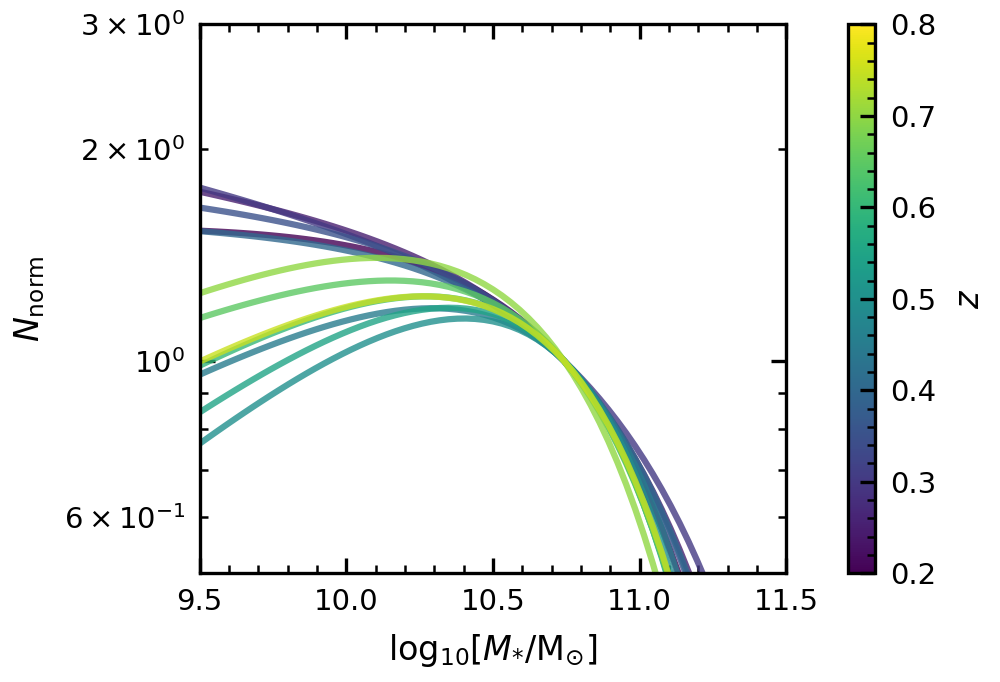}
    \caption{The single Schechter models to the 12 total redshift binned composite SMFs. Each model has been normalised by a common log stellar mass, $\log_{10}[M_*/\mathrm{M_{\odot}}] = 10.75$, in order to emphasise the evolution of the low-mass slope. The colour bar is used to represent the model evolution with redshift.}
    \label{fig:CompSMF}
\end{figure}

The total composite cluster SMF exhibits a trend in which the low-mass slope becomes marginally less negative with increasing redshift. We do observe that $\alpha$ does not appear to vary monotonically with redshift. However, $\alpha$ is not that well constrained at $z > 0.6$, so any fluctuations at high-$z$ is within errors and not significant. We obtain values of $\alpha = -0.99 \pm 0.04$ at $z = 0.2$ and $\alpha = -0.77 \pm 0.12$ at $ z = 0.8$. This indicates a very weak evolution ($ 1.7\sigma$ difference) of $\alpha$ from $ z = 0.8$ to $ z = 0.2$. The redshift evolution of $\alpha$ is much more pronounced for $z < 0.55$, where there is a significant decrease ($4.1 \sigma$ difference) in low-mass slope values, from $\alpha = -0.66 \pm 0.07$ at $z = 0.55$ to $\alpha = -0.99 \pm 0.04$ at $z = 0.2$. The weaker evolutionary trend of the low-mass slope between high and low redshift appears to be driven by a poorly constrained $\alpha$ at high-$z$. To emphasise the evolution of $\alpha$ across the entire redshift interval, a comparison of the 12 single Schechter models to the total composite cluster SMFs (normalised by $\log _{10}[M_*/\mathrm{M_{\odot}}] = 10.75$) is shown in Fig. \ref{fig:CompSMF}.

Meanwhile, there is a very moderate decreasing trend in the characteristic stellar mass with increasing redshift. At low-$z$ ($z = 0.2$), we obtain $M^* = 11.06 \pm 0.03$, while at high-$z$ ($z = 0.8$), it decreases to a value $M^* = 10.90 \pm 0.11$. This reveals a lack of statistically significant evolution ($1.4 \sigma$ growth) in $M^*$ from $z = 0.8$ to $z = 0.2$. However, as observed for $\alpha$, there is a more considerable growth ($3.3 \sigma$) of $M^*$, from $M^* = 10.87 \pm 0.05$ at $z = 0.55$ to $M^* = 11.06 \pm 0.03$ at $z = 0.2$.

Similar trends in $M^*$ and $\alpha$ are observed for the two cluster mass subsets. For each mass subset, we do not observe a considerable evolution ($\lesssim 1\sigma$ difference) of $M^*$ and $\alpha$ from $z = 0.8$ to $z = 0.2$.  However, as observed for the total cluster sample, both subsets exhibit a more notable evolution of $M^*$ and $\alpha$ from $z = 0.55$ to $z = 0.2$ (see Table \ref{tab:smf_table}), although the difference in parameter values for both subsets is not significant ($<3 \sigma$). In comparison with the total cluster SMF per redshift bin, the value of $M^*$ differs by less than $\approx 1 \sigma$ for the low-cluster-mass subset and less than $\approx 2 \sigma$ for the high-cluster-mass subset. For values of $\alpha$, we find a difference of $\lesssim 1 \sigma$ for the low-cluster-mass subset and $\lesssim 1.5 \sigma$ for the high-cluster-mass subset. In fact, the largest differences between the mass subset SMFs and the total SMF occur in the first two redshift bins. 

\begin{figure}
	
	\includegraphics[width=\columnwidth]{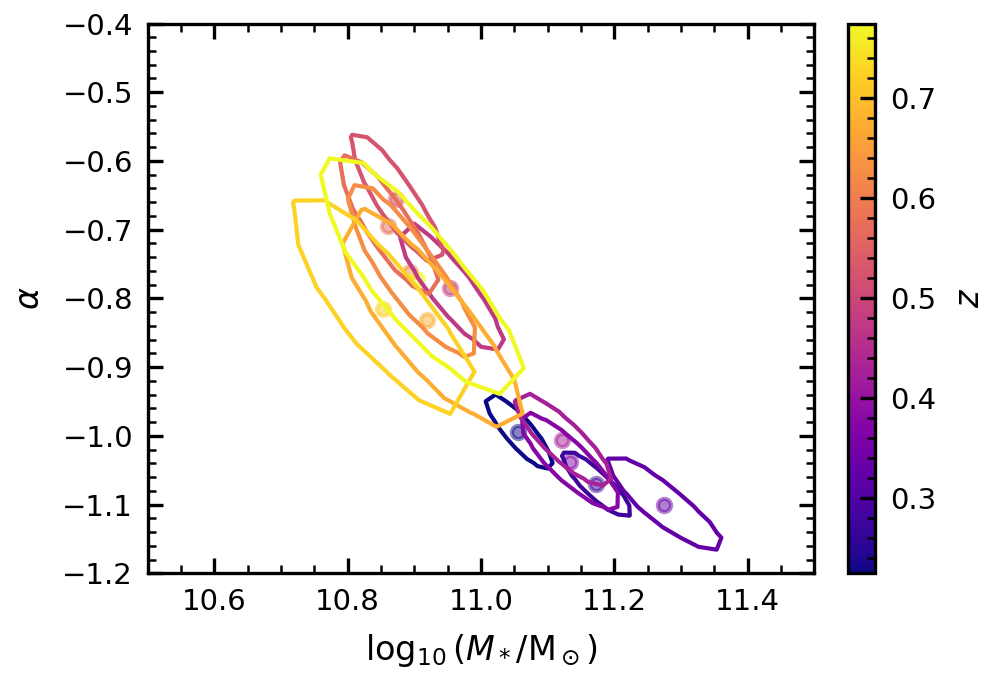}
    \caption{The best-fitting low-mass slope ($\alpha$) and characteristic stellar mass ($M^* = \log_{10} [M^*_*/\mathrm{M_{\odot}]}$) values for the composite cluster SMF. Each point is plotted at the midpoint of the corresponding redshift bin, distinguished by a colour bar. The $1 \sigma$ contours, obtained from the MCMC posterior samples, are plotted for each point. An anti-correlation is observed between both parameters (i.e. a higher $M^*$ value corresponds to a more negative $\alpha$ value and vice versa).}
    \label{fig:MstarvsAlpha}
\end{figure}

A brief inspection of $M^*$ and $\alpha$ values as a function of redshift (shown in Fig. \ref{fig:MstarvsAlpha}) reveals a degeneracy between both parameters. At low-$z$ ($z < 0.45$), the higher $M^*$ values correspond to a steeper $\alpha$, while the opposite is true at higher redshifts ($z > 0.45$).

A phenomenon observed in previous studies of the SMF is a steep rise in the number density of low-mass galaxies that deviate beyond a single Schechter function at the low-mass end, and is commonly referred to as an ``upturn" (e.g., \citealp{Annunziatella2016}; \citealp{Tomczak_2014}; \citealp{Adams2021}). Previous works have found that a double Schechter function is a good solution to model this upturn and the bimodality of the galaxy population in the SMF at low-$z$ (e.g., \citealp{Ilbert2013}; \citealp{COSMOS2015}; \citealp{COSMOS2020}). For instance, \cite{Drory2009} detected an upturn at $\sim 3 \times 10^{9} \rm{M_{\odot}}$ in the SMF measured for COSMOS field galaxies probed at $z \leq 1$. They used a double Schechter function to model this more complicated behaviour. This could suggest that our adopted stellar mass limit ($M_{*} = 10^{9.5} \mathrm{M_{\odot}}$) is too shallow to observe such a trend in our composite SMFs. Furthermore, as confirmed by $\chi^2_{\nu}$ values (see Table \ref{tab:smf_table}), the single Schechter model does a good job at describing the low-mass end. Hence, we find no motivation to use a double Schechter function to model our SMFs.

We display the redshift evolution of $M^*$ and $\alpha$ in Fig. \ref{fig:M,alphavsz}. For comparison, we also include the single Schechter parameter values from previous studies on the SMF in clusters and in field galaxies. The only exception is the field SMF study by \cite{Ilbert2013}, to which we use the equivalent parameter that describes the low-mass slope.

\begin{table*}
        \centering
	\caption{The best-fitting single Schechter function parameters, where $M^* =$ log$_{10}[M^*_*/\mathrm{M_\odot}]$, for the 12 redshift binned composite SMFs. We include the derived fitting parameters for composite SMFs binned by cluster mass. We do not provide uncertainties for the normalisation parameter ($\phi^*$) since it is not held as a free parameter in the fitting procedure. The final three columns include the $\chi^2$, $\chi^2_\nu$, and AIC values.}
	\label{tab:smf_table}
        \renewcommand{\arraystretch}{1.3}
	\begin{tabular}{lcccccccr}
		\hline
		Redshift range & Sample & $N_{\mathrm{clus}}$ &$\phi^*$ & $M^*$ & $\alpha$ & $\chi^2$ &$\chi_{\nu}^2$ & AIC\\
		\hline
        
		$0.20 < z < 0.25$ & All & 30 & 305.36 & $11.06 \pm 0.03$ & $-0.99 \pm 0.04$ & 14.57 & 0.73 & 18.57\\
            & Low-cluster-mass & 9 & 44.83 & $11.15 \pm 0.08$ & $-1.08 \pm 0.08$ & 7.35 & 0.41 & 11.35 \\
            & High-cluster-mass & 9 & 168.63 & $10.95 \pm 0.05$ & $-0.93 \pm 0.06$ & 10.03 & 0.50 & 14.03\\
            \hline

            $0.25 < z < 0.30$ & All & 51 & 338.65 & $11.17 \pm 0.04$ & $-1.07 \pm 0.03$ & 44.45 & 2.22 & 48.45\\
            & Low-cluster-mass & 16 & 70.81 & $11.25 \pm 0.07$ & $-1.10 \pm 0.06$ & 10.32 & 0.52 & 14.32\\
            & High-cluster-mass & 16 & 207.12 & $11.06 \pm 0.05$ & $-0.98 \pm 0.05$ & 20.10 & 0.96 & 24.10\\
            \hline
    
            $0.30 < z < 0.35$ & All & 43 & 165.06 & $11.27 \pm 0.06$ & $-1.10 \pm 0.05$ & 24.34 & 1.22 & 28.34\\
            & Low-cluster-mass & 14 & 41.25 & $11.30 \pm 0.11$ & $-1.12 \pm 0.09$ &  11.27 & 0.56 & 15.27\\
            & High-cluster-mass & 14 & 83.75 & $11.18 \pm 0.09$ & $-1.04 \pm 0.08$ & 11.89 & 0.59 & 15.89\\
            \hline

            $0.35 < z < 0.40$ & All & 61 & 255.43 & $11.13 \pm 0.05$ & $-1.04 \pm 0.05$ & 43.26 & 2.16 & 47.26\\
            & Low-cluster-mass & 20 & 61.75 & $11.09 \pm 0.10$ & $-1.03 \pm 0.10$ & 12.04 & 0.60 & 16.04\\
            & High-cluster-mass & 20 & 112.10 & $11.17 \pm 0.08$ & $-1.05 \pm 0.07$ & 23.14 & 1.16 & 27.14\\
            \hline

            $0.40 < z < 0.45$ & All & 61 & 285.20 & $11.12 \pm 0.05$ & $-1.01 \pm 0.05$ & 24.37 & 1.16 & 28.37\\
            & Low-cluster-mass & 20 & 64.59 & $11.12 \pm 0.12$ & $-1.01 \pm 0.10$ & 12.59 & 0.63 & 16.59\\
            & High-cluster-mass & 20 & 128.27 & $11.15 \pm 0.07$ & $-1.03 \pm 0.06$ & 10.09 & 0.46 & 14.09\\
            \hline

            $0.45 < z < 0.50$ & All & 53 & 335.77 & $10.95 \pm 0.05$ & $-0.78 \pm 0.06$ & 28.88 & 1.44 & 32.88\\
            & Low-cluster-mass & 17 & 91.40 & $10.93 \pm 0.11$ & $-0.74 \pm 0.13$ & 9.00 & 0.47 & 13.00\\
            & High-cluster-mass & 17 & 146.71 & $10.92 \pm 0.08$ & $-0.74 \pm 0.10$ & 10.78 & 0.57 & 14.78\\
            \hline

            $0.50 < z < 0.55$ & All & 51 & 373.65 & $10.87 \pm 0.05$ & $-0.66 \pm 0.07$ & 32.38 & 1.62 & 36.38\\
            & Low-cluster-mass & 16 & 104.55 & $10.81 \pm 0.10$ & $-0.59 \pm 0.15$ & 11.08 & 0.55 & 15.08\\
            & High-cluster-mass & 16 & 142.50 & $10.89 \pm 0.08$ & $-0.64 \pm 0.11$ & 10.11 & 0.56 & 14.11\\
            \hline

            $0.55 < z < 0.60$ & All & 50 & 307.17 & $10.86 \pm 0.06$ & $-0.69 \pm 0.08$ & 14.73 & 0.67 & 18.73\\
            & Low-cluster-mass & 16 & 71.04 & $10.84 \pm 0.12$ & $-0.69 \pm 0.16$ & 5.55 & 0.29 & 9.55\\
            & High-cluster-mass & 16 & 140.29 & $10.81 \pm 0.08$ & $-0.63 \pm 0.12$ & 7.66 & 0.43 & 11.66\\
            \hline

            $0.60 < z < 0.65$ & All & 58 & 255.06 & $10.89 \pm 0.06$ & $-0.76 \pm 0.08$ & 10.89 & 0.54 & 14.89\\
            & Low-cluster-mass & 19 & 70.14 & $10.79 \pm 0.14$ & $-0.67 \pm 0.20$ & 4.16 & 0.23 & 8.16\\
            & High-cluster-mass & 19 & 96.40 & $10.93 \pm 0.10$ & $-0.81 \pm 0.12$ & 3.59 & 0.19 & 7.59\\
            \hline

            $0.65 < z < 0.70$ & All & 40 & 138.21 & $10.92 \pm 0.09$ & $-0.83 \pm 0.10$ & 2.90 & 0.16 & 6.90\\
            & Low-cluster-mass & 13 & 32.58 & $10.96 \pm 0.22$ & $-0.89 \pm 0.21$ & 1.77 & 0.11 & 5.77\\
            & High-cluster-mass & 13 & 61.83 & $10.85 \pm 0.14$ & $-0.74 \pm 0.17$ & 0.87 & 0.05 & 4.87\\
            \hline 

            $0.70 < z < 0.75$ & All & 38 & 137.04 & $10.85 \pm 0.10$ & $-0.82 \pm 0.11$ & 4.59 & 0.24 & 8.59\\
            & Low-cluster-mass & 12 & 34.52 & $10.80 \pm 0.29$ & $-0.75 \pm 0.27$ & 1.10 & 0.08 & 5.10\\
            & High-cluster-mass & 12 & 61.38 & $10.81 \pm 0.16$ & $-0.78 \pm 0.19$ & 2.21 & 0.14 & 6.21\\
            \hline

            $0.75 < z < 0.80$ & All & 32 & 104.97 & $10.90 \pm 0.11$ & $-0.77 \pm 0.12$ & 5.02 & 0.26 & 9.02\\
            & Low-cluster-mass & 10 & 26.19 & $10.88\pm 0.24$ & $-0.78 \pm 0.27$ & 2.38 & 0.14 & 6.38\\
            & High-cluster-mass & 10 & 40.66 & $10.89 \pm 0.18$ & $-0.75 \pm 0.22$ & 1.96 & 0.12 & 5.96\\
		\hline
	\end{tabular}

\end{table*}

\begin{figure*}
    \centering
    \subfloat{\includegraphics[width=1.0\columnwidth]{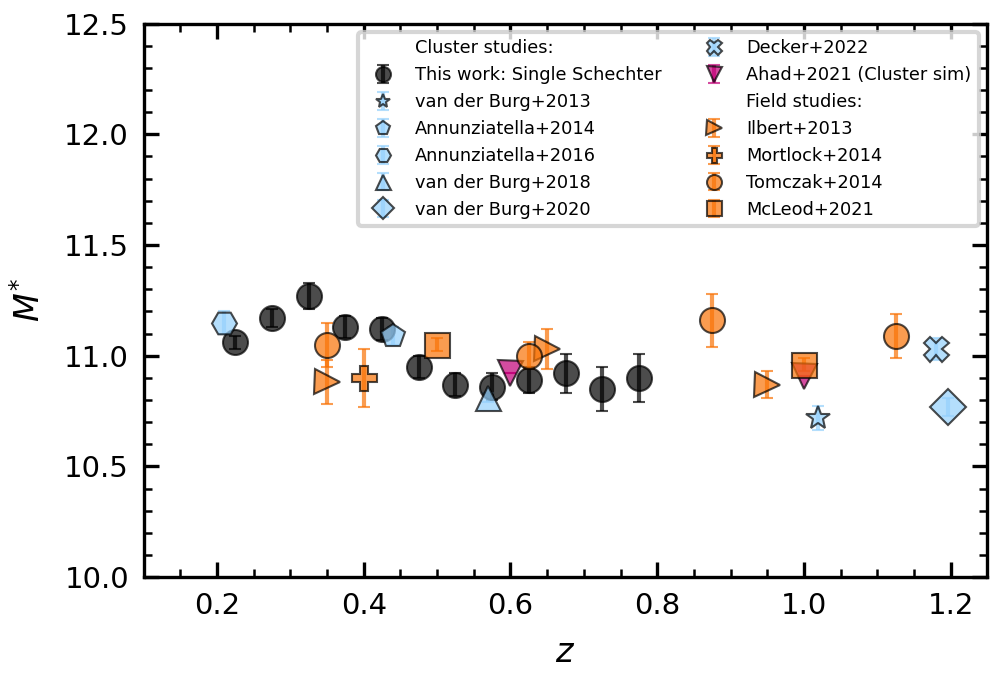}}
    \subfloat{\includegraphics[width=1.0\columnwidth]{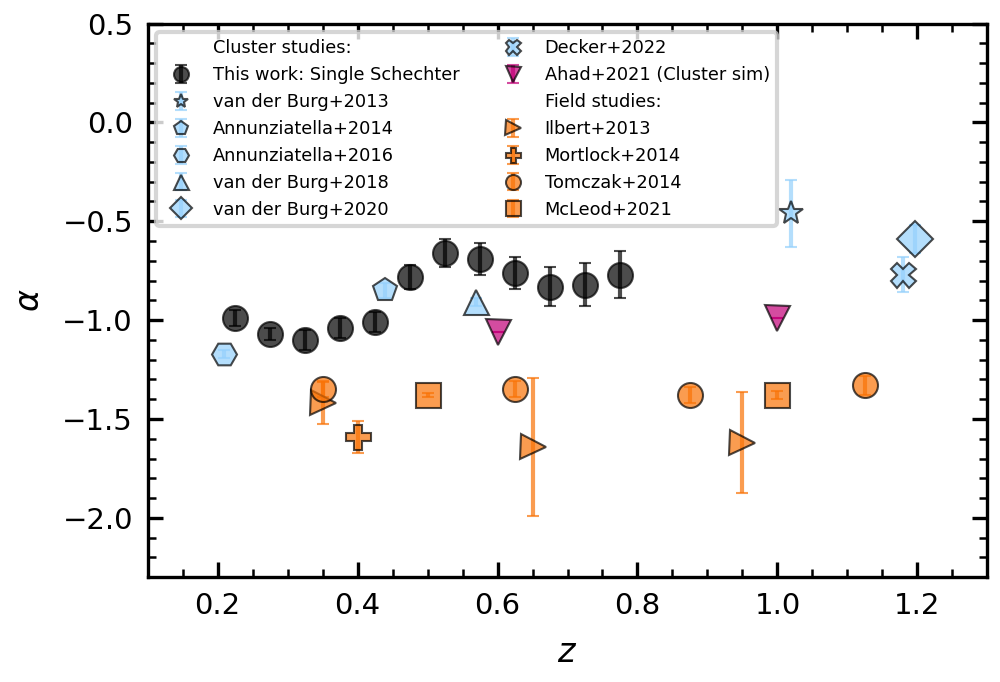}}\\

    \caption{The evolution of the characteristic stellar mass ($M^* = \log_{10}[M^*_*/\mathrm{M_{\odot}}]$) and low-mass slope ($\alpha$) with redshift for each of the 12 redshift bins, which are represented by black circles. The results from previous field (shown as orange markers) and cluster studies (shown as light blue markers) are also included, either plotted at the median redshift of the respective redshift bin (for field studies) or at the median redshift of the cluster sample (for cluster studies). The data points for \citet{Annunziatella2014, Annunziatella2016} are plotted at the cluster redshift, since these are single-cluster studies. It is to be noted that \citet{vanderBurg2013,vanderBurg2018} exclude BCGs in their SMF measurements. $M^*$ and $\alpha$ values from the cluster simulation study by \protect\cite{Ahad2021} are represented by magenta triangle down markers. Each of our values is represented at the median of the redshift bin. Our results indicate a degeneracy between $M^*$ and $\alpha$. }
    \label{fig:M,alphavsz}

\end{figure*}

\subsection{The Schechter+Gaussian model}
\label{subsec:Schechter+Gaussianmodel}

Although we do find a single Schechter function to be an adequate fit to our composite cluster SMFs, we have observed deviations in stellar mass from the Schechter function at the high-mass end (see Fig. \ref{fig:SinglSchechterSMFs}). As noted in Section \ref{subsec:InclusionofBCGs}, similar observations of an ``excess'' in stellar mass at the high-mass end have been made in previous cluster SMF studies (e.g., \citealp{Vulcani2013}; \citealp{vanderBurg2018}).

A large portion of the excess in stellar mass at the high-mass end we observe could be a result of the inclusion of BCGs in our data (e.g., \citealp{Calvi2013}; \citealp{vanderBurg2018}). BCGs are the most massive galaxies in their respective cluster, and as a result, occupy the highest stellar mass bins in our mass range, i.e. $10^{11.0} - 10^{12.5} \mathrm{M_\odot}$. Previous observations show that the luminosity distribution of BCGs tends to follow a Gaussian distribution (e.g., \citealp{Postman1995}; \citealp{Hansen_2005}; \citealp{deFilippis2011}). As an example, a recent study by \cite{Cuillandre2025} using early release Euclid observations of the Perseus galaxy cluster ($z = 0.0167$) found that a Schechter+Gaussian function was effective in modelling the LF of cluster galaxies. Thus, analogous to the cluster LF, the total composite SMF of member galaxies in clusters is the summation of a Schechter and a Gaussian function \citep{WenHan2015}.

Therefore, we model the total composite cluster stellar mass function by using a combined Schechter+Gaussian function. For the Gaussian component, we adopt a similar functional form as described by \cite{WenHan2015},

\begin{equation}
    \phi_\mathrm{g} (M) \ dM = \frac{\phi_{\mathrm{MG}}}{\sqrt{2 \pi} \sigma_{\mathrm{MG}}} \ \exp \left[- \frac{(M - M_{\mathrm{MG}})^2}{2\sigma_{\mathrm{MG}}^2} \right] \ dM ,
	\label{eq:Gaussian}
\end{equation}
where $\phi_{\mathrm{MG}}$ is the Gaussian amplitude or normalisation, $M_{\mathrm{MG}} = \log_{10}[\bar{M}_{\mathrm{MG},*}/\mathrm{M_\odot}]$ is the median log stellar mass of the distribution, and $\sigma_{\mathrm{MG}}$ is the width of the stellar mass distribution. We add this function to the Schechter function described by Eqn. \ref{eq:SingleSchechter} to give the combined model,

\begin{equation}
    \phi_{\mathrm{tot}} (M) \ dM = \left[ \phi_\mathrm{s} (M) + \phi_\mathrm{g}(M) \right] \ dM .
	\label{eq:Schechter+Gaussian}
\end{equation}

\begin{figure*}
    \centering
    \subfloat{\includegraphics[width=1.0\columnwidth]{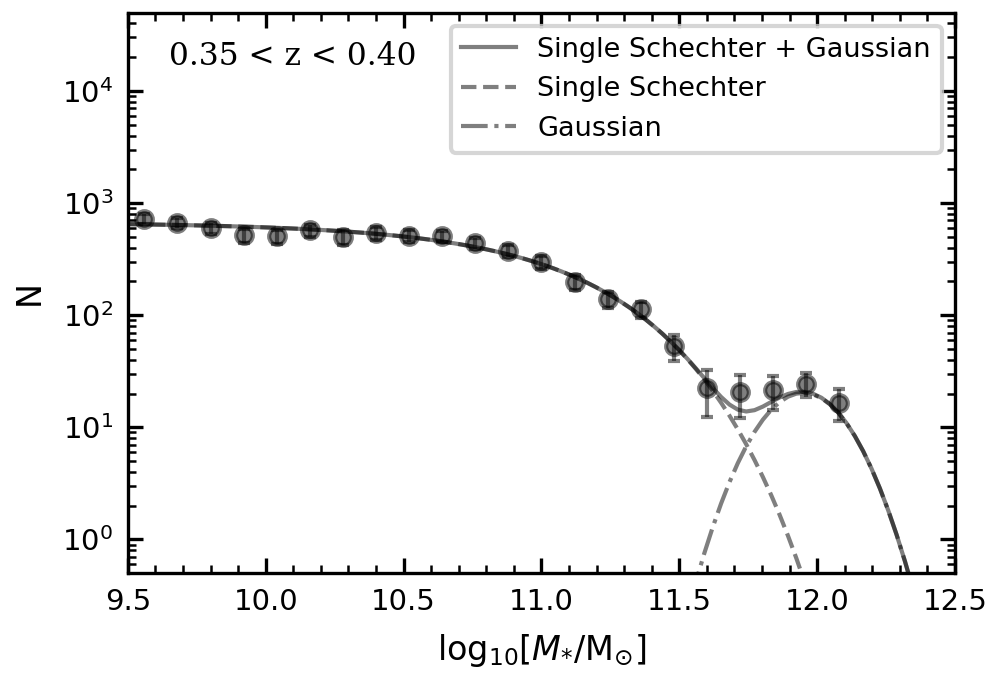}}
    \subfloat{\includegraphics[width=1.0\columnwidth]{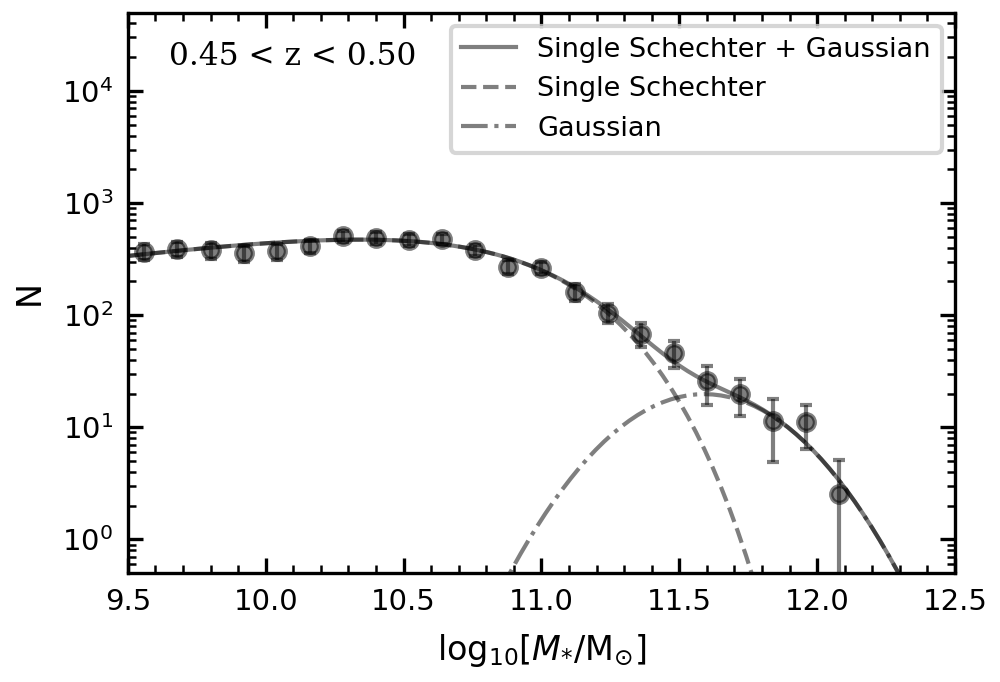}}\\
    
    \caption{The composite cluster SMFs for two redshift bins ($\langle z \rangle= 0.375$ and $\langle z \rangle = 0.475$), modelled with a Schechter+Gaussian function. The excess in stellar mass at the high-mass end, above the Schechter function, is appropriately described by the Gaussian model. Parameter values for the best-fit Schechter+Gaussian model are presented in Table \ref{tab:Schechter+Gaussian_table}.}
    \label{fig:ExampleSchechter+GaussianSMFs}
\end{figure*}

We use 50,000 MCMC iterations to estimate the best-fitting Schechter+Gaussian parameters ($M^*, \alpha, \phi_{\mathrm{MG}}, M_{\mathrm{MG}}$, and $\sigma_{\mathrm{MG}}$) and their uncertainties. The final product values for each parameter must satisfy the convergence condition $\hat{R}-1 <0.01$. The estimated Schechter+Gaussian parameters and their uncertainties are listed in Table \ref{tab:Schechter+Gaussian_table}. As observed for the single Schechter model (see Section \ref{subsec:SingleSchechter}), the Schechter+Gaussian parameters of the cluster-mass subsets are within $ \approx 2\sigma$ of the full cluster sample. Since there is an absence of any notable evolution of the model parameters with cluster mass, we only examine the Schechter+Gaussian fit to the composite SMF for the total cluster sample per redshift bin.  We show the resulting fits to the composite SMF for two redshift bins in Fig. \ref{fig:ExampleSchechter+GaussianSMFs}, and the remainder in Fig. \ref{fig:Schechter+GaussianSMFs}.

A visual inspection of the Schechter+Gaussian model fits to the composite SMFs (shown in Fig. \ref{fig:Schechter+GaussianSMFs}) reveals that the model is successful in accounting for the excess in the stellar mass at the high-mass end. The Schechter+Gaussian parameters, $M^*$ and $\alpha$, duplicate the degeneracy we observed for the single Schechter parameter values. There is evidence of modest evolution in $M^*$ ($2.1 \sigma$ decline) and $\alpha$ ($2 \sigma$ increase) from $z = 0.8$ to $z = 0.2$. Analogously to the single Schechter model (see Section \ref{subsec:SingleSchechter}), we find a stronger evolution of $M^*$ ($4.6 \sigma$ difference) and $\alpha$ ($4.7\sigma$ difference) from $0.20 < z < 0.55$. Compared with estimated parameters for the single Schechter model (see Table \ref{tab:smf_table}), the Schechter+Gaussian $M^*$ and $\alpha$ values only differ by $ \approx 1.4 \sigma$ and $\approx 1\sigma$, respectively, across all redshift bins. The AIC model values (see Table \ref{tab:smf_table} and \ref{tab:Schechter+Gaussian_table}) indicate that the combined Schechter+Gaussian function is the better overall model for the total composite cluster SMF from $0.25 < z < 0.65$, which is $67$ per cent of the redshift bins. However, we find higher AIC values for the Schechter+Gaussian model, compared to the single Schechter model, in the last three redshift bins ($0.65 < z < 0.80$). This is further indicated by $\Delta$AIC (see Table \ref{tab:Schechter+Gaussian_table}), which represents the difference in AIC values between the single Schechter and Schechter+Gaussian models. A value of $\Delta$AIC $=-5.99 (-9.21)$ corresponds to relative likelihood of 5 per cent (1 per cent) between the models being compared. The negative $\Delta$AIC values in the highest redshift bins imply that the Schechter+Gaussian model is not required at high-$z$, and that a single Schechter fit is adequate in describing the composite cluster SMF. The first redshift bin ($\langle z \rangle = 0.225$) has a $\Delta$AIC $=0.79$, which is below the minimum threshold required to select the best model. For this bin, a single Schechter model is perfectly adequate in describing the cluster SMF for the total bin sample.

The Gaussian model component provides valuable insights into the distribution of massive galaxies across different redshift bins. The median stellar mass of the Gaussian distribution ($M_{\mathrm{MG}}$) at the high-mass end has a downward trend with increasing redshift (left panel of Fig. \ref{fig:M_MG,sigma_MGvsz}), significantly decreasing by $4 \sigma$ from $M_{\mathrm{MG}} = 12.01 \pm 0.13$ at $z = 0.2$ to $M_{\mathrm{MG}} = 11.19 \pm 0.16$ at $z = 0.8$, for the total cluster SMF. This indicates that the distribution of massive cluster galaxies is centred around a higher stellar mass ($M_{*} \approx 10^{12} \mathrm{M_{\odot}}$) at low-$z$ ($z < 0.4$). For both cluster-mass subsets, there is an equally strong evolution ($\approx 4\sigma$) of $M_{\mathrm{MG}}$ with redshift.

The redshift evolution of the width of the Gaussian distribution ($\sigma_{\mathrm{MG}}$) reveals only a slight increase ($1 \sigma$) from $\sigma_{\mathrm{MG}} = 0.25 \pm 0.08$ at $z \approx 0.2$ to $\sigma_{\mathrm{MG}} = 0.40 \pm 0.12$ at $z \approx 0.6$, with a moderate flattening at higher redshift ($z \gtrsim 0.6$). However, $\sigma_{\mathrm{MG}}$ is poorly constrained, particularly at high-$z$. The redshift evolution of $\sigma_{\mathrm{MG}}$ can be visualised in the right panel of Fig. \ref{fig:M_MG,sigma_MGvsz}.

While we do attribute the main component of the ``bump'' in stellar mass to the inclusion of BCGs, there could also be additional contributors. For instance, \cite{Cuillandre2025} still found that the Schechter+Gaussian function described the LF of cluster galaxies really well, even when the three brightest central cluster galaxies were excluded. They suggest that the Gaussian component detected in their LF and SMF is likely due to massive galaxies dominating the faint/high-mass end.  Additionally, the use of photo-$z$s to determine cluster membership could have caused contamination by interlopers \citep{Vulcani2013}. Foreground galaxies tend to appear brighter, and hence more massive, when assumed to be at the cluster redshift. This can artificially inflate the number of galaxies contributing to the high-mass end, resulting in this observed bump. However, such interlopers in the SMF are unlikely since we use $P_{\mathrm{mem}}$, calculated from the full $p(z)$ distribution for each galaxy, as a weighting in our cluster membership.

\begin{table*}
        \centering
        \setlength{\tabcolsep}{4pt}
	\caption{The best-fitting Schechter+Gaussian function parameters, where $M^* =$ log$_{10}[M^*_*/\mathrm{M_\odot}]$, for the 12 redshift and cluster-mass binned composite SMFs. It is to be noted that we do not provide uncertainties for the Schechter function normalisation parameter ($\phi^*$) since it is not held as a free parameter in the fitting procedure. The Gaussian model components are represented as follows: $\phi_{\mathrm{MG}}$ is the Gaussian model normalisation, $M_{\mathrm{MG}} = \log_{10}[\bar{M}_{\mathrm{MG},*}/\mathrm{M_\odot}]$ is the median log stellar mass of the Gaussian model, and $\sigma_{\mathrm{MG}}$ is the width of the Gaussian component. The final three columns include the $\chi^2_\nu$, AIC and $\Delta$AIC values. $\Delta$AIC is the difference between the AIC statistic of the single Schechter model and the Schechter+Gaussian model. A negative $\Delta$AIC indicates a higher AIC value for the Schechter+Gaussian model.}
	\label{tab:Schechter+Gaussian_table}
        \renewcommand{\arraystretch}{1.3}
	\begin{tabular}{lccccccccccr}
		\hline
		Redshift range & Sample & $\phi^*$ & $M^*$ & $\alpha$ & $\phi_{\mathrm{MG}}$ & $M_{\mathrm{MG}}$ & $\sigma_{\mathrm{MG}}$ & $\chi_{\nu}^2$ & AIC & $\Delta$AIC\\
		\hline
        
		$0.20 < z < 0.25$ & All & 312.66 & $11.05 \pm 0.03$ & $-0.99 \pm 0.04$ &$2.15 \pm 0.66$ & $12.01 \pm 0.13$ & $0.25 \pm 0.08$ & 0.46 & 17.78 & 0.79 \\
        
        & Low-cluster-mass & 48.47 & $11.11 \pm 0.08$ & $-1.06 \pm 0.08$ & $1.50 \pm 0.73$ & $11.99 \pm 0.14$ & $0.28 \pm 0.08$ & 0.33 & 14.97 & -3.62\\
        & High-cluster-mass & 172.21 & $10.94 \pm 0.05$ & $-0.92 \pm 0.06$ & $1.58 \pm 0.70$ & $11.95 \pm 0.15$ & $0.29 \pm 0.08$ & 0.35 & 16.03 & -2.00\\
        \hline
        
        $0.25 < z < 0.30$ & All & 379.69 & $11.12 \pm 0.03$ & $-1.03 \pm 0.03$ & $6.90 \pm 1.11$& $11.88 \pm 0.06$ & $0.19 \pm 0.07$ & 1.05 & 27.88 & 20.57 \\

        & Low-cluster-mass & 86.21 & $11.15 \pm 0.07$ & $-1.04 \pm 0.07$ & $3.32 \pm 1.60$ & $11.77 \pm 0.11$ & $0.35 \pm 0.10$ & 0.37 & 16.26 & -1.94\\
        & High-cluster-mass & 231.35 & $11.01 \pm 0.05$ & $-0.94 \pm 0.05$ & $3.97 \pm 1.63$ & $11.70 \pm 0.09$ & $0.27 \pm 0.09$ & 0.46 & 18.19 & 5.91\\
        
        \hline
    
        $0.30 < z < 0.35$ & All & 205.73 & $11.17 \pm 0.06$ & $-1.03 \pm 0.05$ & $6.51 \pm 1.36$ & $11.85 \pm 0.08$ & $0.22 \pm 0.09$ & 0.52 & 18.85 & 9.49\\

        & Low-cluster-mass & 59.88 & $11.12 \pm 0.10$ & $-1.01 \pm 0.10$ & $3.92 \pm 1.63$ & $11.79 \pm 0.11$ & $0.37 \pm 0.10$ & 0.31 & 15.22 & 0.05\\
        & High-cluster-mass & 103.28 & $11.08 \pm 0.08$ & $-0.97 \pm 0.08$ & $4.20 \pm 1.73$ & $11.79 \pm 0.11$ & $0.32 \pm 0.11$ & 0.31 & 15.31 & 0.58\\

        \hline

        $0.35 < z < 0.40$ & All & 278.33 & $11.09 \pm 0.05$ & $-1.01 \pm 0.05$ & $7.13 \pm 0.89$ & $11.95 \pm 0.04$ & $0.14 \pm 0.04$ & 0.64 & 20.84 & 26.42\\

        & Low-cluster-mass & 71.98 & $11.01 \pm 0.09$ & $-0.97 \pm 0.10$ & $3.43 \pm 1.58$ & $11.85 \pm 0.11$ & $0.35 \pm 0.12$ & 0.35 & 15.92 & 0.12\\
        & High-cluster-mass & 135.31 & $11.08 \pm 0.07$ & $-0.99 \pm 0.07$ & $5.95 \pm 1.53$ & $11.89 \pm 0.09$ & $0.27 \pm 0.10$ & 0.52 & 18.84 & 8.30\\

        \hline

        $0.40 < z < 0.45$ & All & 312.36 & $11.08 \pm 0.05$ & $-0.98 \pm 0.05$ & $6.34 \pm 1.45$ & $11.84 \pm 0.10$ & $0.27 \pm 0.10$ & 0.65 & 21.67 & 6.70\\

        & Low-cluster-mass & 80.33 & $11.01 \pm 0.10$ & $-0.93 \pm 0.10$ & $3.78 \pm 1.59$ & $11.79 \pm 0.11$ & $0.37 \pm 0.10$ & 0.35 & 15.92 & 0.67\\
        & High-cluster-mass & 144.81 & $11.09 \pm 0.07$ & $-0.99 \pm 0.07$ & $3.71 \pm 1.71$ & $11.75 \pm 0.11$ & $0.34 \pm 0.10$ & 0.30 & 15.69 & -1.60\\
        
        \hline    

        $0.45 < z < 0.50$ & All & 403.53 & $10.85 \pm 0.05$ & $-0.69 \pm 0.07$ & $12.89 \pm 3.47$ & $11.59 \pm 0.10$ & $0.26 \pm 0.05$ & 0.53 & 19.01 & 13.87\\

        & Low-cluster-mass & 114.02 & $10.80 \pm 0.11$ & $-0.62 \pm 0.15$ & $4.12 \pm 2.96$ & $11.50 \pm 0.19$ & $0.27 \pm 0.07$ & 0.25 & 13.93 & -0.93\\
        & High-cluster-mass & 172.92 & $10.82 \pm 0.09$ & $-0.65 \pm 0.11$ & $5.75 \pm 3.75$ & $11.52 \pm 0.17$ & $0.27 \pm 0.07$ & 0.30 & 14.80 & -0.02\\
        
        \hline
            
        $0.50 < z < 0.55$ & All & 428.55 & $10.78 \pm 0.05$ & $-0.57 \pm 0.08$ & $14.27 \pm 3.71$ & $11.56 \pm 0.15$ & $0.38 \pm 0.08$ &  0.66 & 21.28 & 15.10\\

        & Low-cluster-mass & 117.89 & $10.73 \pm 0.10$ & $-0.51 \pm 0.15$ & $4.14 \pm 2.11$ & $11.63 \pm 0.21$ & $0.39 \pm 0.09$ & 0.22 & 13.82 & 1.26\\
        & High-cluster-mass & 167.33 & $10.79 \pm 0.09$ & $-0.53 \pm 0.13$ & $6.50 \pm 3.57$ & $11.49 \pm 0.21$ & $0.35 \pm 0.10$ & 0.33 & 15.00 & -0.89\\
        
        \hline
            
        $0.55 < z < 0.60$ & All & 348.89 & $10.78 \pm 0.06$ & $-0.62 \pm 0.08$ & $11.47 \pm 4.52$ & $11.42 \pm 0.16$ & $0.36 \pm 0.09$ & 0.30 & 15.11 & 3.62\\

        & Low-cluster-mass & 85.64 & $10.71 \pm 0.13$ & $-0.58 \pm 0.18$ & $4.53 \pm 3.31$ & $11.31 \pm 0.21$ & $0.37 \pm 0.10$ & 0.18 & 12.85 & -3.30\\
        & High-cluster-mass & 162.09 & $10.71 \pm 0.09$ & $-0.54 \pm 0.14$ & $6.47 \pm 4.31$ & $11.31 \pm 0.20$ & $0.36 \pm 0.10$ & 0.29 & 14.36 & -2.70\\

        \hline
           
        $0.60 < z < 0.65$ & All & 287.48 & $10.82 \pm 0.07$ & $-0.70 \pm 0.09$ & $10.33 \pm 4.65$ & $11.41 \pm 0.19$ & $0.40 \pm 0.12$ & 0.18 & 13.01 & 1.88\\

        & Low-cluster-mass & 80.29 & $10.69 \pm 0.13$ & $-0.59 \pm 0.20$ & $4.69 \pm 3.18$ & $11.34 \pm 0.22$ & $0.43 \pm 0.13$ & 0.15 & 12.22 & -4.06\\
        & High-cluster-mass & 113.90 & $10.82 \pm 0.12$ & $-0.74 \pm 0.14$ & $5.48 \pm 4.23$ & $11.26 \pm 0.20$ & $0.40 \pm 0.13$ & 0.12 & 11.94 & -4.35\\

        \hline

        $0.65 < z < 0.70$ & All & 153.18 & $10.85 \pm 0.09$ & $-0.79 \pm 0.11$ & $6.61 \pm 4.35$ & $11.26 \pm 0.18$ & $0.40 \pm 0.13$ & 0.14 & 12.15 & -5.25\\

        & Low-cluster-mass & 41.33 & $10.79 \pm 0.21$ & $-0.80 \pm 0.23$ & $4.60 \pm 3.36$ & $11.24 \pm 0.19$ & $0.40 \pm 0.13$ & 0.13 & 11.72 & -5.95\\
        & High-cluster-mass & 71.48 & $10.74 \pm 0.14$ & $-0.67 \pm 0.19$ & $4.48\pm 3.29$ & $11.24 \pm 0.19$ & $0.40 \pm 0.13$ & 0.04 & 10.56 & -5.69\\

        \hline

        $0.70 < z < 0.75$ & All & 157.44 & $10.76 \pm 0.11$ & $-0.75 \pm 0.12$ & $6.42 \pm 4.26$ & $11.24 \pm 0.17$ & $0.37 \pm 0.13$ & 0.19 & 13.08 & -4.49 \\

        & Low-cluster-mass & 42.69 & $10.64 \pm 0.22$ & $-0.64 \pm 0.27$ & $5.09 \pm 3.74$ & $11.33 \pm 0.20$ & $0.41 \pm 0.14$ & 0.13 & 11.44 & -6.34\\
        & High-cluster-mass & 76.94 & $10.65 \pm 0.16$ & $-0.66 \pm 0.21$ & $5.58 \pm 3.68$ & $11.27 \pm 0.19$ & $0.41 \pm 0.13$ & 0.11 & 11.37 & -5.16\\

        \hline

        $0.75 < z < 0.80$ & All & 122.88 & $10.79 \pm 0.12$ & $-0.70 \pm 0.14$ & $7.61 \pm 4.39$ & $11.19 \pm 0.16$ & $0.39 \pm 0.13$ & 0.21& 13.38 & -4.36\\

        & Low-cluster-mass & 34.81 & $10.67 \pm 0.23$ & $-0.64 \pm 0.28$ & $3.86 \pm 2.69$ & $11.12 \pm 0.16$ & $0.39 \pm 0.13$ & 0.14 & 11.90 & -5.52\\
        & High-cluster-mass & 51.21 & $10.71 \pm 0.19$ & $-0.63 \pm 0.24$ & $4.95 \pm 3.47$ & $11.14 \pm 0.16$ & $0.39 \pm 0.13$ & 0.09 & 11.31 & -5.35\\
            
		\hline
	\end{tabular}

\end{table*}

\begin{figure*}
    \centering
    \subfloat{\includegraphics[width=1.0\columnwidth]{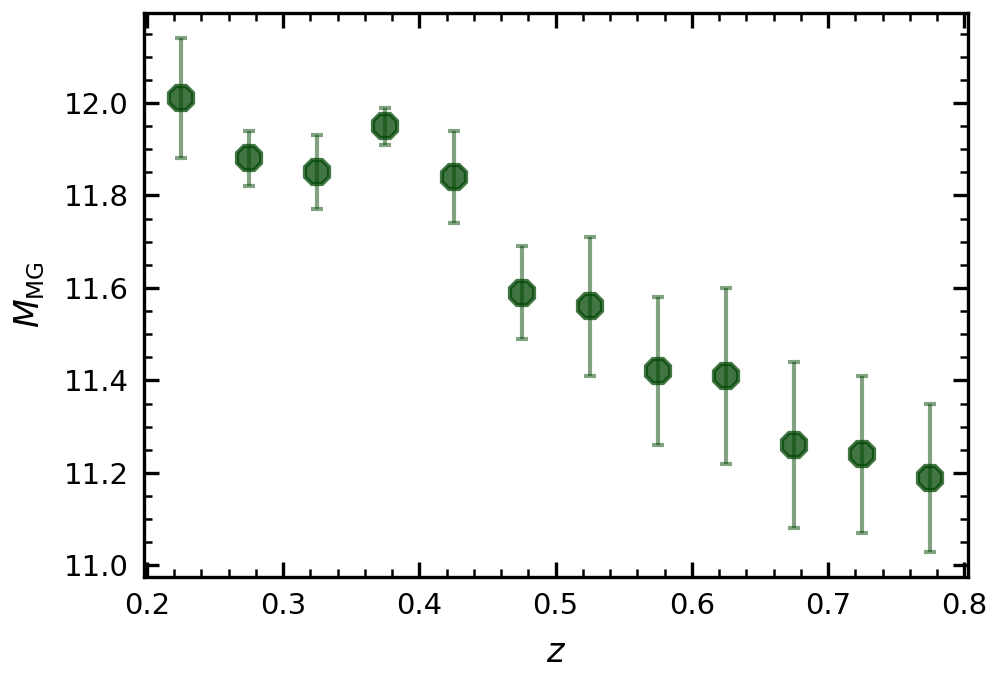}}
    \subfloat{\includegraphics[width=1.0\columnwidth]{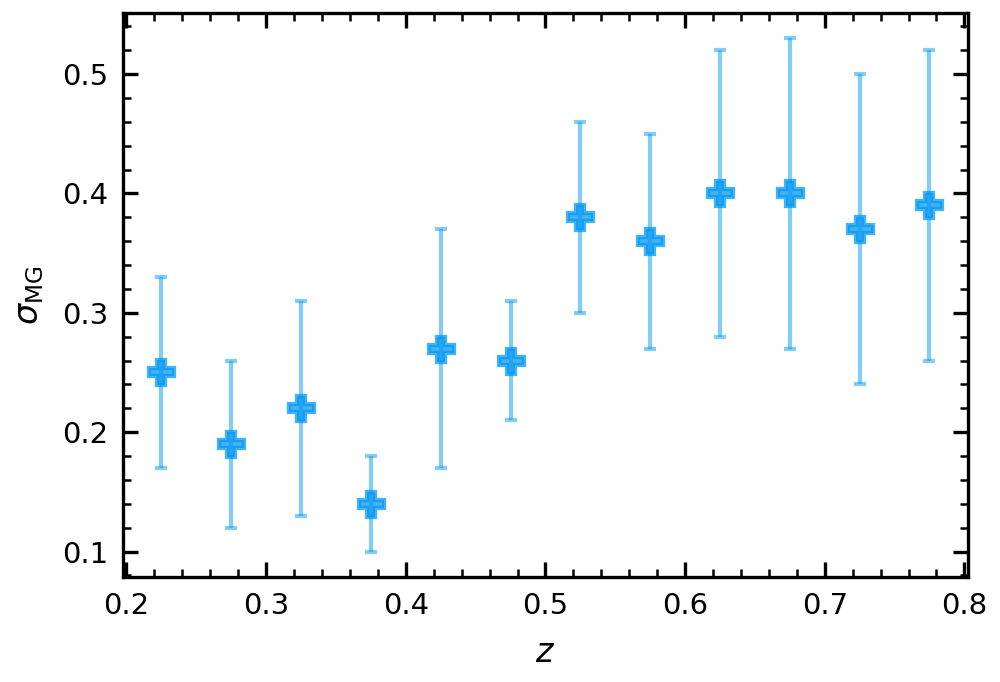}}\\

    \caption{The evolution of the Schechter+Gaussian model parameters: $M_{\mathrm{MG}} = \log_{10}[\bar{M}_{\mathrm{MG},*}/\mathrm{M_\odot}]$ and $\sigma_{\mathrm{MG}}$, with redshift. Error bars represent the $1\sigma$ uncertainties estimated from parameter posterior distributions obtained from MCMC sampling. }
    \label{fig:M_MG,sigma_MGvsz}
\end{figure*}

\subsection{Cluster stellar mass - halo mass scaling relation}
\label{subsec:ClusterMass}

The relationship between the stellar mass content of galaxy clusters and their total halo mass reflects the complete history of galaxy formation and assembly. In order to further examine the stellar mass growth in our cluster sample, we examine the scaling relationship between cluster stellar mass and cluster mass ($M_{\mathrm{200m}}$). To estimate cluster stellar mass within $R_{\mathrm{200m}}$, we take the sum of cluster galaxy (cg) stellar masses multiplied by the $p(z)$ probability that the galaxy is a cluster member, $P_{\mathrm{mem}}$. This is given by the following equation:

\begin{equation}
    M^{\mathrm{cg}}_{*} = \sum_i M_{*,i} \times P_{\mathrm{mem},i},
	\label{eq:Clusterstellarmass}
\end{equation}
where $M_{*,i}$ is the stellar mass of the $i$th cluster galaxy member and $P_{\mathrm{mem},i}$ is the $p(z)$ probability of the $i$th cluster galaxy member. 

$M^{\mathrm{cg}}_{*}$ does not refer to the total cluster stellar mass since we do not consider other stellar contributions, such as the intracluster light (ICL). Instead, $M^{\mathrm{cg}}_{*}$ is the total stellar mass from cluster galaxies above $M_{*} = 10^{9.5} \mathrm{M_{\odot}}$ and within $R_{\mathrm{200m}}$. Since the ICL and BCG in clusters are so closely linked, disentangling the stellar mass contribution from BCGs and the ICL is difficult (e.g., \citealp{Rudick_2011}; \citealp{Burke2015}; \citealp{Furnell2021}). Measuring the stellar mass contributions from the ICL is beyond the scope of this work. 

\begin{figure*}
    \centering
    \subfloat{\includegraphics[width=0.70\columnwidth]{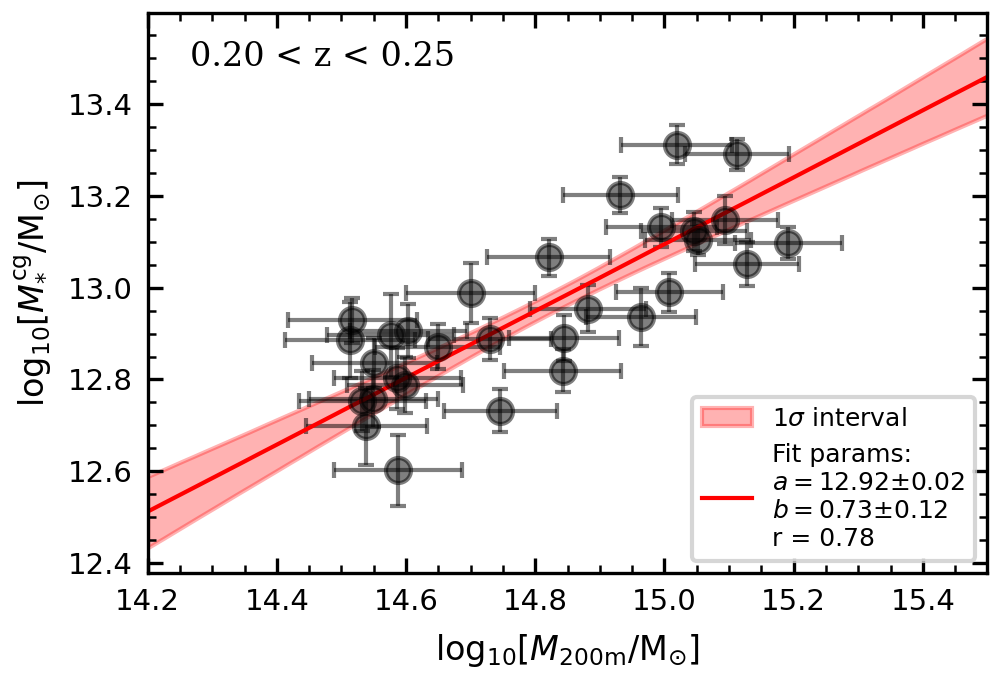}}
    \subfloat{\includegraphics[width=0.70\columnwidth]{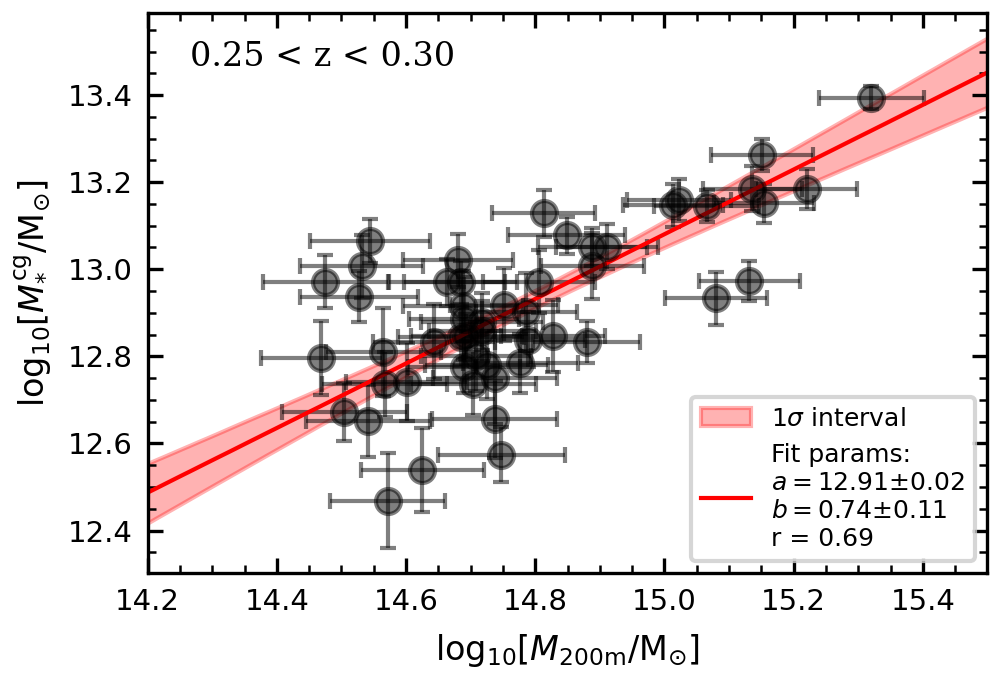}}
    \subfloat{\includegraphics[width=0.70\columnwidth]{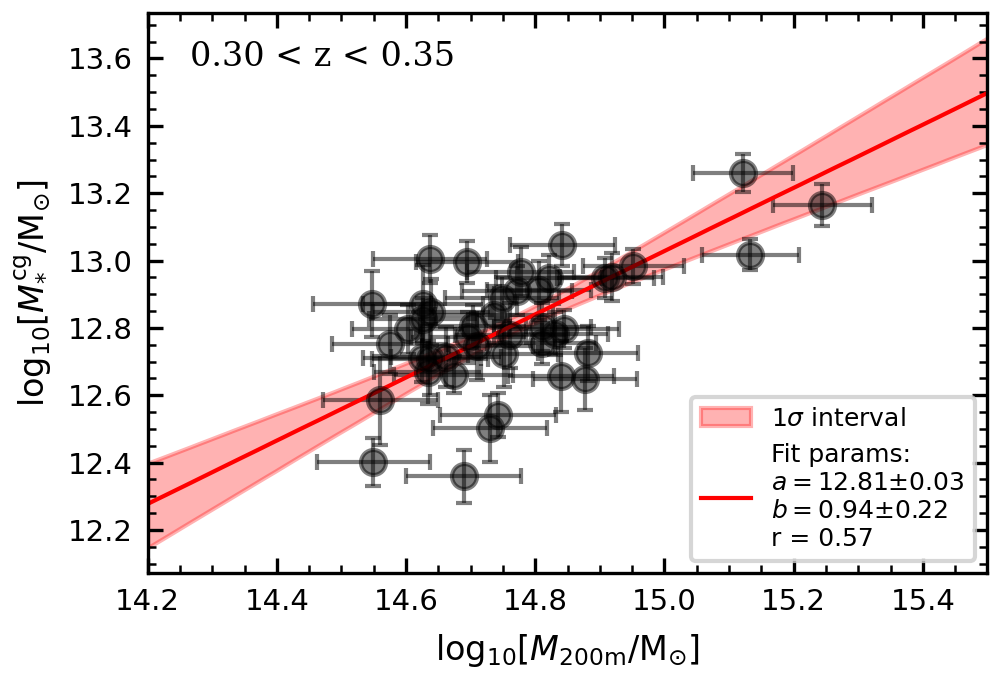}}\\
    \subfloat{\includegraphics[width=0.70\columnwidth]{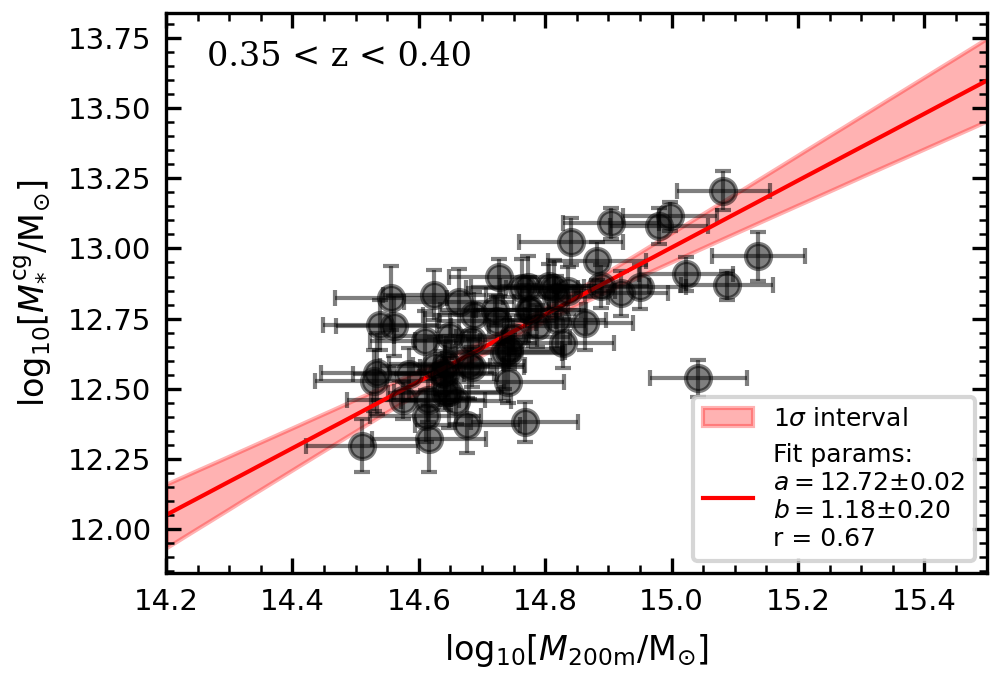}}
    \subfloat{\includegraphics[width=0.70\columnwidth]{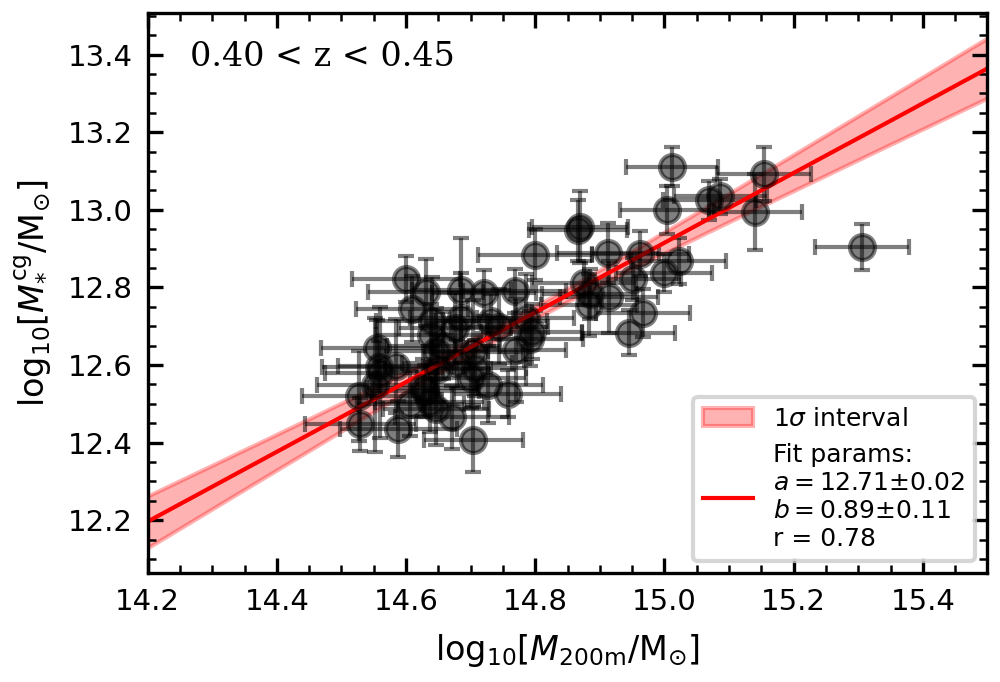}}
    \subfloat{\includegraphics[width=0.70\columnwidth]{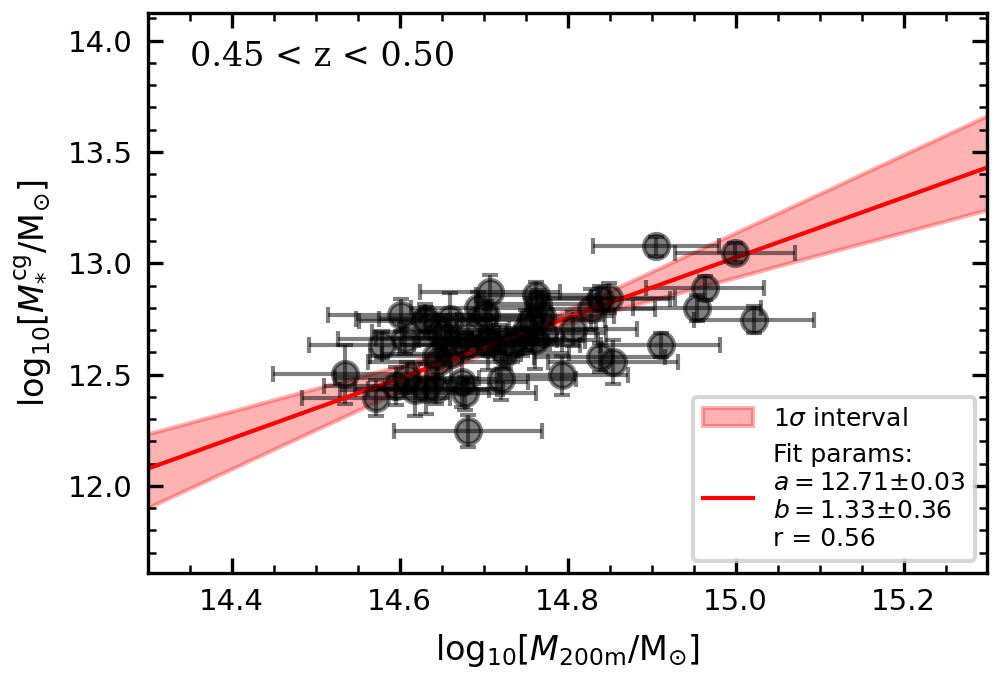}}\\
    \subfloat{\includegraphics[width=0.70\columnwidth]{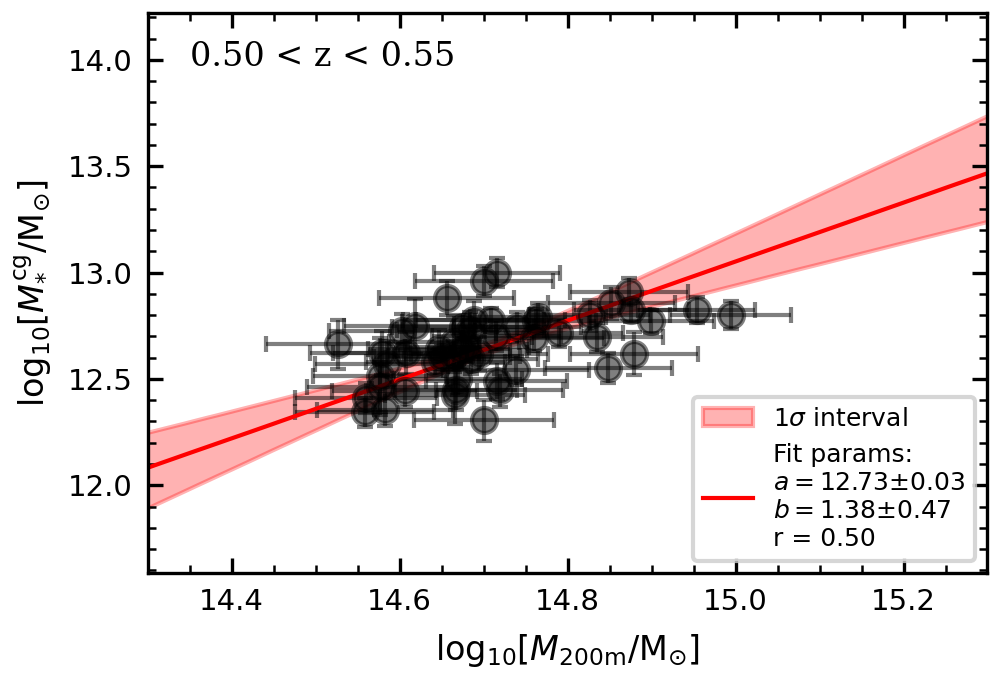}}
    \subfloat{\includegraphics[width=0.70\columnwidth]{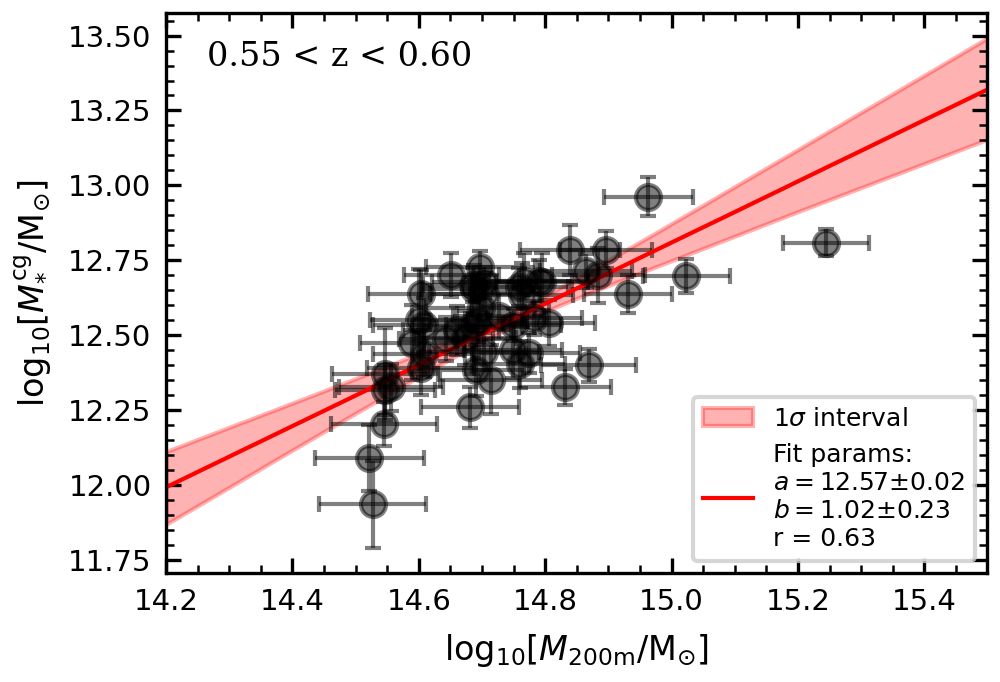}}
    \subfloat{\includegraphics[width=0.70\columnwidth]{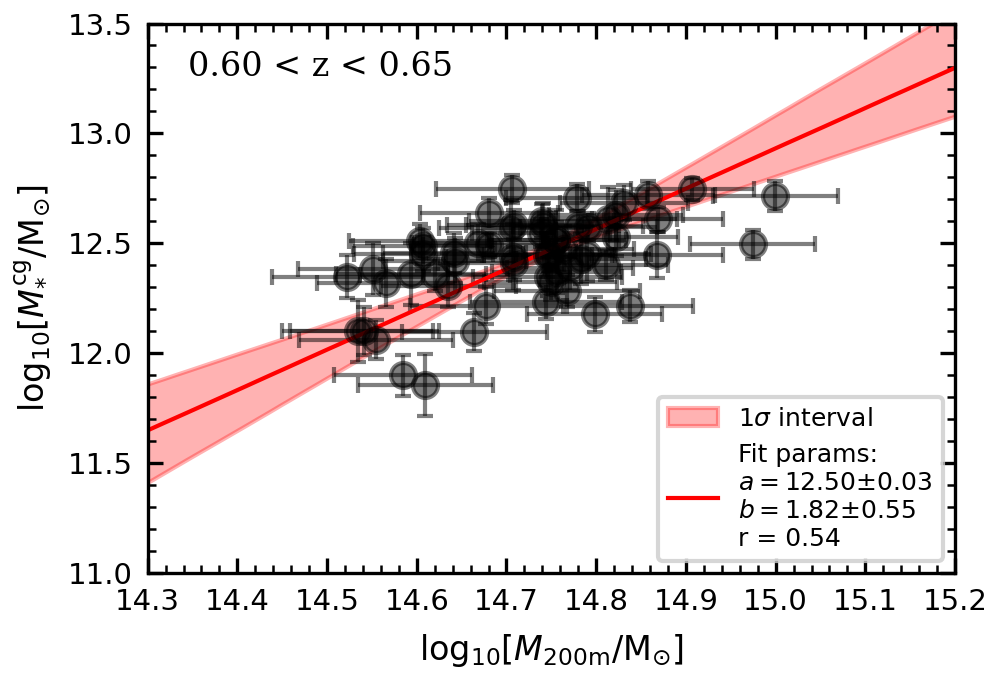}}\\
    \subfloat{\includegraphics[width=0.70\columnwidth]{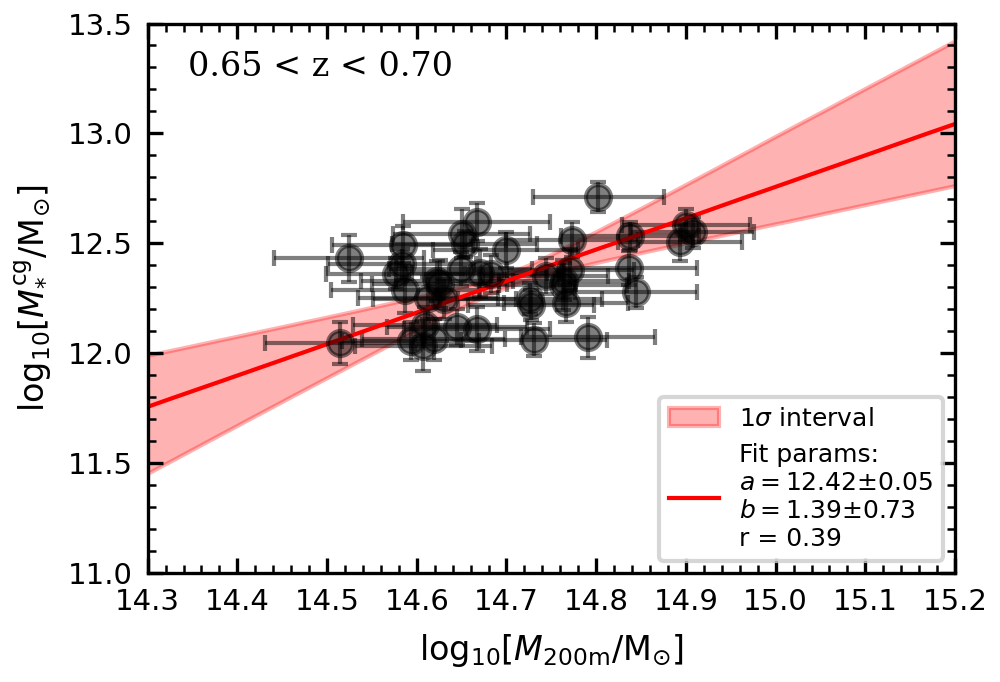}}
    \subfloat{\includegraphics[width=0.70\columnwidth]{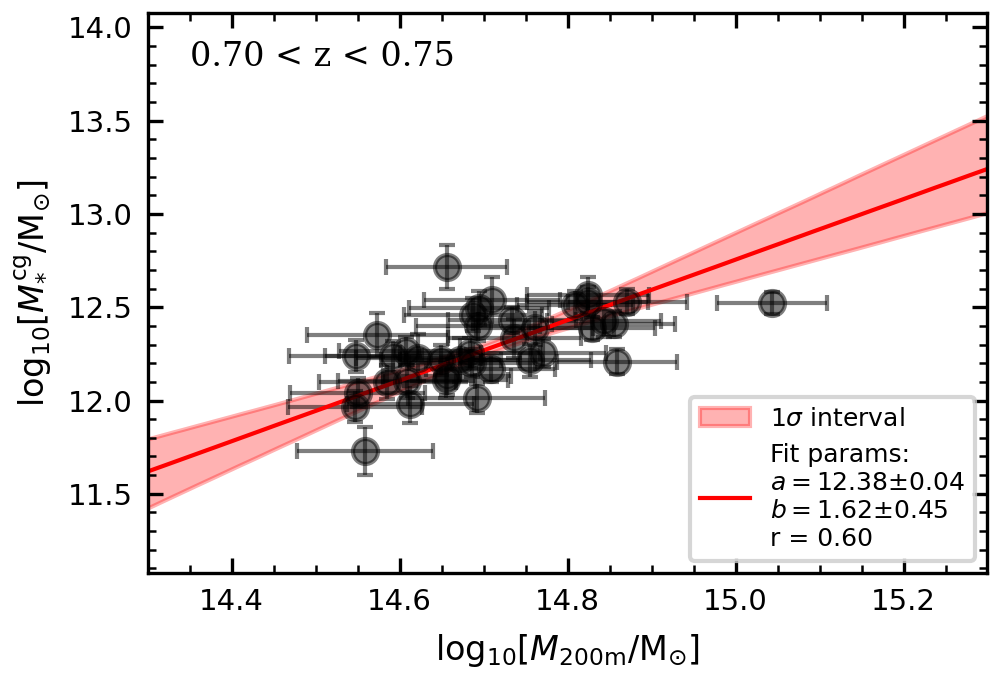}}
    \subfloat{\includegraphics[width=0.70\columnwidth]{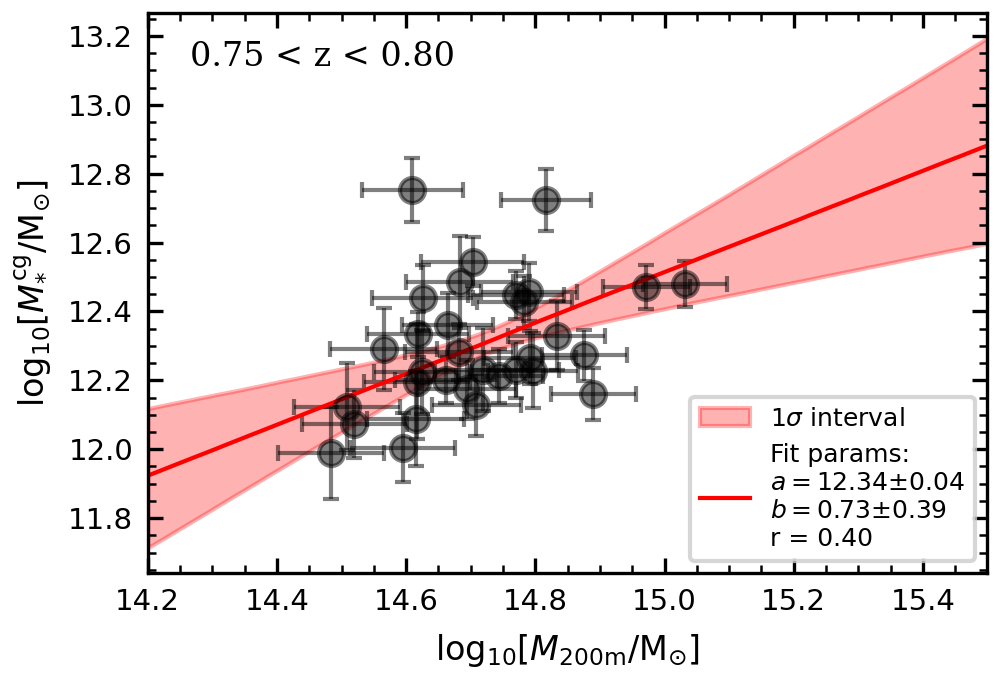}}

    \caption{The scaling relationship between cluster stellar mass ($M^{\mathrm{cg}}_{*}$) and cluster mass ($M_{\mathrm{200m}}$), represented in log-scale, for all 12 redshift bins. The solid red line represents the power-law fit (given by Eqn. \ref{eq:Powerlaw}), with the shaded region showing $1\sigma$ scatter. Time evolution of the slope of the power-law fit ($b$) shows that the $M^{\mathrm{cg}}_{*}$-$M_\mathrm{200m}$ relation does not evolve significantly since $z=0.8$.}
    \label{fig:SMvsCM}
\end{figure*}

The scaling relation between cluster stellar mass and cluster mass can be represented as a power-law relation (e.g., \citealp{Chiu2018}; \citealp{Ahad2021}). $M^{\mathrm{cg}}_{*}$ can be defined as

\begin{equation}
    M^{\mathrm{cg}}_{*} = 10^{a} \cdot \left( \frac{M_{\mathrm{200m}}}{M_{\mathrm{piv}}} \right)^{b} ,
	\label{eq:Powerlaw}
\end{equation}
where $M_{\mathrm{200m}}$ is the cluster mass, $M_{\mathrm{piv}}$ is the pivot mass, $a$ is the intercept and $b$ is the slope of the power-law fit. We fix the pivot mass to the mean $M_{\mathrm{200m}}$ of the total cluster sample, which is $M_{\mathrm{piv}} = 5.83 \times 10^{14} \mathrm{M_{\odot}}$. To perform the power-law model fit, we use the \textsc{linmix} \footnote[3]{\url{https://linmix.readthedocs.io/en/latest/index.html}} \citep{Kelly2007} Python package. \textsc{linmix} uses a Bayesian approach to perform linear regression. In Fig. \ref{fig:SMvsCM}, we present our resulting power-law fit to $M^{\mathrm{cg}}_*-M_{\mathrm{200m}}$ relation for all 12 redshift bins. Table \ref{tab:SMvsCM_table} shows the power-law fit parameters and the Pearson's correlation coefficient ($r$). 

There is a steady increase in the scaling power-law factor ($a$) by $\approx 4.7$ per cent from $z =0.8$ to $z = 0.2$. There is no evidence for the evolution of the power-law fit slope ($b$). However, $b$ is poorly constrained with increasing redshift. We also find a sub-linear power law slope ($b <1$) at $z < 0.3$.

\begin{table}
        \centering
	\caption{The best-fitting parameters of the power-law fit for the $M^{\mathrm{cg}}_{*}-M_{\mathrm{200m}}$ relation for all 12 redshift bins. The final column shows the Pearson correlation coefficient.}
	\label{tab:SMvsCM_table}
        \renewcommand{\arraystretch}{1.3}
	\begin{tabular}{lcccccr}
		\hline
		Redshift range & $N_{\mathrm{clus}}$ & $a$ & $b$ & $r$ \\
		\hline
        
		$0.20 < z < 0.25$ & 30 & $12.92 \pm 0.02$ & $0.73 \pm 0.12$ & 0.78\\
            
            $0.25 < z < 0.30$ & 51 & $12.91 \pm 0.02$ & $0.74 \pm 0.11$ & 0.69\\
            
            $0.30 < z < 0.35$ & 43 & $12.81 \pm 0.03$ & $0.94 \pm 0.22$ & 0.57\\
            
            $0.35 < z < 0.40$ & 61 & $12.72 \pm 0.02$ & $1.18 \pm 0.20$ & 0.67 \\
            
            $0.40 < z < 0.45$ & 61 & $12.71 \pm 0.02$ & $0.89 \pm 0.11$ & 0.78\\
            
            $0.45 < z < 0.50$ & 53 & $12.71 \pm 0.03$ & $1.33 \pm 0.36$ & 0.56\\

            $0.50 < z < 0.55$ & 51 & $12.73 \pm 0.03$ & $1.38 \pm 0.47$ & 0.50\\

            $0.55 < z < 0.60$ & 50 & $12.57 \pm 0.02$ & $1.02 \pm 0.23$ & 0.63\\

            $0.60 < z < 0.65$ & 58 & $12.50 \pm 0.03$ & $1.82 \pm 0.55$ & 0.54\\
            
            $0.65 < z < 0.70$ & 40 & $12.42 \pm 0.05$ & $1.39 \pm 0.73$ & 0.39\\
            
            $0.70 < z < 0.75$ & 38 & $12.38 \pm 0.04$ & $1.62 \pm 0.45$ & 0.60\\
        
            $0.75 < z < 0.80$ & 32 & $12.34 \pm 0.04$ & $0.73 \pm 0.39$ & 0.40\\
            
		\hline
	\end{tabular}

\end{table}

\section{DISCUSSION}
\label{sec:Discussion}

In this section, we discuss the results we have obtained for the composite cluster SMFs in Section \ref{sec:Results}, and the implications of these results with respect to galaxy stellar mass growth within clusters. We additionally compare our results to SMF measurements from previous studies of both field and cluster galaxies.

\subsection{SMF dependence on cluster mass}
\label{subsec:Clustermassdependence}

The results presented in Section \ref{subsec:SingleSchechter} indicate that there are no significant differences ($< 2 \sigma$) between the Schechter parameters ($M^*$ and $\alpha$) of the cluster composite SMFs of the total cluster sample and both cluster mass subsets. This result implies that the shape of the composite cluster SMF is independent of cluster mass above $M_{\mathrm{200m}} \approx 3 \times 10^{14} \mathrm{M_{\odot}}$. 

A study of the dependence of the total SMF on cluster mass was performed by \cite{Vulcani2011}, which also concluded that cluster mass does not significantly impact the shape of the cluster SMF. A study by \cite{Ahad2021} on the cluster SMF for Hydrangea (\citealp{Bahe2017}) simulated clusters have also found no clear dependence in $M^*$ and $\alpha$ on the cluster mass ($M_{\mathrm{500c}}$) at $z \sim 0.6$.

The low-cluster-mass subset appears to be in moderately better agreement ($\lesssim 1 \sigma$) with the total cluster SMF than the high-cluster-mass subset, suggesting that low-mass galaxy clusters ($M_{\mathrm{200m}} < 4.8 \times 10^{14} \mathrm{M_{\odot}}$) tend to follow a more uniform cluster SMF trend than more massive clusters ($M_{\mathrm{200m}} > 5.5 \times 10^{14} \mathrm{M_{\odot}}$). Additionally, we generally observe a flatter $\alpha$ for high-mass clusters compared to the general cluster sample. This suggests a slightly greater abundance of massive galaxies in high-mass clusters, which aligns with galaxy formation models and previous cluster cosmology studies (e.g., \citealp{Stevens2003}; \citealp{Lin_2004}; \citealp{McIntosh2008}).

\subsection{Redshift evolution of $M^*$ and $\alpha$}
\label{subsec:M,alphavsz}

Describing the shape of the SMF by means of the Schechter parameters, $M^*$ and $\alpha$, is crucial to gain a comprehensive view of stellar mass growth. In Section \ref{subsec:SingleSchechter}, we presented the results of fitting a single Schechter function to the composite cluster SMFs. From the results, an anti-correlation is observed between $M^*$ and $\alpha$ as they evolve with redshift (see Fig. \ref{fig:MstarvsAlpha}). Such a degeneracy implies that the inferred evolution of the SMF shape should be interpreted with caution, as changes in one parameter may be reflected in the evolution of the other, rather than a true physical evolution. This correlation or degeneracy has been commonly observed in previous cluster SMF studies (e.g., \citealp{vanderBurg2020}; \citealp{Ahad2021}). In the case of the SMF of simulated clusters at $z \sim 0.6$, \cite{Ahad2021} also found a degeneracy and that a higher $M^*$ correlates with a lower $\alpha$. 

The relatively flat evolution of $M^*$ with redshift (see Section \ref{subsec:SingleSchechter} and Fig. \ref{fig:M,alphavsz}) implies that the typical stellar mass scale in clusters is already established by $z \sim 0.8$. This is consistent with the cosmic downsizing scenario (e.g., \citealp{Pozzetti2010}), in which massive galaxies formed early and have grown only slowly overall thereafter. The observed increase ($\approx 3.3 \sigma$) in $M^*$ from $z = 0.55$ to $z = 0.2$ (see Table \ref{tab:smf_table} and Fig. \ref{fig:M,alphavsz}) suggests that whatever growth occurs (e.g., BCG mergers) is more important at low-$z$. This growth could be a result of the accretion of intermediate-to-high mass ($M_* > 10^{10.75} \mathrm{M_{\odot}}$) galaxies by clusters at $z < 0.55$.

We find our $M^*$ values to be in good agreement ($\lesssim 1.5 \sigma$ difference) with previous cluster SMF studies (\citealp{Annunziatella2014, Annunziatella2016}; \citealp{vanderBurg2018}) and differ by $\lesssim 1.7 \sigma$ for a few previous field SMF studies (\citealp{Mortlock2014}; \citealp{Tomczak_2014}; \citealp{McLeod2021}) at $z < 0.8$. The general (flat) trend observed for our $M^*$ values for $z \gtrsim 0.55$ could be extrapolated to $z \sim 1.2$, where there seems to be a lack of evolution in $M^*$ (\citealp{vanderBurg2013}; \citealp{Ilbert2013}; \citealp{vanderBurg2020}; \citealp{McLeod2021}; \citealp{Decker_2022}). 

The choice to include or exclude BCGs can have an impact on the characteristic stellar mass, either shifting $M^*$ to higher stellar masses or producing lower $M^*$ values. For a consistent comparison with the previous literature, a majority of the studies on the cluster SMF we compare to include BCGs (\citealp{Annunziatella2014, Annunziatella2016}; \citealp{vanderBurg2020}). However, \cite{vanderBurg2013} and \cite{vanderBurg2018} do not include BCGs in their analysis. Despite this, our $M^*$ value at $\langle z \rangle =0.575$ agrees within $\approx 0.8 \sigma$ of $M^* = 10.81 \pm 0.02$ measured by \cite{vanderBurg2018} for the composite SMF of 21 clusters at a median redshift of $z = 0.57$. Additionally, there is only a difference of $\approx 1.4 \sigma$ between our highest redshift bin ($\langle z \rangle = 0.775$) $M^*$ and the value of $M^* = 10.72^{+0.09}_{-0.02}$ measured by \cite{vanderBurg2013} for the composite SMF of 10 red-sequence selected clusters centred at $z = 1.02$.

We also find a consensus for $M^*$ with the cluster simulation study by \cite{Ahad2021}, who noted a similar slight decrease in $M^*$ with redshift. In Fig. \ref{fig:M,alphavsz}, we only include comparisons to two redshift bins from \cite{Ahad2021}. They measured values of $M^* = 10.92$ at $z \sim 0.6$, and $M^* = 10.91$ at $z \sim 1.0$ for their simulated clusters. The value at lower redshift ($z \sim 0.6$) is consistent ($1 \sigma$ difference) with the $M^*$ value we find for our redshift bin at a similar redshift ($\langle z \rangle = 0.625$).

The evolution of the low-mass slope of the cluster SMF provides key information about the abundance of low-mass/faint galaxies in clusters as they evolve. In Section \ref{subsec:SingleSchechter}, we find that the mass slope of the composite cluster SMF remains remarkably flat ($\alpha \gtrsim -1$) throughout the redshift interval $0.2 < z < 0.8$. At lower redshifts ($z < 0.55$), $\alpha$ is more negative (see Table \ref{tab:smf_table} and Fig. \ref{fig:M,alphavsz}), which indicates an abundance of low-mass galaxies in the less massive clusters. At higher redshifts ($z \geq 0.55$), $\alpha$ increases and becomes shallower, implying that there are fewer low-mass galaxies compared to high-mass galaxies in clusters during this epoch. This suggests a later infall of low-mass galaxies into clusters, consistent with downsizing (\citealp{Vulcani2011}; \citealp{vanderBurg2015}). For instance, \cite{Vulcani2011} found a flat low-mass slope at $z \sim 0.8$ for the SMF of cluster galaxies with $M_* \lesssim 10^{10.8} \mathrm{M_{\odot}}$.

In direct comparison with more recent work, we find that our $\alpha$ values moderately agree (within $\approx 3\sigma$) with the single cluster SMF study on the MACS J1206.2-0847 cluster at $z =0.44$ \citep{Annunziatella2014}, and the stacked cluster SMF study of 21 Planck clusters at $<\langle z \rangle = 0.57$ \citep{vanderBurg2018}. We do find a strong disagreement ($4\sigma$) between our low-mass slope value at $\langle z \rangle = 0.225$ and that found by \cite{Annunziatella2016} for the SMF of the Abell 209 cluster at $z = 0.209$. A possible factor to be considered for the large difference in $\alpha$ is the cluster sample size. \cite{Annunziatella2016} only perform their measurement of the SMF on a single cluster as opposed to our composite cluster SMF measurements. The SMF can vary from cluster to cluster depending on different cluster galaxy populations and dynamical states. 

The flattening trend in $\alpha$ we observe could be extrapolated to $z \sim 1.2$, where \cite{vanderBurg2020} have documented a similar shallow low-mass slope for their SMF study of 11 GOGREEN clusters from $1.0 < z < 1.4$. Similarly, a recent study on the composite LF and SMF of 12 clusters at $\langle z \rangle = 1.18$, from the Massive and Distant Clusters of WISE Survey (MaDCoWS), by \cite{Decker_2022} found a low-mass slope of $\alpha  = -0.77 \pm 0.09$. We find the same slope value for the composite SMF in our highest redshift bin ($\langle z \rangle = 0.775$).\cite{vanderBurg2013} have also found a very shallow low-mass slope of $\alpha = -0.46^{+0.08}_{-0.26}$ for the total galaxy population of the cluster SMF measured for 10 GCLASS clusters in the redshift interval $0.86 < z < 1.34$. 

A notable difference ($\approx 3.8 \sigma$) is observed between our low-mass slope values and those obtained from the cluster simulation study by \cite{Ahad2021}. This study found a smaller increase ($14$ per cent) in their predicted low-mass slope values at $0.6 < z <1.3$, compared to the larger ($35$ per cent) increase in $\alpha$ during this epoch found by previous cluster SMF observations (\citealp{vanderBurg2013}; \citealp{vanderBurg2020}). The overprediction of low-mass galaxies ($M_* < 10^{10.5} \mathrm{M_{\odot}}$) may point to a failure in simulations at disrupting low-mass satellite galaxies at $z \sim 1$, or highly efficient star formation in low-mass galaxies (\citealp{Ahad2021}). \cite{Cuillandre2025} also find that the measured faint-end/low-mass end slopes ($\alpha = -1.2$ to $-1.3$) of the LF/SMF of galaxies in the Perseus cluster are much flatter than the slope of the dark matter halo mass function predicted by cosmological simulations (e.g., \citealp{Somerville&Dave2015}).

Our slope values are much shallower than previous SMF studies of field galaxies (e.g., \citealp{Ilbert2013}; \citealp{Mortlock2014}; \citealp{Tomczak_2014}; \citealp{McLeod2021}). There is a clear lack of evolution in $\alpha$ with redshift for previous field studies. The difference in $\alpha$ between field and cluster implies that the SMF is dependent on environment - i.e. galaxies evolve differently in field environments compared to dense cluster environments. This is supported by studies that have examined the impact of the environment on the SMF (e.g., \citealp{vanderBurg2013}; \citealp{Etherington2017}; \citealp{vanderBurg2020}). For example, \cite{vanderBurg2013} observed a notable difference in the shape of the total galaxy SMF between the clusters and the field environment at $z \sim 1$. They attribute this difference to a result of a larger proportion of quiescent galaxies in high-density environments such as clusters. A study by \cite{Etherington2017} on the environmental dependence of the SMF using a large galaxy dataset from DES, found a larger fraction of massive galaxies in dense regions than in low-density environments for $z < 0.75$. In addition, simulation studies also point towards a significant difference between the low-mass slope of the field SMF and cluster SMF at $z < 1$ (\citealp{Furlong2015}; \citealp{Ahad2021}). While the field SMF is found to evolve towards a moderately steeper $\alpha$ with increasing redshift, the cluster SMF evolves towards a flatter low-mass slope at high-$z$ \citep{Ahad2021}.

The absence of a strong faint/low-mass end upturn in the composite cluster SMFs (see Section \ref{subsec:SingleSchechter}) is identified by the minimal/lack of deviation of low-mass galaxy number counts from the single Schechter function. This could suggest that clusters are still assembling at $0.2 < z < 0.8$, and many satellite galaxies are recently accreted from the field. This supports the idea of downsizing in quenching, as massive galaxies are quenched earlier, while low-mass galaxies quench later. However, as noted in Section \ref{subsec:SingleSchechter}, a caveat to this analysis is that our shallow stellar mass limit may be disguising a clear upturn that is observed at $M_* < 10^{9.5} \rm{M_{\odot}}$ in field SMF studies (e.g., \citealp{Adams2021}).

\subsection{The growth of massive galaxies}
\label{subsection:GrowthofMGs}

The usage of narrow redshift and stellar mass bins, which was enabled by our large cluster sample, offers the advantage of revealing subtle SMF trends that would otherwise be smoothed over in wider bins. Such a trend is the peak/excess in stellar mass we observed and modelled in Section \ref{subsec:Schechter+Gaussianmodel}. 

The $M^*$ and $\alpha$ values obtained from the Schechter+Gaussian model are not significantly different ($\lesssim 2\sigma$) from the single Schechter values, so in this section we focus on the time evolution of the peak of the Gaussian distribution, $M_{\mathrm{MG}}$, and the width of the distribution, $\sigma_{\mathrm{MG}}$ (see Fig. \ref{fig:M_MG,sigma_MGvsz}).

As noted in Section \ref{subsec:Schechter+Gaussianmodel}, there is a decline in the median stellar mass of the Gaussian distribution ($M_{\mathrm{MG}}$) with increasing redshift. The change in the value indicating growth in stellar mass of the high-mass galaxy population with time ($\Delta M_{\mathrm{MG}}= 0.82$ dex) suggests significant growth ($4 \sigma$) in the massive galaxy population within galaxy clusters since $z = 0.8$. This implies that at lower redshifts ($z \approx 0.2$), galaxies are more massive since they have had more time to grow, while they are less massive earlier in cosmic time ($z \approx 0.8$). A fraction of galaxies that contribute to this excess in stellar mass is the BCG population. Hence, we can infer that the observed declining trend of $M_{\mathrm{MG}}$ with redshift is consistent with the hierarchical formation of BCGs, where these massive galaxies assemble their mass over time through mergers and accretion (e.g., \citealp{Lidman2012}; \citealp{Lin_2013}; \citealp{Ragone-Figueroa2018}). 

The moderate increase of the Gaussian dispersion, $\sigma_{\mathrm{MG}}$, with redshift (see Fig. \ref{fig:M_MG,sigma_MGvsz}), implies that the Gaussian model component is more likely necessary at low-$z$ as the high-mass excess becomes more prominent due to galaxy growth. This is further supported by the AIC statistic comparison between the Schechter+Gaussian and single Schechter models (see Table \ref{tab:Schechter+Gaussian_table} and Section \ref{subsec:Schechter+Gaussianmodel}). However, since $\sigma_{\mathrm{MG}}$ is poorly constrained at high-$z$, our interpretation of the redshift evolution of this parameter is limited.

\subsection{Evolution of galaxy stellar mass content in clusters}
\label{subsec:stellarmasscontent}

\begin{figure}
	\includegraphics[width=\columnwidth]{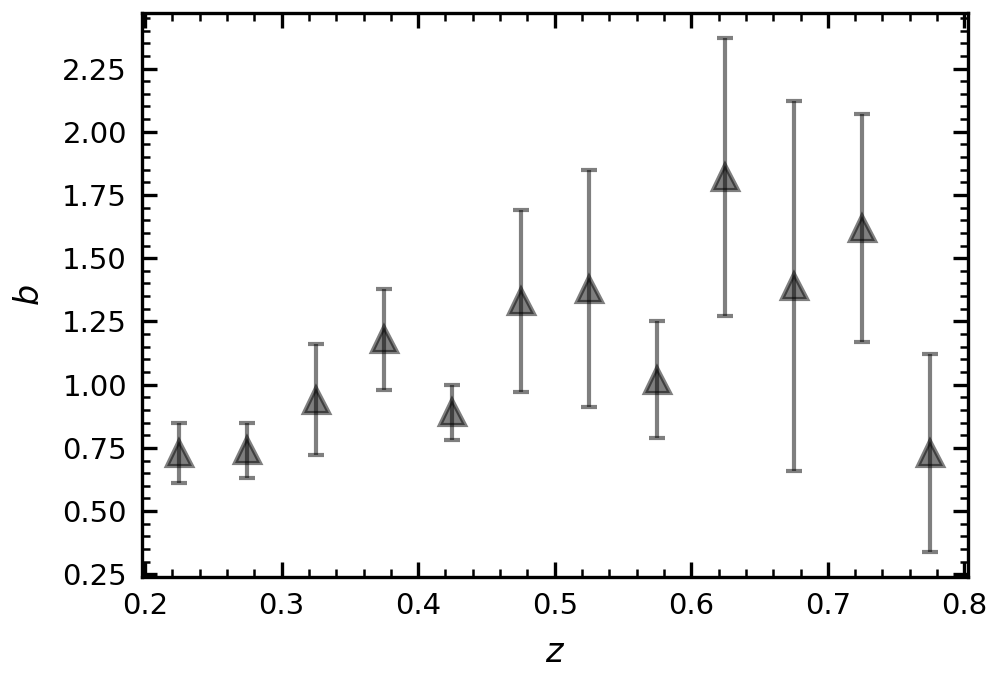}
    \caption{The redshift evolution of the power-law slope ($b$) for the $M_*^{\mathrm{cg}}-M_{\mathrm{200m}}$ relation, represented as black triangles, for each of the 12 redshift bins. A moderate increasing trend for the slope is observed with increasing redshift. However, $b$ is poorly constrained at high-$z$.}
    \label{fig:slopevsz}
\end{figure}

In Section \ref{subsec:ClusterMass}, we presented the scaling relationship between cluster stellar mass ($M^{\mathrm{cg}}_*$) and cluster mass ($M_{\mathrm{200m}}$) for each redshift bin (see Table \ref{tab:SMvsCM_table} and Fig. \ref{fig:SMvsCM}). The evidence of a sub-linear slope ($b <1$) at low-$z$ ($z < 0.3$) suggests that as halo mass increases, star formation efficiency decreases. A sub-linear slope for the cluster stellar mass and halo mass relationship has also been reported in the literature (e.g., \citealp{Lin_2004}; \citealp{Kravtsov2018}). For example, \cite{Kravtsov2018} found that the stellar mass of cluster galaxies correlates with halo mass ($M_{\mathrm{500}}$) with a sub-linear slope of $\sim 0.6 \pm 0.1$. We present the redshift evolution of $b$ in Fig. \ref{fig:slopevsz}. 

Our next step is to use the information from the $M^{\rm{cg}}_*$ - $M_{\rm{200m}}$ relationship to study the stellar content contributions from cluster members towards the total mass budget of galaxy clusters.

\subsubsection{Stellar mass fractions}
\label{subsubsec:stellarmassfractions}

To gain a complete picture of how stellar mass is assembled within galaxy clusters, we must also consider the ratio of stellar mass to total halo mass, i.e. the galaxy cluster stellar mass fraction. Recent studies based on simulations have consistently shown that the stellar mass fraction within clusters is inversely correlated with halo mass, and evolves with redshift, indicative of hierarchical structure formation and changing star formation efficiency (e.g., \citealp{Pillepich2018}; \citealp{Behroozi2019}). To probe this further, we measure stellar mass fractions for our clusters, only considering the stellar mass contribution from cluster galaxies within $R_{\mathrm{200m}}$ and with stellar mass $M_* > 10^{9.5} \mathrm{M_{\odot}}$. The cluster stellar mass fraction for the $i$th cluster in each redshift bin, represented as a percentage, is calculated using the following equation:

\begin{equation}
    f_{*,i}^{\mathrm{cg}} = \frac{ M^{\mathrm{cg}}_{*,i}}{M_{\mathrm{200m},i}} \times 100,
	\label{eq:StellarMassFracs}
\end{equation}
where $M^{\mathrm{cg}}_{*,i}$ is the cluster stellar mass and $M_{\mathrm{200m},i}$ is the total cluster mass of the $i$th cluster in the redshift bin. We generate 5,000 bootstrap samples by resampling with replacement to estimate the average $f^{\mathrm{cg}}_{*}$ for the total cluster sample in each redshift bin and the $1 \sigma$ errors. To study the dependence of $f^{\rm{cg}}_{*}$ on halo mass, we also compute the average fraction for the low- and high-cluster-mass subsets that are defined relative to the halo mass distribution per redshift bin. These are the same mass bins we use for our composite SMF measurements (see Section \ref{sec:method}). Table \ref{tab:Stellarmassfrac_table} lists the estimated average cluster stellar mass fractions in each redshift and cluster mass bin. The left panel of Fig. \ref{fig:stellarmassfracs} shows a plot of the average cluster stellar mass fractions against redshift and lookback time, in Gyr. 

The overall trend for all clusters reveals that cluster stellar mass fractions decrease with increasing redshift (i.e., with lookback time). At low-$z$ ($z = 0.2$), we measure a stellar mass fraction of $f_{*}^{\mathrm{cg}} = 1.49 \pm 0.08$ per cent, which then decreases to $f^{\mathrm{cg}}_{*} = 0.43 \pm 0.04$ per cent at $z = 0.8$. This indicates an approximately $1$ per cent increase in the average stellar mass fraction within the halo population at $z = 0.8 $ to the fraction in the halo population at $z = 0.2$. There also appears to be a considerable difference ($5\sigma$) in $f_{*}^{\mathrm{cg}}$ from $z = 0.6$ to $z = 0.5$, which is where we begin to observe significant evolution of the shape of the SMF (see Section \ref{subsec:M,alphavsz}). 

For the cluster mass subsets, we recover similar trends for $f^{\rm{cg}}_{*}$ with redshift, as observed for the total cluster sample. However, the stellar mass fractions in low-mass clusters are higher than those in more massive clusters, per redshift bin. For $z < 0.3$, the stellar mass fractions in low-mass clusters are $f^{\mathrm{cg}}_{*} \approx 2$ per cent, and are $\approx 0.5$ per cent higher than high-mass clusters in this epoch (see Table \ref{tab:Stellarmassfrac_table}).

Further interpretation of the trends for $f^{\rm{cg}}_{*}$ as an evolutionary sequence for individual haloes is more complicated. In a $\Lambda$CDM framework, cluster-scale haloes grow significantly in mass from $z = 0.8$ to $z = 0.2$. However, our current approach compares different halo populations at each epoch, rather than tracing the progenitors and descendants of the same system. The cluster mass subsets also probe the stellar content in clusters that are relatively high-mass or low-mass at each redshift, rather than tracking the evolution of clusters at fixed halo mass. As such, the trends observed for $f^{\rm{cg}}_*$ combine genuine redshift evolution with changes in the underlying halo population. Interpretation of these trends needs to account for potential changes in the underlying halo mass distribution of the cluster sample.

\subsubsection{Evolving halo mass threshold}
\label{Evolvinghalomassthreshold}

To ensure we compare similar populations of haloes in each redshift bin, we repeat the analysis for a cluster sample selected above an evolving halo mass threshold. This threshold allows us to maintain a constant comoving number density across redshift bins.

For this approach, we adopt the method described by \cite{Golden-Marx2023}, who examined the properties and stellar content of BCGs and ICL in a sample of DES-ACT clusters at $0.2 < z < 0.8$. Our first step is to re-bin our cluster sample into 6 redshift bins of width $0.1$ from $0.2 < z < 0.8$. The same binning was done by \cite{Golden-Marx2023}. The reason for reducing the number of bins from 12 to 6 is to increase the statistical power of the cluster sample in each bin since we have a much smaller initial cluster sample compared to \cite{Golden-Marx2023}. We then compute the comoving volumes for each bin. We select the highest redshift bin ($0.7 < z < 0.8$) as a pivot bin, and compare the volumes of each lower redshift bin relative to the volume of the pivot bin. We use the number density of the pivot bin to obtain the ideal, target number of clusters to select in each bin in order to achieve a constant volume density across redshift bins. This defines a halo mass threshold ($M_{\rm{200m}}^{\rm{thresh}}$) in each bin that evolves with redshift.

\begin{figure}
	\includegraphics[width=\columnwidth]{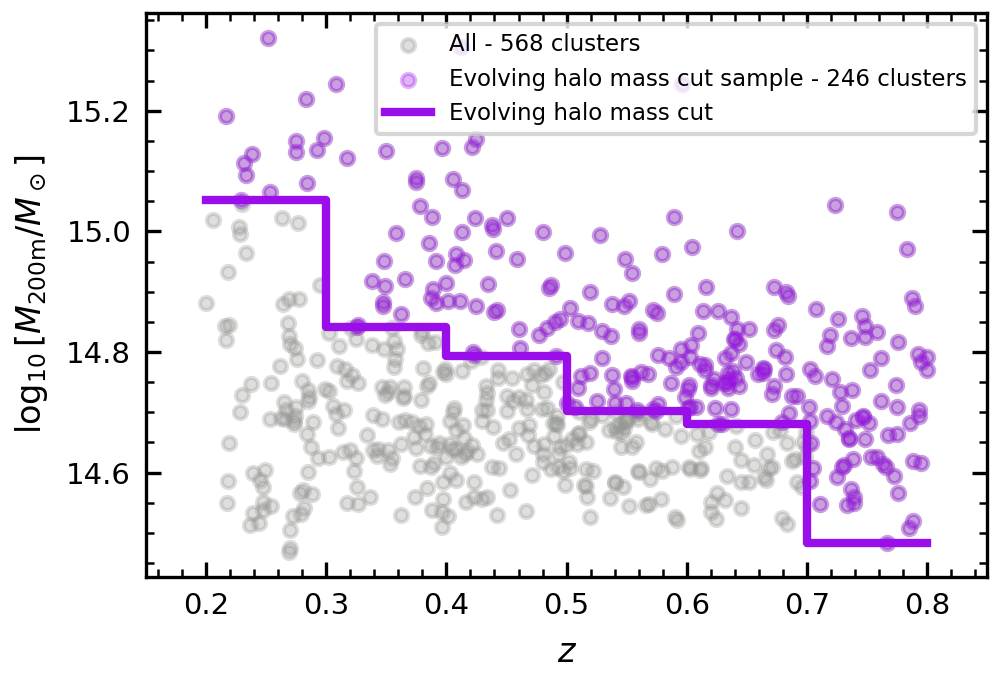}
    \caption{The $\log$ halo mass ($\log_{10} [M_{\rm{200m}}/\rm{M_{\odot}}]$) distribution, as a function of redshift, for all 568 ACT DR5 clusters (shown as grey shaded circles) in our sample. The thick, violet line indicates the evolving $\log$ halo mass cut ($\log_{10} [M_{\rm{200m}}^{\rm{thresh}}/\rm{M_{\odot}}]$) across 6 redshift bins, applied to mitigate the effects of halo mass evolution on our cluster stellar mass fractions. $f^{\rm{cg}}_{*, \rm{thresh}}$ is measured for clusters above the violet line for each redshift bin (see Table \ref{tab:Masscut_table}). The resulting cluster sample (246 clusters) selected above $\log_{10} [M_{\rm{200m}}^{\rm{thresh}}/\rm{M_{\odot}}]$ is shown as violet shaded circles.}
    \label{fig:Evolvingmasscut}
\end{figure}

\begin{figure*}
	\includegraphics[width=2.0\columnwidth]{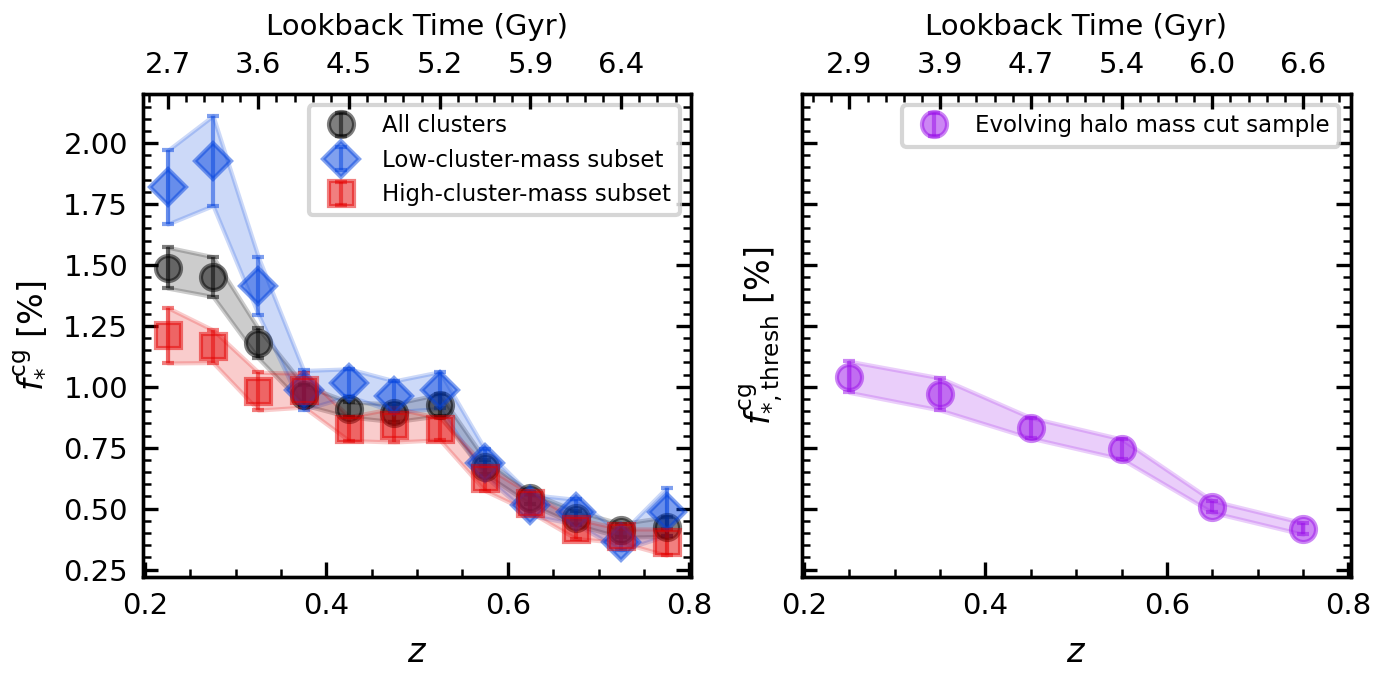}
    \caption{\textit{Left panel:} a comparison of the average cluster stellar mass fractions ($f^{\mathrm{cg}}_*$) at fixed redshift for the entire cluster sample (black circles), the low-cluster-mass subset (blue diamonds), and the high-cluster-mass subset (red squares). \textit{Right panel:} the average stellar mass fractions ($f^{\rm{cg}}_{*, \rm{thresh}}$) measured for a cluster sample (violet circles) selected above an evolving halo mass threshold per redshift bin. We have omitted the fractions for the low- and high-cluster-mass subsets of the evolving halo mass cut sample since they are within $1\sigma$ difference of $f^{\rm{cg}}_{*, \rm{thresh}}$ for the full sample per bin. Shaded regions represent $1\sigma$ scatter for each sample. All fractions are observed to be increasing towards lower redshift.}
    \label{fig:stellarmassfracs}
\end{figure*}

We lose a large number of low-mass clusters in the lower redshift bins due to higher $M_{\rm{200m}}^{\rm{thresh}}$ values, as a result of the fixed volume density selection. Hence, our cluster sample is reduced from $568$ to $246$ clusters for this analysis. A visualisation of the evolving mass threshold is shown in Fig. \ref{fig:Evolvingmasscut}, where all the clusters above the violet line are selected. Finally, we re-measure the average cluster stellar mass fractions ($f^{\rm{cg}}_{*, \rm{thresh}}$) for the reduced sample of clusters in each of the 6 redshift bins using Equation \ref{eq:StellarMassFracs}, and compute their errors using bootstrap resampling. We also repeat the analysis for our cluster mass subsets per bin, which are defined in the same manner as described in Section \ref{sec:method}. A summary of our results and the evolving mass cut for all 6 bins is shown in Table \ref{tab:Masscut_table}. A display of $f^{\rm{cg}}_{*, \rm{thresh}}$, as a function of redshift and lookback time, for the total cluster sample is shown in the right panel of Fig. \ref{fig:stellarmassfracs}.

For the evolving halo mass cut sample, we find a growth trend of $f^{\rm{cg}}_{*, \rm{thresh}}$ with decreasing redshift and lookback time. We observe a stellar mass fraction of $f^{\rm{cg}}_{*, \rm{thresh}} = 0.42 \pm 0.02$ per cent at $z = 0.8$, which increases by $\approx 0.6$ per cent to $f^{\rm{cg}}_{*,\rm{thresh}} = 1.04 \pm 0.06$ per cent at $z = 0.2$. This overall evolutionary trend is consistent with the redshift trend of $f^{\rm{cg}}_*$ (see Section \ref{subsubsec:stellarmassfractions}). Additionally, there is significant growth ($\approx 5 \sigma$ difference) of $f^{\rm{cg}}_{*, \rm{thresh}}$ observed over a short time-scale, from $z =0.7$ to $z = 0.5$. A similar difference is observed for $f^{\rm{cg}}_{*}$ at a similar redshift but occurs over a shorter period of cosmic time. The consistency with our $f^{\rm{cg}}_*$ measurements suggests that our observed trends for the stellar mass fractions are not primarily driven by halo mass evolution. These trends suggest that the stellar mass in clusters builds up over time, consistent with hierarchical growth models (e.g., \citealp{Cole2000}; \citealp{Springel2005}; \citealp{Baugh2006}). The growth of $f^{\mathrm{cg}}_{*, \rm{thresh}}$, as a function of cosmic time, also implies ongoing galaxy growth or late infall of satellite galaxies, which is consistent with the evolution of the cluster SMF low-mass slope (see Section \ref{subsec:M,alphavsz}). Infalling galaxies experience mechanisms such as quenching, stripping and merging, affecting the baryon content in the cluster. 

For both cluster-mass subsets of the evolving halo mass threshold sample, we find that $f^{\rm{cg}}_{*, \rm{thresh}}$ similarly increases with decreasing redshift. The sample size for each subset has been drastically reduced, particularly at lower redshifts, due to our constant-volume number-density selection. However, the mass dependence identified through the low- and high-cluster-mass subsets is largely preserved within the evolving halo mass cut sample. Once again, we find consistently higher fractions of stellar mass in low-mass clusters compared to high-mass clusters from $z = 0.8$ to $z =0.2$. This indicates a higher concentration of galaxies in low-mass clusters, pointing towards the increased star formation efficiency in low-mass haloes. Our result is also in alignment with previous studies, such as \cite{Andreon2010}, which found that low-mass clusters contain more stellar content per unit of cluster mass. Environmental processes such as mergers, tidal stripping and the formation of the ICL  can affect the distribution of stellar mass within clusters. Previous studies show that a large fraction of the total stellar content in clusters, particularly low-mass haloes, may be attributed to the ICL and BCGs (e.g., \citealp{Gonzalez2013}).

The high-mass clusters have the lowest stellar mass fractions across the redshift range. At $z = 0.8$, the stellar mass fraction in high-mass clusters is $f^{\mathrm{cg}}_{*, \rm{thresh}}=0.38 \pm 0.03$ per cent which then increases to $f^{\mathrm{cg}}_{*, \rm{thresh}} =0.98 \pm 0.07$ per cent at $z = 0.2$. This is in agreement with the inversely proportional scaling relation between stellar mass fraction and total cluster mass, which is a result of decreasing star formation efficiency in massive haloes (\citealp{Lin_2004}; \citealp{Moster2013}; \citealp{Pillepich2018}; \citealp{Behroozi2019}). For example, a study by \cite{Chiu2018} on the baryon content in 84 SZ-selected galaxy clusters from the South Pole Telescope (SPT-SZ) survey at $0.2 < z < 1.25$, observed that stellar mass fractions decrease with increasing cluster mass for cluster masses $M_{\mathrm{500}} >  2.5 \times 10^{14} \mathrm{M_{\odot}}$. They found a stellar mass fraction of $f_{*} = 0.83 \pm 0.06$ per cent at the sample median mass of $M_{\mathrm{500}} = 4.8 \times 10^{14} \mathrm{M_{\odot}}$ and median redshift $z = 0.6$. This is comparable ($\approx 1 \sigma$ difference) to the fraction of $f^{\rm{cg}}_{*, \rm{thresh}} = 0.74 \pm 0.04$ we find at similar redshift ($\langle z \rangle = 0.55$).

In comparison with previous works, \cite{Hilton2013} examined the stellar content in 14 SZ-selected clusters from ACT at $0.27 < z < 1.07$ and found a median stellar mass fraction of $f_* \approx 2.2$ per cent, which is significantly higher than the median stellar mass fraction of $f^{\rm{cg}}_{*, \rm{thresh}} \approx 0.78$ per cent we find for our evolving halo mass cut cluster sample. This major difference is likely due to the usage of the \cite{Salpeter1955} IMF by \cite{Hilton2013}, which results in $\approx 0.24$ dex higher stellar mass estimates compared to using a \cite{Chabrier2003} IMF. \cite{Decker_2022} measured the stellar mass fractions for 12 infrared-selected clusters from MaDCoWS at $\langle z\rangle = 1.18$ and found a decreasing trend in $f_*$ with total mass, which is in agreement with our results for the high-cluster mass subsets, and indicative of the mass-dependent processes slowing down star formation at high-$z$. However, they found that large uncertainties in SZ cluster mass measurements are the dominant source of error on $f_*$.

There are several sources of uncertainty that may affect our measured stellar mass fractions. These include systematic uncertainties such as the choice of IMF used to estimate galaxy stellar masses, uncertainties in cluster mass estimates, and the stellar mass completeness limit. Additionally, the reduced number of clusters in each redshift bin for our evolving mass threshold sample can introduce statistical uncertainties in the average stellar mass fractions, especially for cluster mass subsets. However, the overall redshift trend we find for $f^{\rm{cg}}_*$ is consistent between the full sample, cluster mass-subsets, and the evolving halo mass threshold selected sample. This indicates that our main conclusions are robust to potential systematic uncertainties.

\begin{table*}
        \centering
	\caption{The average cluster stellar mass fractions ($f^{\mathrm{cg}}_{*}$), measured using only the stellar mass contributions from cluster galaxies with $M_* > 10^{9.5} \mathrm{M_{\odot}}$. We include the fractions for the low- and high-cluster-mass subsets per redshift bin. For each redshift bin, we represent the average stellar mass fraction as a percentage of the total cluster mass ($M_{\mathrm{200m}}$). All fractions exclude stellar mass contributions from the ICL. Additionally included are the average cluster mass ($\langle M_{\mathrm{200m}} \rangle [10^{14} \mathrm{M_{\odot}}]$) and cluster stellar mass ($\langle M^{\mathrm{cg}}_{*} \rangle [10^{12} \mathrm{M_{\odot}}]$) for each sample per redshift bin. The represented errors for $\langle M_{\rm{200m}} \rangle [10^{14} \rm{M_{\odot}}]$ are the mean symmetric error in $M_{\rm{200m}}$ per bin. For $\langle M^{\mathrm{cg}}_* \rangle [10^{12} \mathrm{M_{\odot}}]$, the uncertainties are calculated using bootstrap resampling. }
    
	\label{tab:Stellarmassfrac_table}
        \renewcommand{\arraystretch}{1.3}
	\begin{tabular}{lcccccr}
		\hline
		Redshift range & Sample & $N_{\mathrm{clus}}$ & $\langle M_{\mathrm{200m}}\rangle [10^{14} \mathrm{M_{\odot}}] $ & $\langle M^{\mathrm{cg}}_* \rangle [10^{12} \mathrm{M_{\odot}}] $ & $f^{\mathrm{cg}}_{*} [\%]$ \\
		\hline
        
		$0.20 < z < 0.25$ & All & 30 & $ 7.12 \pm 1.58$ & $9.62 \pm 1.11$ & $1.49 \pm 0.08$ \\
        & Low-cluster-mass & 9 & $3.56 \pm 0.89$ & $6.40 \pm 1.03$ & $1.82 \pm 0.15$\\
        & High-cluster-mass & 9 & $11.91 \pm 2.48$ & $14.11 \pm 1.33$ & $1.21 \pm 0.11$\\

        \hline
            $0.25 < z < 0.30$ & All & 51 & $6.80 \pm 1.46$ & $ 8.95 \pm 1.24$ & $1.45\pm 0.08$ \\
            & Low-cluster-mass & 16 & $3.80 \pm 0.89$ & $7.19 \pm 1.15$ & $1.93 \pm 0.18$ \\
            & High-cluster-mass & 16 & $11.41 \pm 2.29$ & $12.92 \pm 1.39$ & $1.17 \pm 0.07$\\
        \hline
         
            $0.30 < z < 0.35$ & All & 43 & $6.20 \pm 1.30$ & $6.95 \pm 1.19$ & $1.18 \pm 0.06$ \\
            & Low-cluster-mass & 14 & $4.13 \pm 0.93$ & $5.83 \pm 1.15$ & $1.41\pm 0.12$ \\
            & High-cluster-mass & 14 & $8.97 \pm 1.77$ & $8.69 \pm 1.32$ & $0.98\pm 0.08$\\
            
        \hline
            
            $0.35 < z < 0.40$ & All & 61 & $5.99 \pm 1.23$ & $5.72 \pm 1.10$ & $0.97\pm 0.04$ \\ 
            & Low-cluster-mass & 20 & $3.98 \pm 0.89$ & $3.88 \pm 0.89$ & $0.99\pm 0.08$ \\
            & High-cluster-mass & 20 & $8.68 \pm 1.65$ & $8.34 \pm 1.41$ & $0.99\pm 0.07$ \\

        \hline  
        
            $0.40 < z < 0.45$ & All & 61 & $6.49 \pm 1.27$ & $5.62 \pm 1.01$ & $0.91\pm 0.03$ \\
            & Low-cluster-mass & 20 & $4.00 \pm 0.87$ & $4.06 \pm 0.86$ & $1.02\pm 0.06$ \\
            & High-cluster-mass & 20 & $10.16 \pm 1.87$ & $8.14 \pm 1.21$ & $0.83\pm 0.05$\\

        \hline
        
            $0.45 < z < 0.50$ & All & 53 & $5.55 \pm 1.11$ & $4.89 \pm 0.85$ & $0.89\pm 0.04$ \\
            & Low-cluster-mass & 17 & $4.16 \pm 0.88$ & $4.01 \pm 0.78$ & $0.96 \pm 0.06$ \\ 
            & High-cluster-mass & 17 & $7.39 \pm 1.40$ & $6.14 \pm 0.90$ & $0.84\pm 0.07$\\
        
        \hline
        
            $0.50 < z < 0.55$ & All & 51 & $5.26 \pm 1.04$ &  $4.77 \pm 0.87$ & $0.93 \pm 0.04$\\
            & Low-cluster-mass & 16 & $3.98 \pm 0.85$ & $3.96 \pm 0.88$ & $0.99 \pm 0.08$\\ 
            & High-cluster-mass & 16 & $6.97 \pm 1.31$ & $5.69 \pm 0.87$ & $0.83 \pm 0.04$\\

        \hline
        
            $0.55 < z < 0.60$ & All & 50 & $5.65 \pm 1.09$ & $3.71 \pm 0.67$ & $0.67\pm 0.03$ \\
            & Low-cluster-mass & 16 & $ 3.94 \pm 0.81$ & $2.77 \pm 0.61$ & $0.69 \pm 0.06$ \\
            & High-cluster-mass & 16 & $7.90 \pm 1.44$ & $4.76 \pm 0.74$ & $0.62 \pm 0.05$\\

        \hline
        
            $0.60 < z < 0.65$ & All & 58 & $5.46 \pm 1.04$ & $2.95 \pm 0.59$ & $0.54 \pm 0.02$ \\
            & Low-cluster-mass & 19 & $4.01 \pm 0.81$ & $2.07 \pm 0.45$ & $0.52\pm 0.04$ \\
            & High-cluster-mass & 19 & $6.95 \pm 1.25$ & $3.62 \pm 0.67$ & $0.52 \pm 0.04$\\

        \hline
        
            $0.65 < z < 0.70$ & All & 40 & $5.11 \pm 0.97$ & $2.31 \pm 0.45$ & $0.46\pm 0.03$\\
            & Low-cluster-mass & 13 & $3.89 \pm 0.78$ & $1.88 \pm 0.40$ & $0.49 \pm 0.05$ \\
            & High-cluster-mass & 13 & $6.65 \pm 1.19$ & $2.79 \pm 0.50$ & $0.42 \pm 0.04$\\

        \hline
        
            $0.70 < z < 0.75$ & All & 38 & $5.24 \pm 0.98$ & $2.14 \pm 0.43$ & $0.41 \pm 0.03$ \\
            & Low-cluster-mass & 12 & $3.88 \pm 0.77$ & $1.41 \pm 0.31$ & $0.36 \pm 0.03$ \\
            & High-cluster-mass & 12 & $6.98 \pm 1.23$ & $2.68 \pm 0.47$ & $0.39\pm 0.03$\\

        \hline
        
            $0.75 < z < 0.80$ & All & 32 & $5.38 \pm 0.99$ & $2.20 \pm 0.47$ & $0.43 \pm 0.04$ \\
            & Low-cluster-mass & 10 & $3.79 \pm 0.76$ & $1.88 \pm 0.47$ & $0.49 \pm 0.10$ \\
            & High-cluster-mass & 10 & $7.34 \pm 1.27$ & $2.58 \pm 0.49$& $0.36 \pm 0.05$ \\
            
		\hline
	\end{tabular}

\end{table*}

\begin{table*}
        \centering
	\caption{The average cluster stellar mass fractions ($f^{\mathrm{cg}}_{*, \rm{thresh}}$) for all, low-cluster-mass and high-cluster-mass samples measured using an evolving halo mass threshold ($M_{\rm{200m}}^{\rm{thresh}}$) across 6 redshift bins. $N_{\mathrm{clus}}^{\rm{thresh}}$ shows the number of clusters above $M_{\rm{200m}}^{\rm{thresh}}$ in each redshift bin, for each cluster sample.}
	\label{tab:Masscut_table}
        \renewcommand{\arraystretch}{1.3}
	\begin{tabular}{lcccccr}
		\hline
		Redshift range & Sample & $N_{\mathrm{clus}}^{\rm{thresh}}$ & $M_{\rm{200m}}^{\rm{thresh}} [10^{14} M_{\odot}]$ & $f^{\mathrm{cg}}_{*, \rm{thresh}} [\%]$ \\
		\hline
        
		$0.20 < z < 0.30$ & All & 13 & $11.27$ & $1.04 \pm 0.06$\\
        & Low-cluster-mass & 4 & $11.27$ & $1.04 \pm 0.10$\\
        & High-cluster-mass & 4 & $14.27$ & $0.98 \pm 0.07$\\

        \hline
            
            $0.30 < z < 0.40$ & All & 23 & $6.94$ & $0.97 \pm 0.07$ \\
            & Low-cluster-mass & 7 & $6.94$ & $0.95 \pm 0.12$\\
            & High-cluster-mass & 7 & $11.03$ & $0.84 \pm 0.13$\\

            \hline
            
            $0.40 < z < 0.50$ & All & 35 & $6.22$ & $0.83 \pm 0.04$ \\
            & Low-cluster-mass & 11 & $6.22$ & $0.88 \pm 0.08$\\
            & High-cluster-mass & 11 & $9.98$ & $0.82 \pm 0.07$\\

            \hline

            $0.50 < z < 0.60$ & All & 47 & $5.04$ & $0.74 \pm 0.04$ \\
            & Low-cluster-mass & 15 & $5.04$ & $0.81 \pm 0.09$\\
            & High-cluster-mass & 15 & $7.11$ & $0.70 \pm 0.06$\\

            \hline
            
            $0.60 < z < 0.70$ & All & 58 & $4.79$ & $0.51 \pm 0.02$ \\
            & Low-cluster-mass & 19 & $4.79$ & $0.57 \pm 0.05$\\
            & High-cluster-mass & 19 & $6.35$ & $0.50 \pm 0.04$\\

            \hline

            $0.70 < z < 0.80$ & All & 70 & $3.05$ & $0.42 \pm 0.02$ \\
            & Low-cluster-mass & 23 & $3.05$ & $0.43 \pm 0.05$\\
            & High-cluster-mass & 23 & $5.77$ & $0.38 \pm 0.03$\\

		\hline
	\end{tabular}

\end{table*}

\section{Conclusions}
\label{sec:conclusion}

In this paper, we provide stellar mass function (SMF) measurements for a large sample of 568 massive galaxy clusters from ACT DR5 at $0.2 < z < 0.8$. Thanks to deep multi-band photometric data from DECaLS DR10, we are able to probe the cluster SMF down to $M_{*} = 10^{9.5} \mathrm{M_{\odot}}$. We derive composite or stacked cluster SMFs, which makes it easier to identify minor variations and trends in cluster stellar mass distributions. Due to our large cluster sample and wide redshift range, we can bin clusters into 12 narrow redshift bins of width 0.05 and by cluster mass over the log stellar mass range: $9.5 \leq \log_{10} [M_{*}/\mathrm{M_{\odot}}] \leq 12.5$. This allows us to track the evolution of the cluster SMF during this epoch. We use these SMFs to also study the galaxy stellar mass content in clusters by measuring cluster stellar mass fractions. Our main results are summarised as follows:

(i) There is evidence for the evolution of the total galaxy stellar mass distribution in the virialised regions ($R < R_{\mathrm{200m}}$) of galaxy clusters for stellar mass $M_* > 10^{9.5} \mathrm{M_{\odot}}$. The characteristic stellar mass ($M^*$) of the cluster SMF evolves moderately in the redshift interval $0.2 < z < 0.8$, implying that the population of massive cluster galaxies ($M_* \gtrsim 10^{10.75}  \mathrm{M_{\odot}}$) is well established by $z \sim 0.8$. There is a more significant growth of $M^*$ at $z < 0.55$, indicating a late-time accretion of intermediate-to-high mass galaxies by clusters. Galaxy clusters at high-$z$ are found to contain more massive galaxies than at low-$z$, indicated by a less negative low-mass slope ($\alpha$) at $z \geq 0.55$. The low-mass slope steepens at $z < 0.55$, suggesting a late infall of galaxies into clusters.  The redshift evolution of $M^*$ and $\alpha$ is consistent with the cosmic downsizing scenario.

(ii) A clear disagreement between our composite cluster SMFs and previous studies of the SMF in field galaxies (Fig. \ref{fig:M,alphavsz}) points towards an environmental dependence of the SMF. This confirms the findings of previous studies of the role of the environment on the SMF (e.g., \citealp{vanderBurg2020}). Due to a significantly shallow low-mass slope ($\alpha \gtrsim -1$) of the composite cluster SMF, we also find more massive galaxies in clusters compared to what is observed in field environments at this epoch. 

(iii) A Schechter+Gaussian model is better suited to account for an excess in stellar mass observed at the massive end of the composite cluster SMF from $0.25 \leq z \leq 0.65$ (Table \ref{tab:Schechter+Gaussian_table}). We attribute a large component of the excess stellar mass distribution to BCGs. Despite an improved model fit, we do not find significant differences for $M*$ and $\alpha$ values between the single Schechter and Schechter+Gaussian models. The redshift evolution of the median stellar mass of the Gaussian distribution ($M_{\mathrm{MG}}$) shows a strong growth in massive galaxies since $z = 0.8$ (Fig. \ref{fig:M_MG,sigma_MGvsz}). 

(iv) The evolution of the shape of the composite cluster SMF (i.e. $M^*$ and $\alpha$) is independent of cluster mass for $M_{\mathrm{200m}} > 2.9 \times10^{14} \mathrm{M_{\odot}}$. Additionally, the composite SMF of massive clusters ($M_{\mathrm{200m}} > 5.5 \times 10^{14} \mathrm{M_{\odot}}$) is found to have a flatter low-mass slope ($\alpha$) than for the low-mass cluster composite SMF, signalling a high fraction of massive galaxies in larger haloes.

(v) Accounting for the impact of the growth of cluster-scale haloes with cosmic time, we find that cluster stellar mass fractions ($f^{\rm{cg}}_{*, \rm{thresh}}$), measured from cluster galaxies within $R_{\mathrm{200m}}$ and with $M_* > 10^{9.5} \mathrm{M_{\odot}}$, have grown by a factor of $2.5$ from $z = 0.8$ to $z = 0.2$ (Fig. \ref{fig:stellarmassfracs}). This implies ongoing galaxy growth or a late infall of galaxies into clusters, which is also in agreement with the evolutionary trends of the total composite cluster SMF.

In summary, we find that galaxy stellar mass within clusters evolves significantly from $0.20 < z < 0.55$. The disparity between the SMFs in cluster and field galaxies suggests that the environment plays a key role in the distribution of galaxy stellar mass. The dominance of massive galaxies ($M_* > 10^{10.75} \mathrm{M_{\odot}}$), particularly BCGs, has a strong impact on the shape of the cluster SMF. However, the shape of the cluster SMF appears to be independent of cluster mass for $M_\mathrm{{200m}} > 2.9 \times 10^{14} \mathrm{M_{\odot}}$. Finally, the significant growth of cluster stellar mass fractions since $z \sim 0.8$ reveals the ongoing galaxy growth taking place in clusters. These results emphasise the complexities of structure formation during this epoch.

\section*{Acknowledgements}

The authors thank the anonymous referee for providing useful and constructive comments that led to the improvement of this paper.
We also thank Roberto de Propris for his suggestions and comments to improve this paper.

Support for ACT was through the U.S.~National Science Foundation through awards AST-0408698, AST-0965625, and AST-1440226 for the ACT project, as well as awards PHY-0355328, PHY-0855887 and PHY-1214379. Funding was also provided by Princeton University, the University of Pennsylvania, and a Canada Foundation for Innovation (CFI) award to UBC. ACT operated in the Parque Astron\'omico Atacama in northern Chile under the auspices of the Agencia Nacional de Investigaci\'on y Desarrollo (ANID). The development of multichroic detectors and lenses was supported by NASA grants NNX13AE56G and NNX14AB58G. Detector research at NIST was supported by the NIST Innovations in Measurement Science program. Computing for ACT was performed using the Princeton Research Computing resources at Princeton University, the National Energy Research Scientific Computing Center (NERSC), and the Niagara supercomputer at the SciNet HPC Consortium. SciNet is funded by the CFI under the auspices of Compute Canada, the Government of Ontario, the Ontario Research Fund–Research Excellence, and the University of Toronto. We thank the Republic of Chile for hosting ACT in the northern Atacama, and the local indigenous Licanantay communities whom we follow in observing and learning from the night sky.

A special thanks to the University of the Witwatersrand for their support for this work. Funding for this work was provided by the South African Astronomical Observatory (SAAO). 

DCR, US and MH acknowledge support from the National Research Foundation of South Africa (grant nos. 137975, 97792, CPRR240513218388). KM also acknowledges support from the National Research Foundation of South Africa.
TM acknowledges support from the Agencia Estatal de Investigaci\'on (AEI) and the Ministerio de Ciencia, Innovaci\'on y Universidades (MICIU) Grant ATRAE2024-154740 funded by MICIU/AEI//10.13039/501100011033. TM is also partly supported by the Spanish program Unidad de Excelencia María de Maeztu CEX2020-001058-M, financed by MCIN/AEI/10.13039/501100011033, and by the MaX-CSIC Excellence Award MaX4-SOMMA-ICE.
JPH, the George A. and Margaret M. Downsbrough Professor of Astrophysics at Rutgers, acknowledges the Downsbrough heirs and the estate of George Atha Downsbrough for their support.

A large portion of the computations in this paper were performed on the Hippo high-performance computing facility at the University of KwaZulu-Natal. The remainder of the computations were performed on the Center for High Performance Computing (project ASTR1534), which is part of the National Integrated CyberInfrastructure System (NICIS) and is supported by the South African Department of Science and Innovation (DSI). We thank both computing facilities for access to their computing nodes.

The Legacy Surveys consist of three individual and complementary projects: the Dark Energy Camera Legacy Survey (DECaLS; Proposal ID \#2014B-0404; PIs: David Schlegel and Arjun Dey), the Beijing-Arizona Sky Survey (BASS; NOAO Prop. ID \#2015A-0801; PIs: Zhou Xu and Xiaohui Fan), and the Mayall z-band Legacy Survey (MzLS; Prop. ID \#2016A-0453; PI: Arjun Dey). DECaLS, BASS and MzLS together include data obtained, respectively, at the Blanco telescope, Cerro Tololo Inter-American Observatory, NSF’s NOIRLab; the Bok telescope, Steward Observatory, University of Arizona; and the Mayall telescope, Kitt Peak National Observatory, NOIRLab. Pipeline processing and analyses of the data were supported by NOIRLab and the Lawrence Berkeley National Laboratory (LBNL). The Legacy Surveys project is honored to be permitted to conduct astronomical research on Iolkam Du’ag (Kitt Peak), a mountain with particular significance to the Tohono O’odham Nation.

NOIRLab is operated by the Association of Universities for Research in Astronomy (AURA) under a cooperative agreement with the National Science Foundation. LBNL is managed by the Regents of the University of California under contract to the U.S. Department of Energy.

This project used data obtained with the Dark Energy Camera (DECam), which was constructed by the Dark Energy Survey (DES) collaboration. Funding for the DES Projects has been provided by the U.S. Department of Energy, the U.S. National Science Foundation, the Ministry of Science and Education of Spain, the Science and Technology Facilities Council of the United Kingdom, the Higher Education Funding Council for England, the National Center for Supercomputing Applications at the University of Illinois at Urbana-Champaign, the Kavli Institute of Cosmological Physics at the University of Chicago, Center for Cosmology and Astro-Particle Physics at the Ohio State University, the Mitchell Institute for Fundamental Physics and Astronomy at Texas A\&M University, Financiadora de Estudos e Projetos, Fundacao Carlos Chagas Filho de Amparo, Financiadora de Estudos e Projetos, Fundacao Carlos Chagas Filho de Amparo a Pesquisa do Estado do Rio de Janeiro, Conselho Nacional de Desenvolvimento Cientifico e Tecnologico and the Ministerio da Ciencia, Tecnologia e Inovacao, the Deutsche Forschungsgemeinschaft and the Collaborating Institutions in the Dark Energy Survey. The Collaborating Institutions are Argonne National Laboratory, the University of California at Santa Cruz, the University of Cambridge, Centro de Investigaciones Energeticas, Medioambientales y Tecnologicas-Madrid, the University of Chicago, University College London, the DES-Brazil Consortium, the University of Edinburgh, the Eidgenossische Technische Hochschule (ETH) Zurich, Fermi National Accelerator Laboratory, the University of Illinois at Urbana-Champaign, the Institut de Ciencies de l’Espai (IEEC/CSIC), the Institut de Fisica d’Altes Energies, Lawrence Berkeley National Laboratory, the Ludwig Maximilians Universitat Munchen and the associated Excellence Cluster Universe, the University of Michigan, NSF’s NOIRLab, the University of Nottingham, the Ohio State University, the University of Pennsylvania, the University of Portsmouth, SLAC National Accelerator Laboratory, Stanford University, the University of Sussex, and Texas A\&M University.

The Legacy Surveys imaging of the DESI footprint is supported by the Director, Office of Science, Office of High Energy Physics of the U.S. Department of Energy under Contract No. DE-AC02-05CH1123, by the National Energy Research Scientific Computing Center, a DOE Office of Science User Facility under the same contract; and by the U.S. National Science Foundation, Division of Astronomical Sciences under Contract No. AST-0950945 to NOAO.

\section*{Data Availability}

The galaxy cluster data used in this paper are obtained from the publicly available ACT DR5 cluster catalogue \citep{Hilton_2021}. We use the very recently released ACT DR6 cluster catalogue \citep{Hilton2025} for cluster redshift comparisons. All photometric data is also obtained from the publicly accessible DECaLS DR10 web database (\url{https://www.legacysurvey.org/dr10/}). Access to the SMF data and codes can be obtained on request from the author.




\bibliographystyle{mnras}
\bibliography{references} 




\appendix

\section{Removed clusters}

In comparison with the ACT DR6 cluster catalogue \citep{Hilton2025}, we identified 9 galaxy clusters with newly assigned redshifts, which have a difference of $\Delta z \geq 0.05$ compared to the DR5 catalogue. These clusters are removed from our analysis to negate the impact they may have on individual redshift-binned composite cluster SMFs. A detailed list of these clusters is shown in Table \ref{tab:Removed_clus_table}.

\begin{table*}
        \centering
	\caption{A list of the 9 removed clusters from the ACT DR5 catalogue used in this work. The removal is based on the significant difference ($\Delta z > 0.05$) between the provided redshifts in the DR5 catalogue and those provided in the DR6 catalogue. The last two columns show the different types and sources of the redshifts in DR6. }
	\label{tab:Removed_clus_table}
        \renewcommand{\arraystretch}{1.3}
	\begin{tabular}{lcccccccr}
		\hline
		Cluster name & RA (deg) & DEC (deg) & $z_{\mathrm{DR5}}$ & $z_{\mathrm{DR6}}$ & $\Delta z$ & DR6 redshift type& DR6 redshift source\\
		\hline
        
		ACT-CL J0131.7-5604 & 22.9354 & -56.0793 & $0.7483 \pm 0.0462$ & $0.8046 \pm 0.0162$ & 0.0563 & phot & SPT\\
            
            ACT-CL J0205.1-4125 & 31.2990 & -41.4308 & $0.3962 \pm 0.0103$ & $0.3500 \pm 0.0360$ & 0.0462 & phot & MaDCoWS2\\
            
            ACT-CL J0205.9-0307 & 31.4970 & -3.1220 & $0.4936 \pm 0.0000$ & $1.7000 \pm 0.0250$ & 1.2064 & phot & ACT-DR5-MCMF\\
            
            ACT-CL J0236.8-0607 & 39.2249 & -6.1251 & $0.6856 \pm 0.0144$ & $0.5510 \pm 0.0000$ & 0.1346 & spec & MaDCoWS2\\
            
            ACT-CL J0346.0-3645 & 56.5136 & -36.7526 & $0.3635 \pm 0.0093$ & $0.3100 \pm 0.0350$ & 0.0534 & phot & MaDCoWS2 \\
            
            ACT-CL J0434.5-5726 & 68.6336 & -57.4457 & $0.4802 \pm 0.0081$ & $0.4250 \pm 0.0380$ & 0.0552 & phot & MaDCoWS2\\

            ACT-CL J0512.6-5139 & 78.1651 & -51.6614 & $0.6432 \pm 0.0119$ & $0.5688 \pm 0.0385$ & 0.0744 & phot & SPT\\

            ACT-CL J0619.7-5802 & 94.9260 & -58.0389 & $0.3913 \pm 0.0239$ & $0.5538 \pm 0.0394$ & 0.1625 & phot & SPT\\

            ACT-CL J2228.8-5828 & 337.2196 & -58.4735 & $0.7605 \pm 0.0125$ & $1.1470 \pm 0.0000$ & 0.3865 & spec & MaDCoWS2\\

		\hline
	\end{tabular}

\end{table*}

\section{Composite SMFs}

\begin{figure*}
    \centering
    \subfloat{\includegraphics[width=0.72\columnwidth]{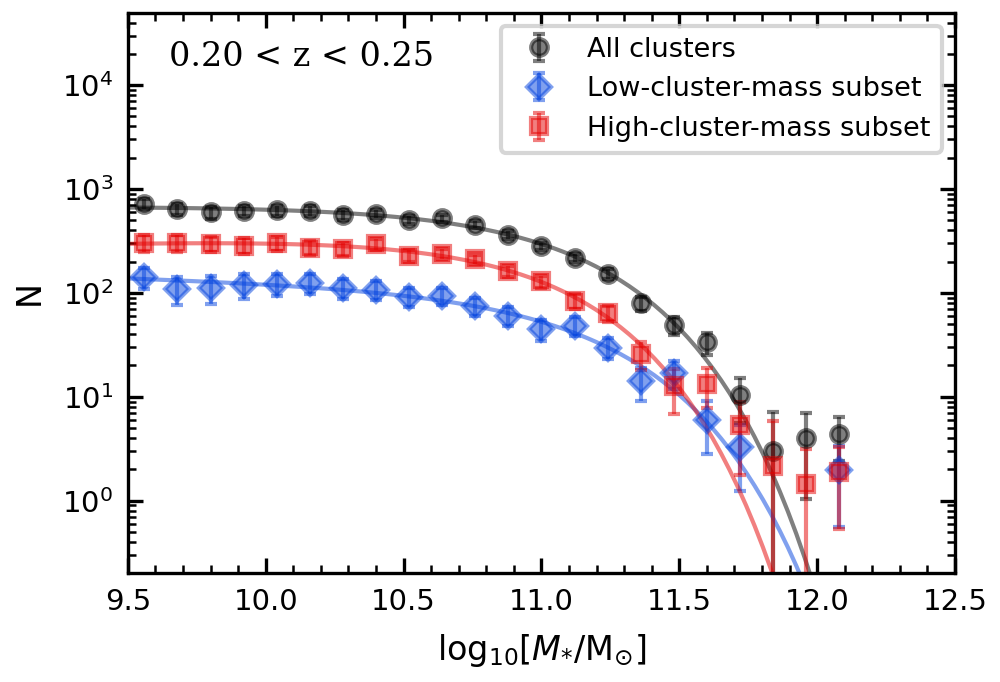}}
    \subfloat{\includegraphics[width=0.72\columnwidth]{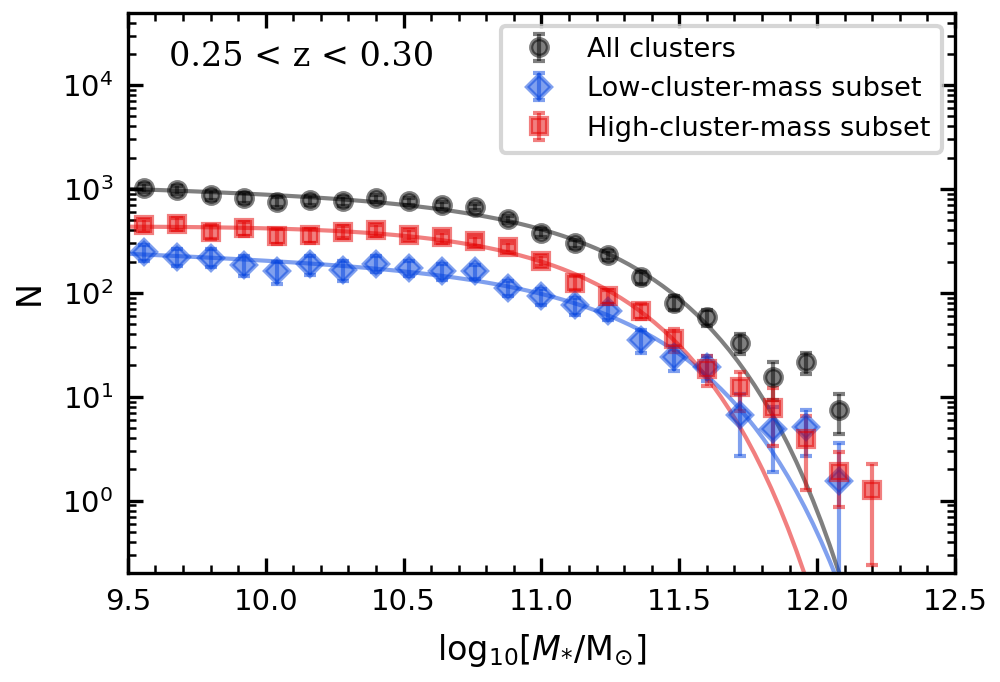}}
    \subfloat{\includegraphics[width=0.72\columnwidth]{SingleSchechterSMFs/CompSMF-MCMCplot_0.3_0.35.png}}\\
    \subfloat{\includegraphics[width=0.72\columnwidth]{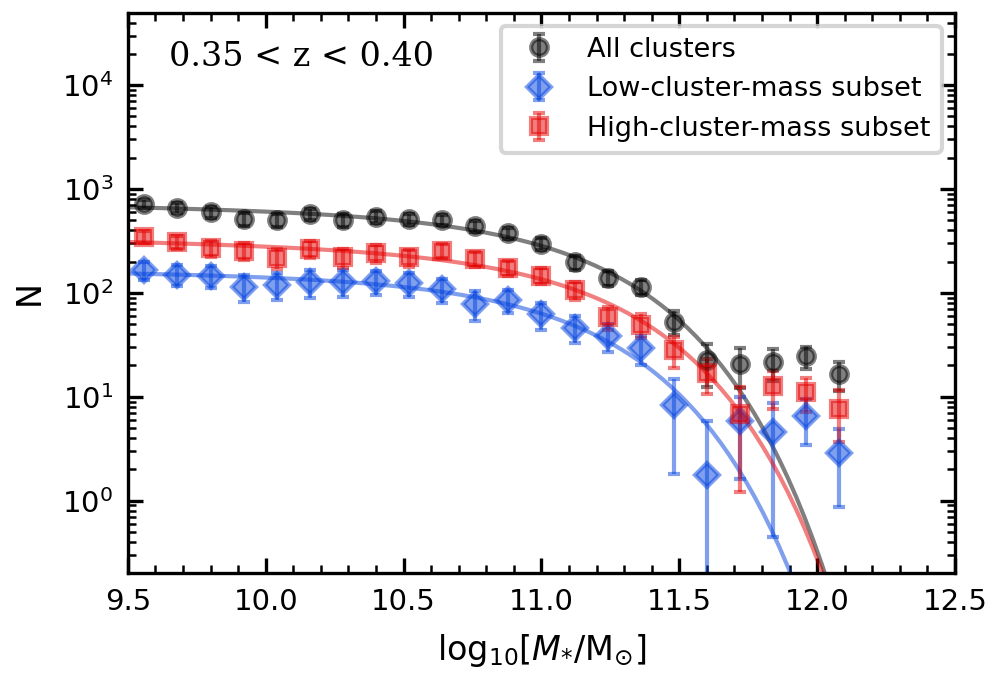}}
    \subfloat{\includegraphics[width=0.72\columnwidth]{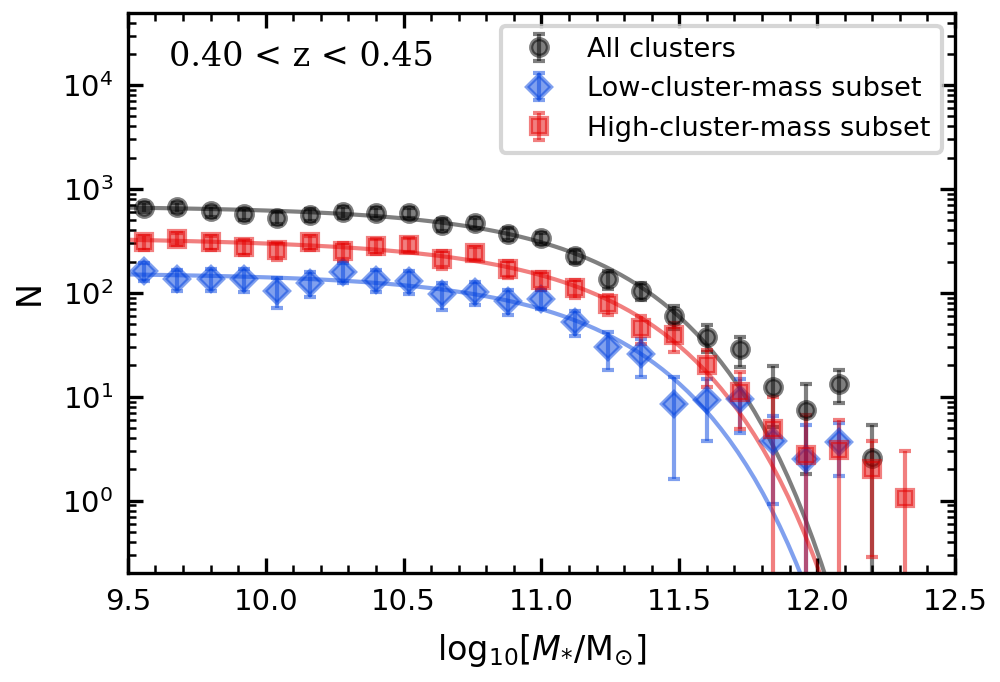}}
    \subfloat{\includegraphics[width=0.72\columnwidth]{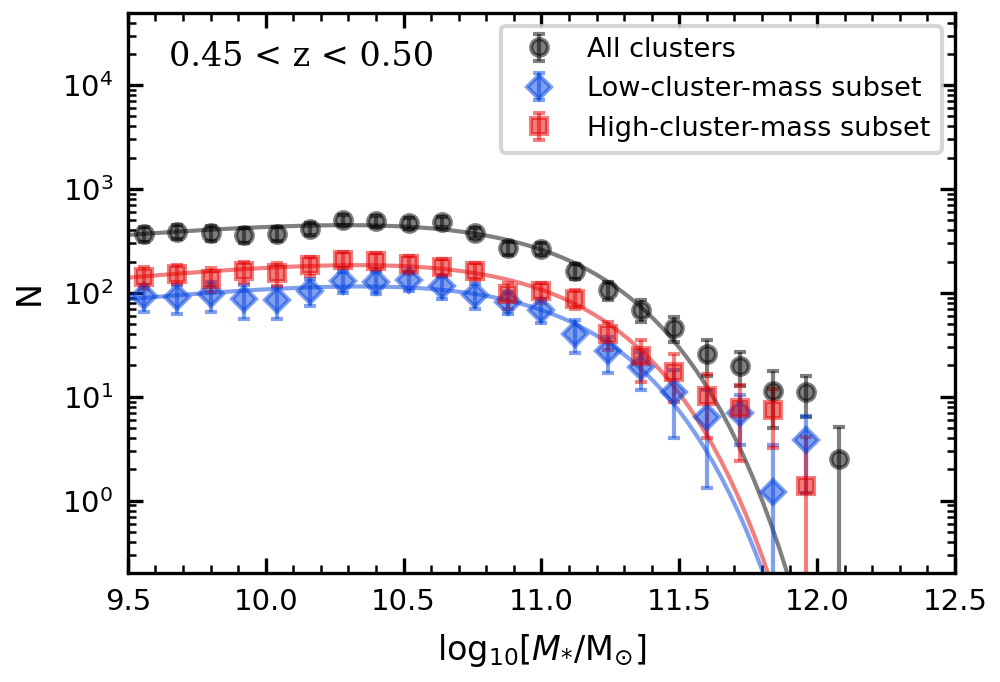}}\\
    \subfloat{\includegraphics[width=0.72\columnwidth]{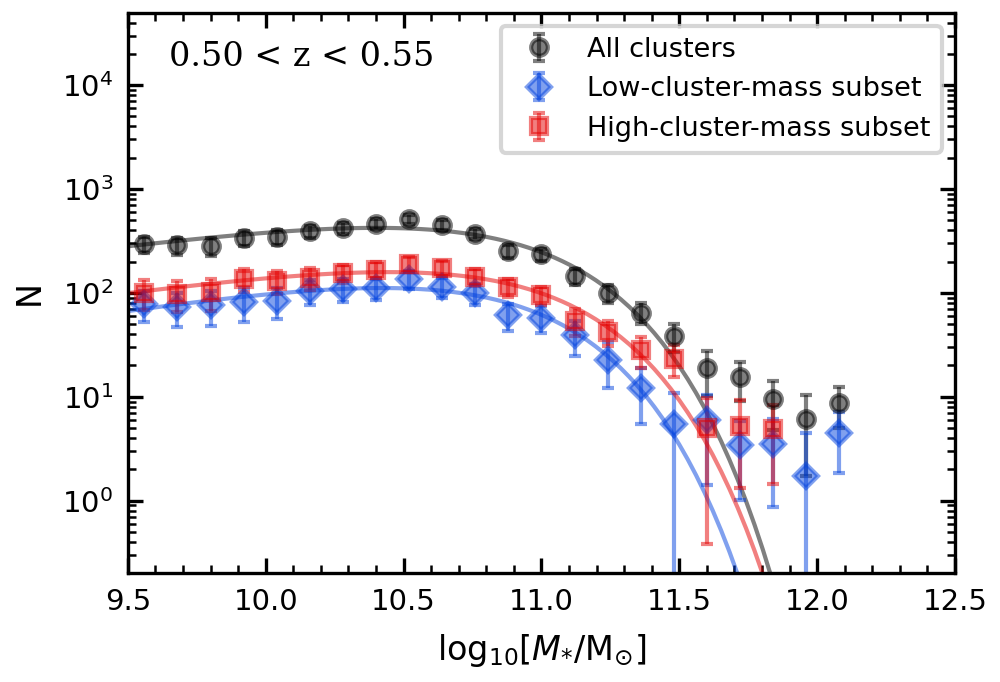}}
    \subfloat{\includegraphics[width=0.72\columnwidth]{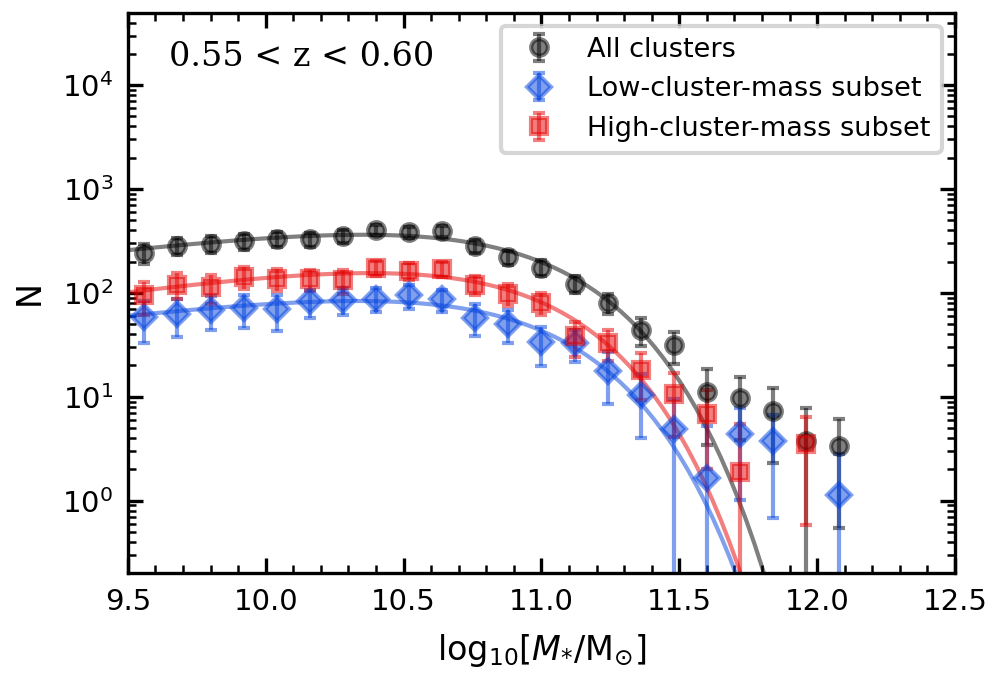}}
    \subfloat{\includegraphics[width=0.72\columnwidth]{SingleSchechterSMFs/CompSMF-MCMCplot_0.6_0.65.png}}\\
    \subfloat{\includegraphics[width=0.72\columnwidth]{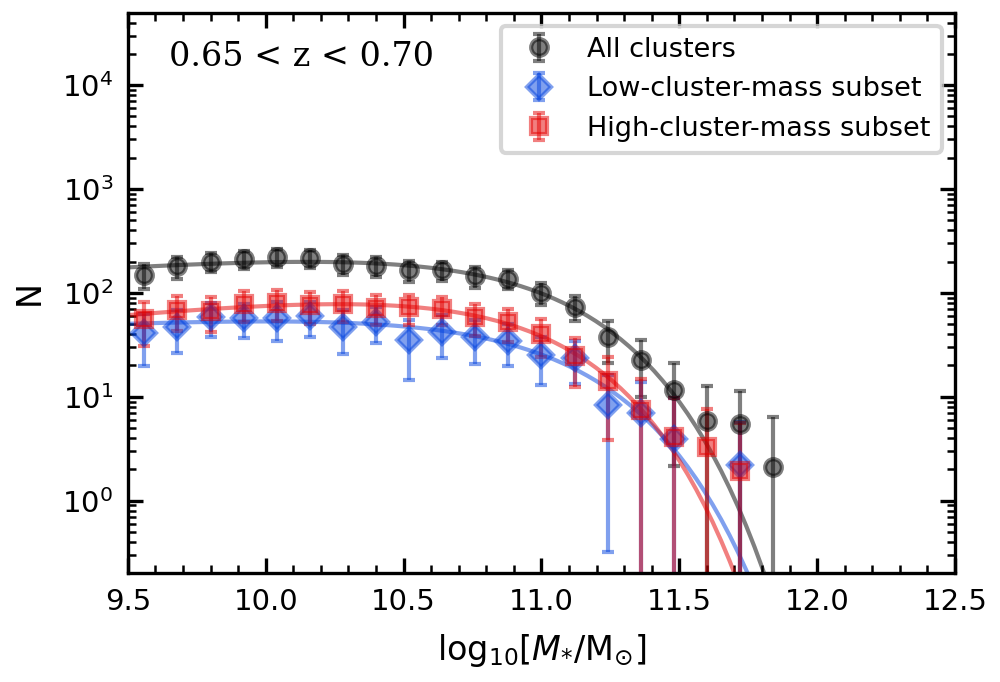}}
    \subfloat{\includegraphics[width=0.72\columnwidth]{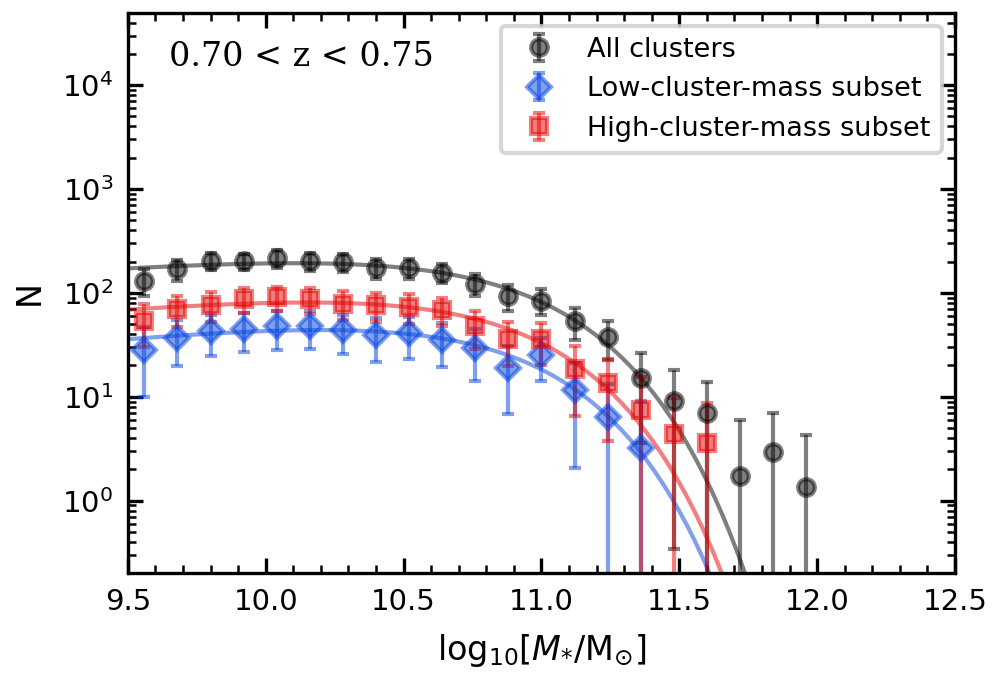}}
    \subfloat{\includegraphics[width=0.72\columnwidth]{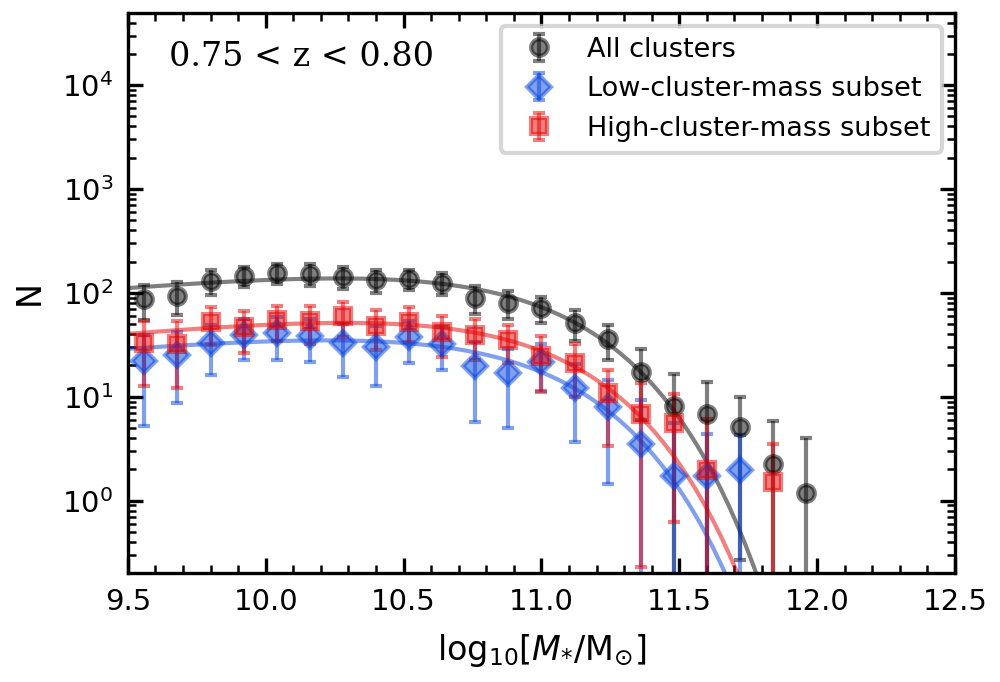}}

    \caption{The best-fit composite cluster SMFs for all 12 redshift bins. The total cluster SMF per redshift bin is shown as black circles. The blue diamonds and red squares represent the low- and high-cluster-mass subset SMFs, respectively. Errors represent the $1\sigma$ uncertainties estimated using bootstrap resampling. The best-fitting single Schechter SMF are shown by a solid line. Parameter values of the best-fit are presented in Table \ref{tab:smf_table}. The low-mass slope evolves marginally, trending towards a flatter shape with increasing redshift. Deviations in stellar mass above the Schechter function are observed at the high-mass end.}
    \label{fig:SinglSchechterSMFs}
\end{figure*}

\begin{figure*}
    \centering
    \subfloat{\includegraphics[width=0.72\columnwidth]{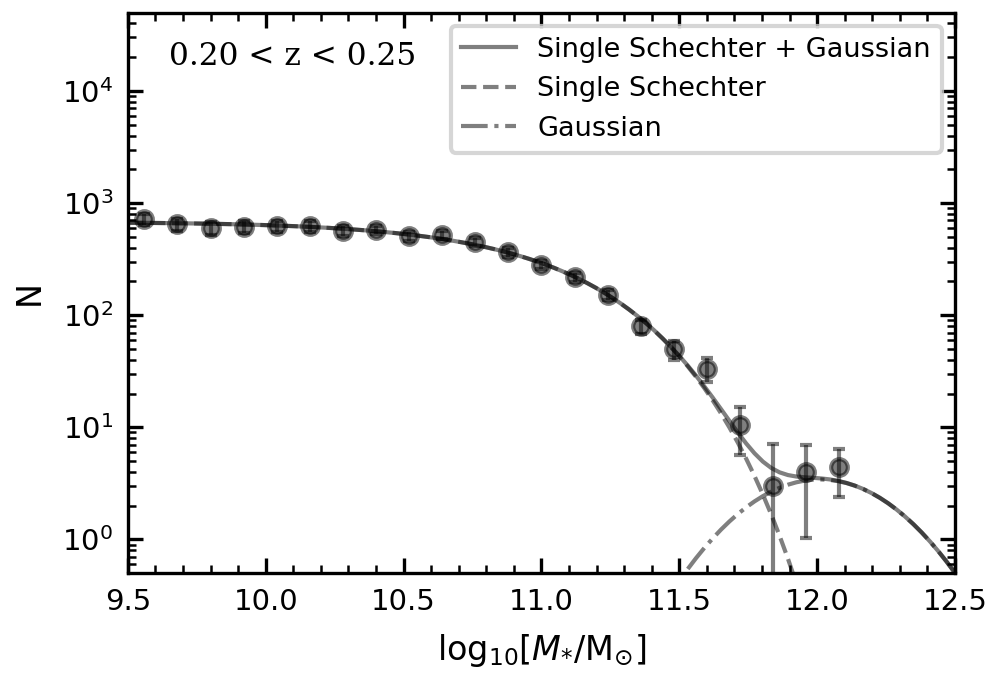}}
    \subfloat{\includegraphics[width=0.72\columnwidth]{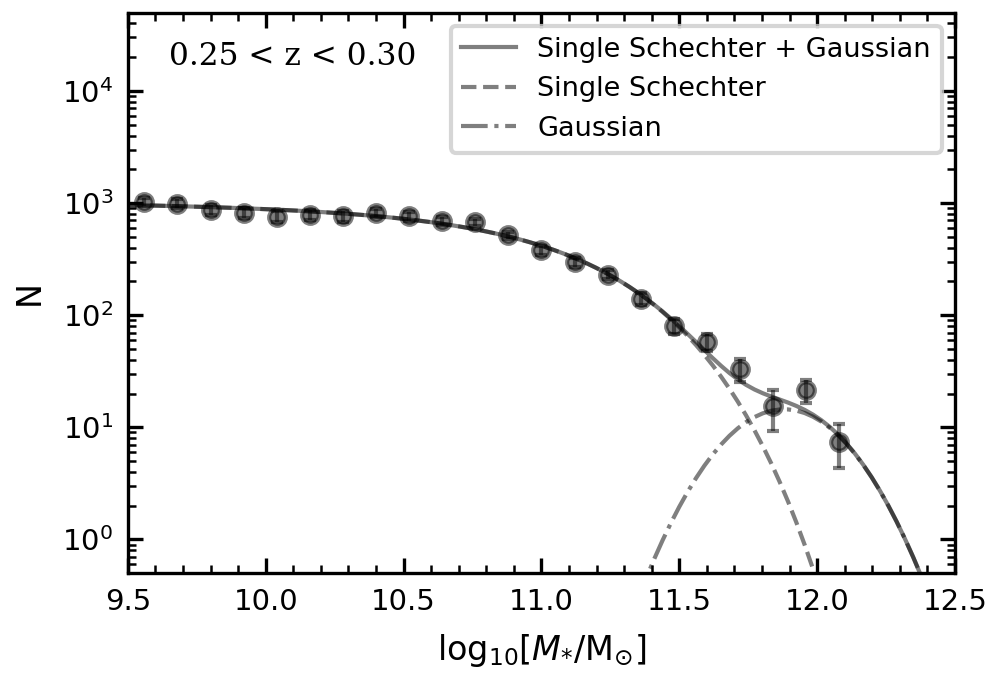}}
    \subfloat{\includegraphics[width=0.72\columnwidth]{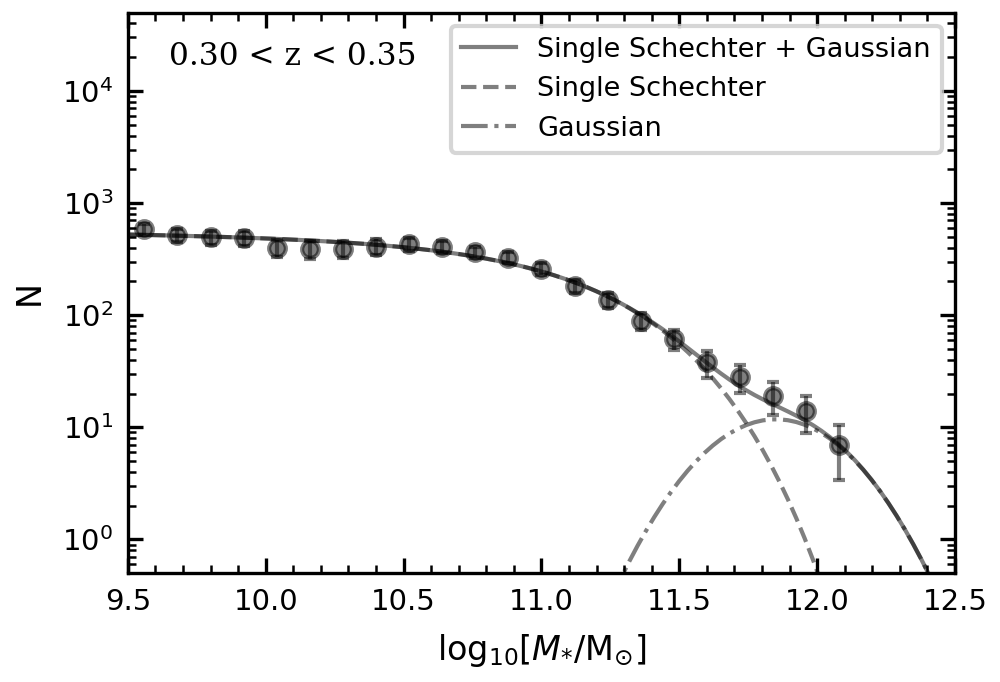}}\\
    \subfloat{\includegraphics[width=0.72\columnwidth]{Schechter+GaussianSMFs/SFG_MCMC_plot_0.35_0.4.png}}
    \subfloat{\includegraphics[width=0.72\columnwidth]{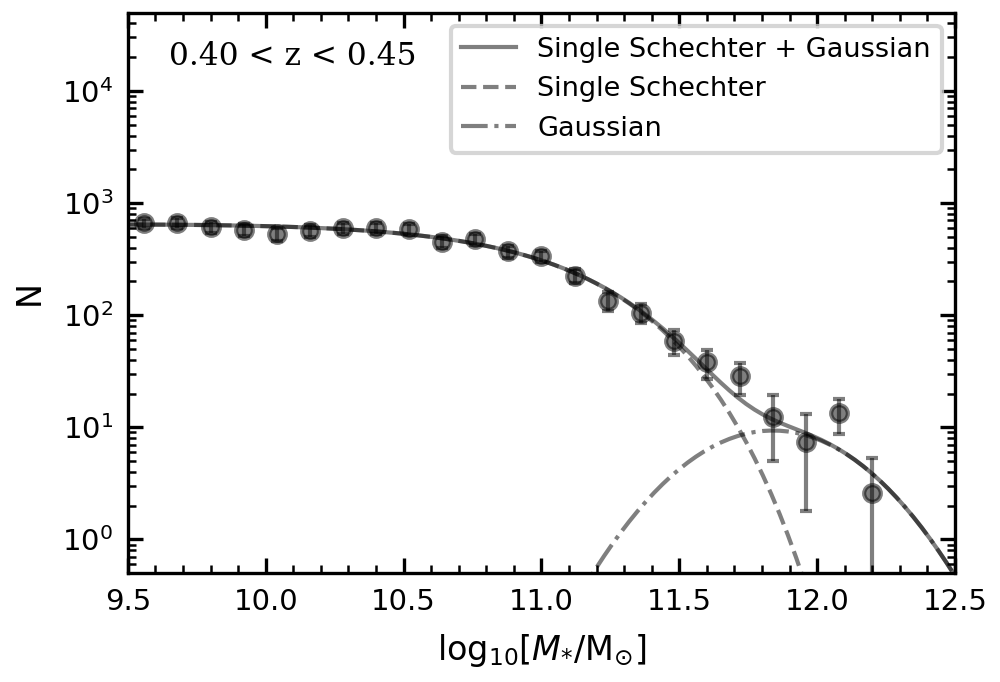}}
    \subfloat{\includegraphics[width=0.72\columnwidth]{Schechter+GaussianSMFs/SFG_MCMC_plot_0.45_0.5.png}}\\
    \subfloat{\includegraphics[width=0.72\columnwidth]{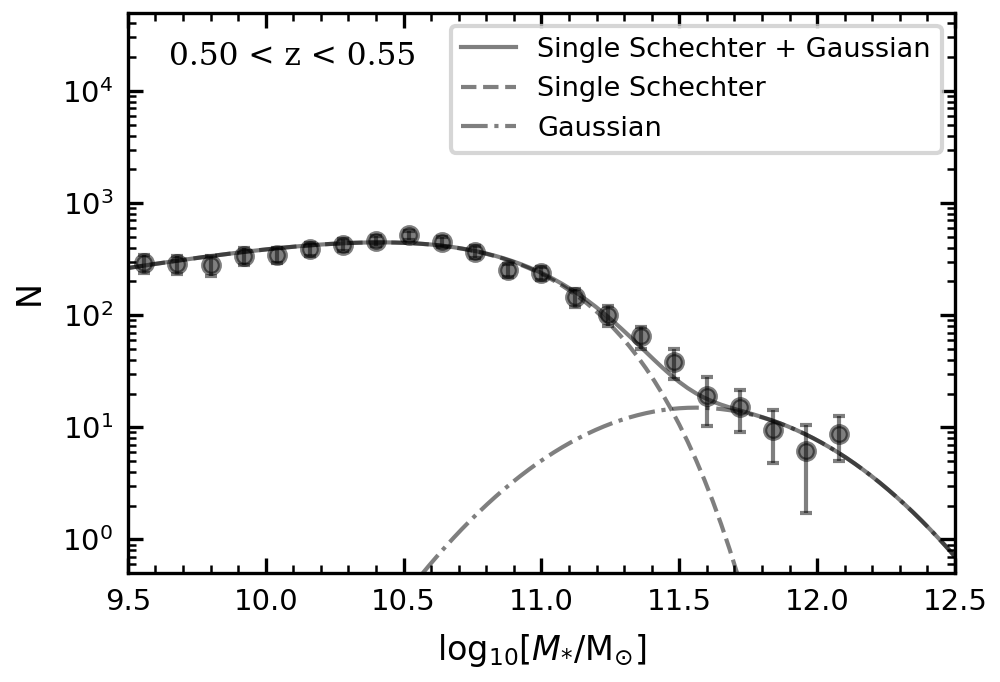}}
    \subfloat{\includegraphics[width=0.72\columnwidth]{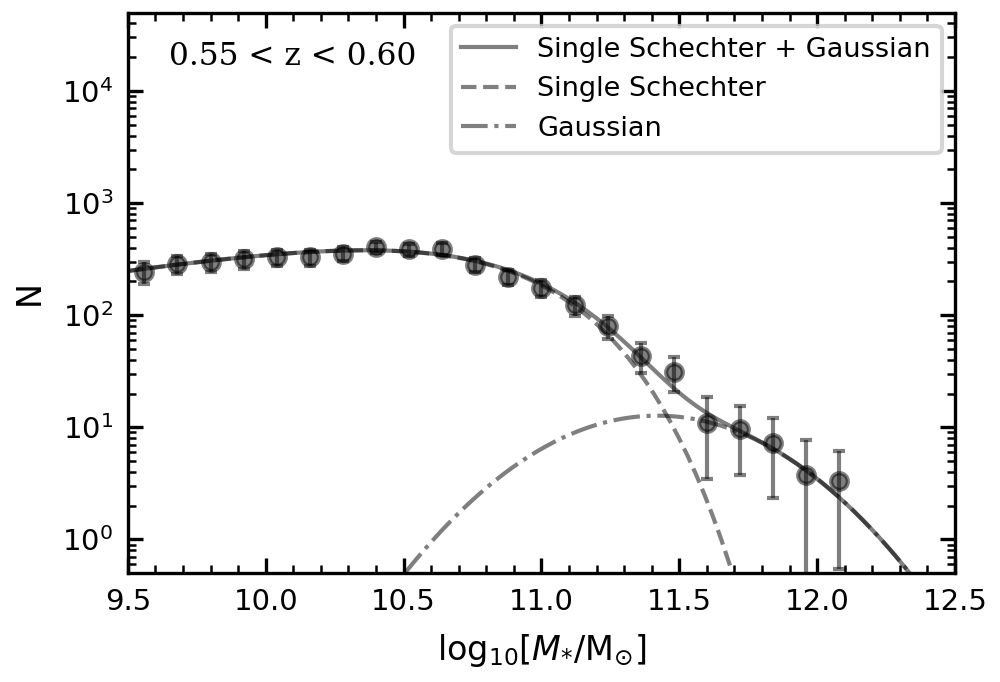}}
    \subfloat{\includegraphics[width=0.72\columnwidth]{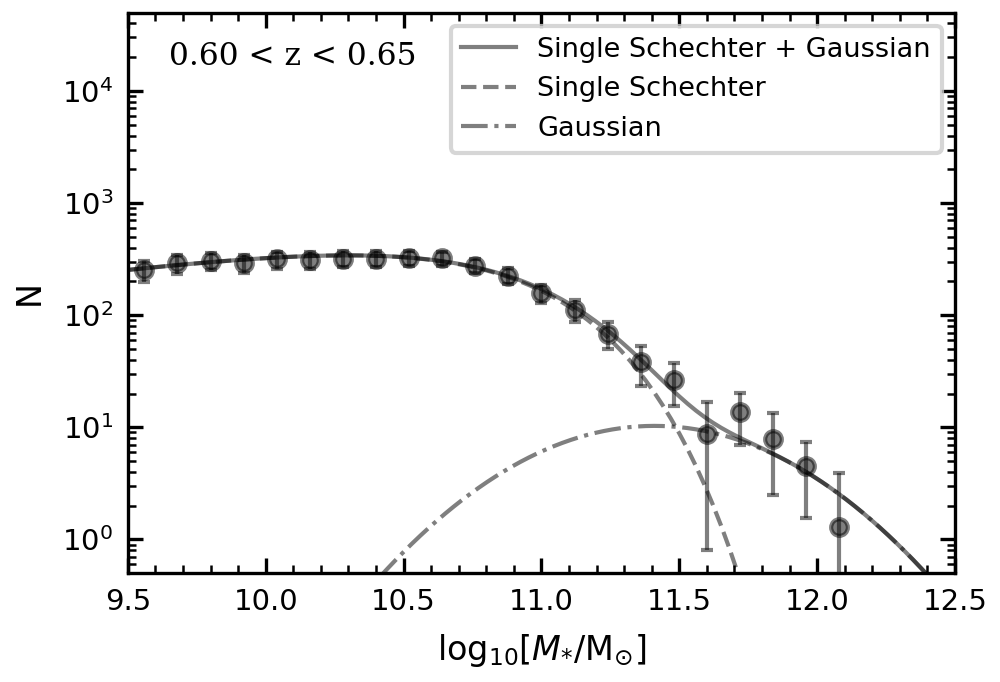}}\\
    \subfloat{\includegraphics[width=0.72\columnwidth]{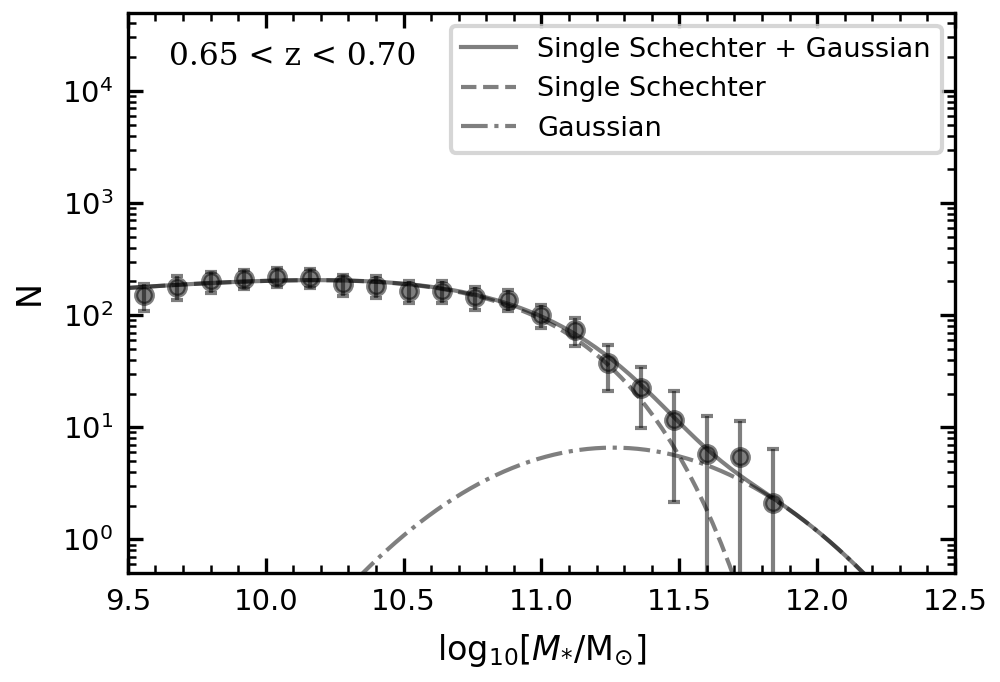}}
    \subfloat{\includegraphics[width=0.72\columnwidth]{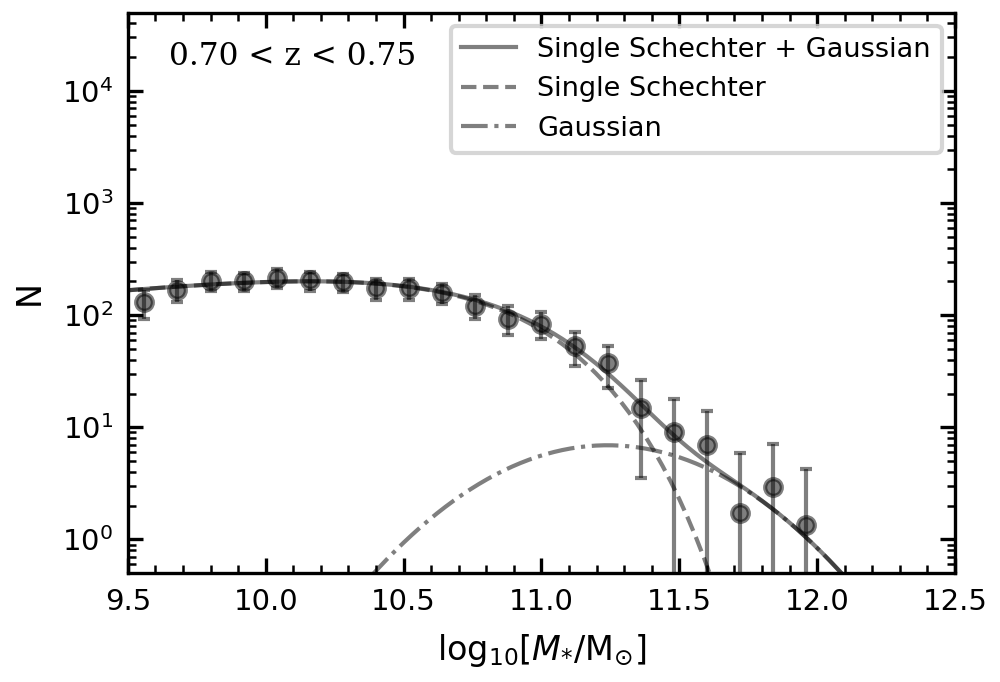}}
    \subfloat{\includegraphics[width=0.72\columnwidth]{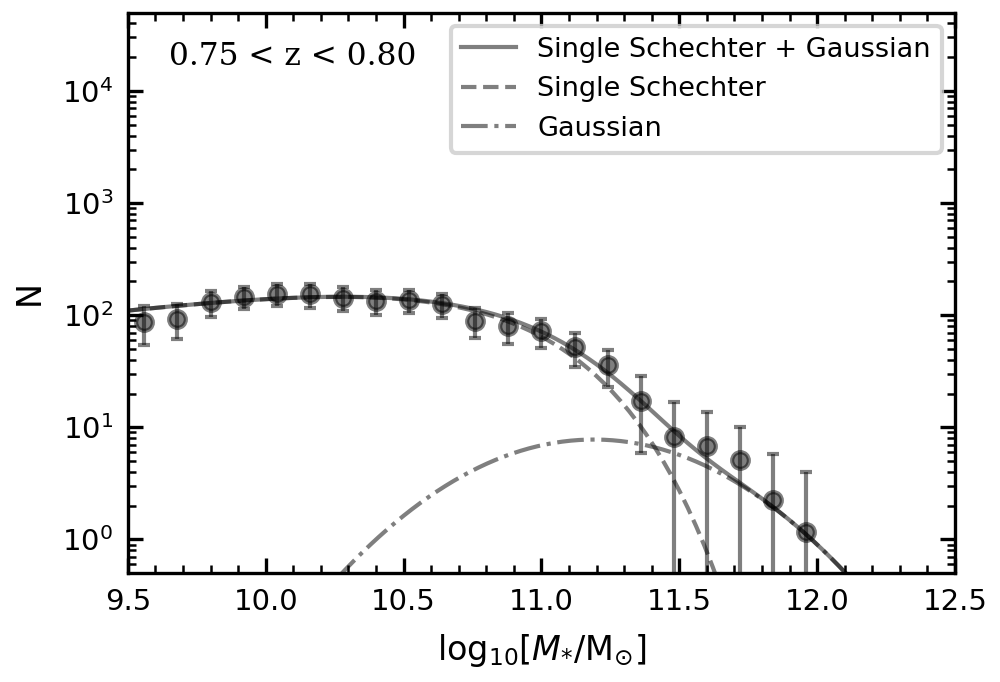}}

    \caption{The composite cluster SMFs for all 12 redshift bins, modelled with a Schechter+Gaussian function. Excess stellar mass at the high-mass end, above the Schechter function, is appropriately described by the Gaussian model. Parameter values for the best-fit Schechter+Gaussian model are presented in Table \ref{tab:Schechter+Gaussian_table}. The width of the Gaussian model component is observed to moderately increase with increasing redshift.}
    \label{fig:Schechter+GaussianSMFs}
\end{figure*}

In Fig. \ref{fig:SinglSchechterSMFs} and \ref{fig:Schechter+GaussianSMFs}, we display the fitted composite cluster SMFs with a Schechter and Schecter+Gaussian model, respectively. Cluster mass subset SMFs are not shown in Fig. \ref{fig:Schechter+GaussianSMFs} since they exhibit very similar trends as observed for the total cluster SMF (see Table \ref{tab:Schechter+Gaussian_table}).

\bsp	
\label{lastpage}
\end{document}